\documentclass[]{jfm}

\usepackage{graphicx}
\usepackage{epstopdf,epsfig}

\usepackage{newtxtext}
\usepackage{newtxmath}
\usepackage{natbib}

\usepackage{longtable} % long table
\usepackage{tabularray} % alternate long table
\usepackage{amsmath}
\usepackage{amssymb,amsbsy,bm}
\usepackage{physics}

\usepackage[dvipsnames]{xcolor}

\usepackage[percent]{overpic} 
\usepackage{tikz}
\usetikzlibrary{arrows.meta}

\usepackage{subfiles}
\usepackage{subfig}

\usepackage{placeins}
\usepackage{mathtools,xparse}

\usepackage{hyperref}
\hypersetup{
    colorlinks = true,
    urlcolor   = blue,
    citecolor  = blue,
    linkcolor=red,
    citecolor=blue,
    filecolor=cyan,
}

\usepackage{cleveref}
\usepackage[normalem]{ulem} % for strike out, i.e. \sout{#}

\crefformat{section}{\S#2#1#3} % see manual of cleveref, section 8.2.1
\crefformat{subsection}{\S#2#1#3}
\crefformat{subsubsection}{\S#2#1#3}

\usepackage{array}

\newcommand{\RomanNumeralCaps}[1]
\newcommand{\vect}[1]{\bm{#1}}

\title{Investigating the parametric dependence of the impact of two-way coupling on inertial particle settling in turbulence}

\author{Soumak Bhattacharjee\aff{1},
Josin Tom\aff{1}\footnote{\small Present address: Research Engineer, Convergent Science, 6400 Enterprise Lane, Madison WI, 53719},
  Maurizio Carbone\aff{2,3},
  Andrew D. Bragg\aff{1}\corresp{\email{andrew.bragg@duke.edu}}
}
\affiliation{
\aff{1}
Department of Civil and Environmental Engineering, Duke University, Durham, NC 27708, USA
\aff{2}
Theoretical Physics I, University of Bayreuth, Universit\"atsstrasse 30,
95447 Bayreuth, Germany
\aff{3}
Max Planck Institute for Dynamics and Self-Organization, Am Fa{\ss}berg 17, 37077 Göttingen, Germany
}

\begin{document}
\maketitle

\begin{abstract}
Tom et al.\ (J.\ Fluid Mech.\ 947, A7, 2022) investigated the impact of two-way coupling (2WC) on particle settling in turbulence. 
For the limited parameter choices explored, it was found that 2WC substantially enhances particle settling compared to the one-way coupled (1WC) case, even at low mass loading $\Phi_m$. 
Moreover, contrary to previous claims, it was demonstrated that preferential sweeping remains the mechanism responsible for the particles settling faster than the Stokes settling velocity in 2WC flows. However, crucial questions remain: 1) how small must $\Phi_m$ be for the effects of 2WC on particle settling to be negligible? 2) does the preferential sweeping mechanism remain relevant in 2WC flows as $\Phi_m$ is increased? To answer these, we explore a much broader portion of the parameter space, and our simulations cover cases where the impact of 2WC on the global fluid statistics ranges from negligible to strong. 
We find that even for $\Phi_m=7.5\times 10^{-3}$, 2WC can noticeably increase the settling for some choices of the Stokes and Froude numbers. 
We also demonstrate that even when $\Phi_m$ is large enough for the global fluid statistics to be strongly affected by the particles, preferential sweeping is still the mechanism responsible for the enhanced particle settling. 
The difference between the 1WC and 2WC cases is that, in the latter the particles are not merely swept around the downward-moving side of vortices, but they also drag the fluid with them as they move down.
\end{abstract}


{\bf MSC Codes }  {\it(Optional)} Please enter your MSC Codes here

\section{Introduction}\label{sec:I}
Turbulent, dispersed, multiphase flows are not yet fully understood due to their complex, multiscale dynamics. The dispersed phase may consist of solid inertial particles, bubbles or gases and their interaction with the carrier fluid depends on a number of factors including their density (relative to the fluid density), their total mass (relative to the total fluid mass) and the nature of the turbulence. Despite their complexity, experimental, computational and theoretical investigations have made significant progress in understand various aspects of such flows, including the dispersion and clustering of particles, droplet breakdown and coalescence, and turbulent modulation \citep[see][for reviews]{balachandar2010,brandt2022,Gustavsson16}. This has led to advances in our understanding of diverse problems including cloud microphysics \citep{pruppacher1997, shaw2003}, aerosol deposition in human lungs \citep{ou2020}, and planetesimal formation \citep{cuzzi2000,birnstiel2016}.

We focus on the problem of the settling velocity of inertial particles in isotropic turbulence for particles that are small compared with the Kolmogorov length scale, and are much denser than the fluid.
Many experimental and numerical studies have observed that such inertial particles settle faster in a turbulent flow than they would in a quiescent fluid \citep{wang1993,aliseda2002,yang2005,good2014,bec14b,rosa16,monchaux2017,momenifar19a,Momenifar19b,dhariwal2019,petersen2019,berk2021}.
Prior to these observations, it had also been shown by \citet{maxey86} that in random flow fields inertial particles settle faster than they would in a quiescent flow. 
\citet{maxey1987} proposed the preferential sweeping mechanism to explain how turbulence (or fluctuations in a random flow) could cause inertial particles to settle faster than they would in a quiescent flow. In particular, Maxey argued that for weakly-inertial settling particles, the combined effects of the structure of the strain-rate and vorticity fields of the flow, particle inertia and gravity cause the particles to be preferentially swept around the downward moving side of vortices in the flow. Hence, the average fluid velocity along their trajectory points down, leading to an enhancement of their settling velocity. Recently, \cite{tom2019} extended the analysis of \citet{maxey1987} to explain how the preferential sweeping mechanism works when the particle inertia is finite, and the role that different turbulent flow scales play in the mechanism. This was called the multi-scale preferential sweeping mechanism and it successfully explained a number of previous results that could not be accounted for by the original analysis of \citet{maxey1987}.

Many of the studies mentioned above focused on the one-way coupled (1WC) regime where the effect of the particles on the flow is ignored, in contrast to the two-way coupled (2WC) regime. 
Other studies have however considered the effect of 2WC on particle settling, including \citet{bosse2006, dejoan2011,monchaux2017,rosa21}. 
These studies all found that 2WC causes the particles to settle faster than the 1WC case. 
The study of \citet{monchaux2017} also emphasized that 2WC can substantially increase the particle settling velocities compared with the 1WC case even in regimes where the particle mass loading is small enough for the effect of the particles on the global fluid statistics to be negligible. 
They also argued that in the presence of 2WC, the preferential sweeping mechanism is not responsible for the enhanced settling speeds due to turbulence. 
Instead, they argued that the dominant contribution comes from a fluid dragging effect, where the particles drag the fluid surrounding them with them as the fall, lowering the drag force on the particles and thereby enhancing their settling velocities. 
\citet{tom2022} investigated this in greater detail, applying the multi-scale preferential sweeping mechanism developed in \cite{tom2019} to the 2WC case. 
In agreement with \citet{monchaux2017}, \citet{tom2022} found that 2WC can substantially increase the particle settling velocities compared with the 1WC case even in regimes where the particle mass loading is small enough for the effect of the particles on the global fluid statistics to be negligible. 
However, in contrast to \citet{monchaux2017}, they demonstrated that preferential sweeping continues to be the mechanism responsible for the enhanced settling speeds due to turbulence. 
In particular they showed that the difference between the 1WC and 2WC cases is that in the 2WC case, the particles are not merely swept around the downward moving side of vortices in the flow but they also drag the fluid down with them in these regions as they fall. 
Hence, the fluid dragging effect does not replace the preferential sweeping mechanism, but acts with it to further enhance the particle settling velocities compared to the 1WC case.

The study of \citet{tom2022} was, however, for a limited portion of the parameter space of the problem.
In particular, while it considered a range of Stokes numbers $St\equiv \tau_p/\tau_\eta\in[0.3,2]$ (where $\tau_p$ is the particle response time and $\tau_\eta$ is the Kolmogorov timescale), it restricted attention to a single Froude number $Fr\equiv u_\eta/(\tau_\eta g)=1$ (where $u_\eta$ is the Kolmogorov velocity scale and $g$ is the gravitational acceleration) and volume fraction $\Phi=1.5\times 10^{-5}$. Due to this, crucial open questions remain. 
In particular, 1) how small must $\Phi$ be for the effects of 2WC on the particle settling to be negligible?, 2) does the preferential sweeping mechanism continue to be relevant in 2WC flows at higher volume fractions? This study aims to answer these crucial questions.

The rest of the paper is organised as follows: In Sec. \ref{sec:PF} we outline the problem of interest, in Sec. \ref{sec:NM} we summarise the numerical methods, in Sec. \ref{sec:R} we present and discuss the observations of our simulations. Finally, in Sec. \ref{sec:C} we summarise our findings.
\section{Problem Formulation}\label{sec:PF}
% \subsection{Background}\label{Background}

We consider a dilute suspension (i.e. the volume-fraction $\Phi_v$ is small enough to ignore particle collisions) of small ($d_p/\eta \ll 1$, where $d_p$ is the particle diameter and $\eta$ is the Kolmogorov length scale), dense ($\rho_p/\rho_f \gg 1$, where $\rho_p$ is the particle density and $\rho_f$ is the fluid density), spherical, inertial particles settling under the force of gravity. Such assumptions are often considered to be suitable for modeling droplet transport in atmospheric clouds \citep{shaw03,grabowski2013}, and dust transport in the atmosphere \citep{RichterBLM2018}.

Assuming that the particle Reynolds number is small, the evolution equation for particles in this regime is given by \citep{maxey1983}
\begin{align}
\ddot{\bm{x}}^p(t)\equiv\dot{\bm{v}}^p(t)=\frac{1}{St\tau_\eta}\Big[\bm{u}(\bm{x}^p(t),t)-\bm{v}^p(t)\Big]+\bm{g},\label{eqn_eom}
\end{align}
where $\bm{x}^p(t)$, $\bm{v}^p(t)$ denote the particle position and velocity, respectively, $\bm{u}(\bm{x}^p(t),t)$ is the undisturbed fluid velocity at the particle position, $\tau_\eta$ is the Kolmogorov time scale, and $\bm{g}$ denotes the gravitational acceleration.

The particles settle in a statistically stationary, homogeneous turbulent flow, which is governed by the incompressible Navier-Stokes equation (written here in rotational form)
\begin{align}
\partial_t \vect{u}  =
\vect{u}\times\vect{\omega}
-\bnabla\left(\frac{P}{\rho_f} + \frac{u^2}{2}\right) + \nu\nabla^2\vect{u} + \vect{F} + \vect{C},\quad \bnabla\bcdot\vect{u} = 0.
\label{eq_NS}
\end{align}
Here, $\vect{u}(\vect{x},t)$ and $\vect{\omega}\equiv\bnabla\times\vect{u}$ are the fluid velocity and vorticity fields respectively, 
$P(\vect{x},t)$ is the pressure, $\nu$ is the dynamic viscosity, $\vect{C}(\vect{x},t)$ is the total momentum feedback of the particles on the carrier fluid,
and $\vect{F}(\vect{x},t)$ is the large-scale forcing required to maintain steady-state turbulence. In our DNS we will consider the case where $\vect{F}$ isotropically forces the flow, such that when $\bm{g}=\bm{0}$ the flow is statistically isotropic, but it is anisotropic for $\bm{g}\neq\bm{0}$ due to the momentum coupling term $\vect{C}$.

We consider the particle motion in both the one-way coupled (1WC) scenario where we set $\vect{C}=\bm{0}$, and the two-way coupled (2WC) scenario where $\vect{C}\neq\bm{0}$. The particle feedback $\vect{C}$ is the sum of the hydrodynamic forces generated by all particles at location $\bm{x}^p(t)=\bm{x}$
\begin{align}
\vect{C}(\vect{x},t)=\frac{1}{\rho_f}\sum_i \bm{f}_i^p(t) \delta(\bm{x}-\bm{x}^p_i(t)), \quad \bm{f}_i^p(t) = -\frac{m_p}{St\tau_\eta}\left[\bm{u}\left(\bm{x}^p_i(t),t\right)-\bm{v}^p_i(t)\right],
\label{eqn_f_feedbk}
\end{align}
where $m_p$ is the mass of each particle,  $\vect{x}^p_i$ and $\vect{v}^p_i$ are the position and velocity of the $i$-th particle, and $\vect{f}^p_i$ is the reaction force generated by the $i$-th particle. As discussed in \citet{tom2022}, when $\vect{C}\neq\bm{0}$ settling particles will generate a mean flow in the direction of $\bm{g}$. 
In order to generate a zero mean-flow velocity in the vertical direction the pressure is accordingly modified, a method that was also used in \citet{maxey2001,bosse2006,monchaux2017,rosa21}.

The ensemble-averaged mean particle settling speed can be obtained from Eq.~\ref{eqn_eom}, assuming statistical homogeneity and stationarity
\begin{align}
\langle{v}_z^p(t)\rangle=\langle{u}_z(\bm{x}^p(t),t)\rangle-St\tau_\eta{g}.\label{eqn_SV}
\end{align}
Here, $St \tau_\eta g$ is the Stokes settling speed (the settling speed the particle would have in a quiescent fluid) and $\langle \bm{u}(\bm{x}^p(t),t) \rangle$ is the ensemble-average vertical fluid velocity at the particle location. 
It is this latter contribution that represents the contribution of turbulence to the settling speed. 
Even when the Eulerian average of the vertical fluid velocity is zero $\langle{u}_z(\bm{x},t)\rangle={0}$, the average along the particle trajectory $\langle{u}_z(\bm{x}^p(t),t)\rangle$ need not be zero because inertial particles will not in general sample the flow field ergodically. 
As a result of this, through the contribution $\langle{u}_z(\bm{x}^p(t),t)\rangle$ turbulence can lead to either enhanced settling ($\langle{v}_z^p(t)\rangle < St\tau_\eta{g}$) or loitering ($\langle{v}_z^p(t)\rangle > St\tau_\eta{g}$).
Results from laboratory experiments, field measurements, and numerical simulations show that for the case of small, dense, inertial particles it is enhanced settling that has been predominantly observed \citep{wang1993, yang2005, good2014, petersen2019,nemes2017, li2021a, li2021b,bec14b,rosa16,ireland2016b,tom2022}.
\section{Direct Numerical Simulations}\label{sec:NM}
\subsection{Numerical method}\label{sec:sim_app}
We adopt an Eulerian-Lagrangian approach to numerically integrate Eq.~\ref{eqn_eom} in conjunction with Eq.~\ref{eq_NS}.
We perform DNS of statistically stationary, homogeneous turbulence in a three-dimensional, periodic domain using a pseudo-spectral scheme on $N^3$ collocation points~\citep{ireland2013}.
A second-order Runge-Kutta time-stepping scheme with an exponential integration is employed for the viscous stress, and a combination of spherical truncation and phase-shifting for the dealiasing schemes.
A seventh-order, B-spline polynomial interpolation scheme (that uses eight points) is used to calculate both the fluid velocities at the particle position and to project the particle momentum feedback onto the fluid at the grid points.
Equation \ref{eqn_eom} is solved using an exponential integrator method \citep{hochbruck2010}, which ensures stability and accuracy for low Stokes numbers particles while allowing the same time step to be used for both the fluid and particle equations of motion. 
We refer the reader to \cite{ireland2013, tom2022} for further details on the numerical solver, and \cite{beylkin95,carbone2019} for the method for computing the momentum coupling term $\bm{C}$, and the interpolation schemes.

\subsection{Simulation parameters}\label{sec:sim_para}

\begin{table}
\centering
\captionsetup{width=\linewidth}
\begin{tabular}{c c c c c c c c c c c} %\hline
$\mathcal{L}$     & $N$       & $\nu$     & $\langle\epsilon\rangle$  & $u_{\eta}$    & $\eta$    & $\dd t$   &  $u'$     &  ${L}$    & $\tau_L$  & $Re_{\lambda}$
\\ \\ %\hline
$2\pi$  & $128$     & $0.005$   &  $0.272$   & $0.192$                   & $0.026$       & $0.002$   & $0.916$  &  $1.456$          & $1.589$  & $88$\\ %\hline
\end{tabular}
\caption{Flow parameters in DNS for the unladen flow. $Re_{\lambda} \equiv u'\lambda/\nu = 2k/\sqrt{5/3\nu \langle\epsilon\rangle}$ is the Taylor microscale Reynolds Number,  $\lambda$ is the Taylor microscale, $\mathcal{L}$ is the domain length, $N$ is the number of grid points in each direction, $\nu$ is the fluid kinematic viscosity, $\langle\epsilon\rangle$ is the mean turbulent kinetic energy dissipation rate, $L$ is the integral length scale, $\eta \equiv (\nu^{3}/\langle\epsilon\rangle)^{1/4} $ is the Kolmogorov length scale, $u' \equiv \sqrt{2\mathcal{K}/3}$ is the r.m.s of fluctuating fluid velocity, $\mathcal{K}$ is the turbulent kinetic energy, $u_{\eta}$ is the Kolmogorov velocity scale, $\tau_L \equiv L/u'$ is the large-eddy turnover time, $\tau_{\eta}$ is the Kolmogorov time scale, $k_{max} = \sqrt{2N/3}$ is the maximum resolved wavenumber, $dt$ is the time step. The small-scale resolution, $k_{max}\eta$ and the total flow kinetic energy measured by $u'$ are approximately constant between the different simulations. These flow statistics are constructed by averaging over the spatial domain and averaging over a time period of $10\tau_L$.\label{tab:Unladen_stats}}
\end{table}
Table \ref{tab:Unladen_stats} summarises the parameters of our DNS, where $\mathcal{L}$ is the domain-length, $N$ is the number of collocation points (the grid-size is thus $N^3$), and $\nu$ is the viscosity. 
These same values are used for all of the 1WC and 2WC DNS cases. 
The table also shows turbulence statistics from the DNS of the unladen flow: the mean dissipation rate $\langle\epsilon\rangle$, 
the root mean square (r.m.s.) fluctuating velocity $u'\equiv\sqrt{2\mathcal{K}/3}$ (where $\mathcal{K}$ is the turbulent kinetic energy), the Kolmogorov velocity and length scales, $u_{\eta}$ and $\eta$, respectively, the integral length scale $L$, and the Taylor microscale Reynolds number $Re_{\lambda} \equiv u'\lambda/\nu$, where $\lambda$ is the Taylor lenghtscale.

For the 1WC and 2WC simulations, the Stokes number $St\equiv\tau_p/\tau_\eta$ and the Froude number $Fr\equiv u_\eta/(\tau_\eta g)$ are defined with respect to the unladen fluid statistics, and we consider Stokes numbers $St=0.3$, $0.5$, $0.7$, $1$, $2$ (referred to as $\rm{St}_1$, $\rm{St}_2$, $\rm{St}_3$, $\rm{St}_4$, and $\rm{St}_5$, respectively), each of which is considered for three different Froude numbers $Fr=0.3$, $1$, and $3$. 
The particle mass-loading $\Phi_m$ is related to the volume-fraction $\Phi$ through $$\Phi_m=(\rho_p/\rho_f)\Phi=(\rho_p/\rho_f) \pi d_p^3 N_p/6,$$ and we consider particles with constant density $\rho_p/\rho_f= 5000$, the same as that used in \citet{bosse2006,monchaux2017,tom2022}. 
Since $\rho_p/\rho_f$ is fixed, varying $\Phi_m$ corresponds to varying $\Phi$, and therefore we will often refer to the effect of varying $\Phi$ even though strictly speaking the momentum coupling depends on $\Phi_m$ and not simply $\Phi$. 
For each choice of $St$ and $Fr$, we consider $\Phi= 1.5~\times~10^{-6}$, $7~\times~10^{-6}$, $1.5~\times~10^{-5}$, $7~\times~10^{-5}$, and $1.5~\times~10^{-4}$ which we refer to by $\Phi_1,\Phi_2,\Phi_3,\Phi_4,\Phi_5$, respectively. 
However, due to computational resources, we were unable to complete the run with $St=2, Fr=0.3, \Phi=1.5~\times~10^{-4}$. 
This makes for a total of 59 different DNS, and due to this we are restricted to cases where the unladen flow has $Re_{\lambda} \simeq 88$. 
The present study significantly expands on the study of \citet{tom2022} where only $Fr=1$ and $\Phi= 1.5~\times~10^{-5}$ were considered, for the same $St$ and $Re_\lambda$ values.

\subsection{Simulation approach}\label{sec:SimApp_PP}

DNS of the unladen flow were run for approximately 20 $\tau_L$ until a statistically stationary state was reached. 
For the 1WC runs, the DNS were initialized using the fluid data obtained at the final time step of the unladen DNS, and using random initial particle positions and initial particle velocities equal to the local fluid velocity plus the Stokes settling velocity. 
For the 2WC simulations, the DNS were initialized using the fluid and particle data obtained at the final time step of the corresponding 1WC DNS (i.e. having the same $St, Fr, \Phi$). 
For both the 1WC and 2WC simulations, we collected statistics over a period of 60 $\tau_L$ to ensure reasonable statistical convergence (this is a smaller window than that used in \cite{tom2022} but tests showed that it was sufficient). 
A table presenting some of the key fluid statistics and parameters from all of the 2WC runs is provided in the Appendix for reference.

\section{Results}\label{sec:R}

\subsection{Discussion on fluid statistics} \label{subsec:fluid_stats}

% \begin{figure}
% \centering
% \captionsetup{width=\linewidth}
% \vspace{0mm}	
% \subfloat
% {\begin{overpic}
%     [trim = 0mm -5mm 0mm 5mm,
%     scale=0.7,clip,tics=20]{Figures/EQp_Qn_diff.eps}
%     \put(-3,78){(a)}
%     \put(-3,50){(b)}
%     \put(-3,22){(c)}
% \end{overpic}} 
% \caption{}
% \label{fig:EQp_Qn_diff}
% \end{figure}

% -------------
\begin{figure}
\centering
\captionsetup{width=\linewidth}
\subfloat
{\begin{overpic}
    [trim = 0mm -5mm -10mm 0mm,
    scale=0.65,clip,tics=20]{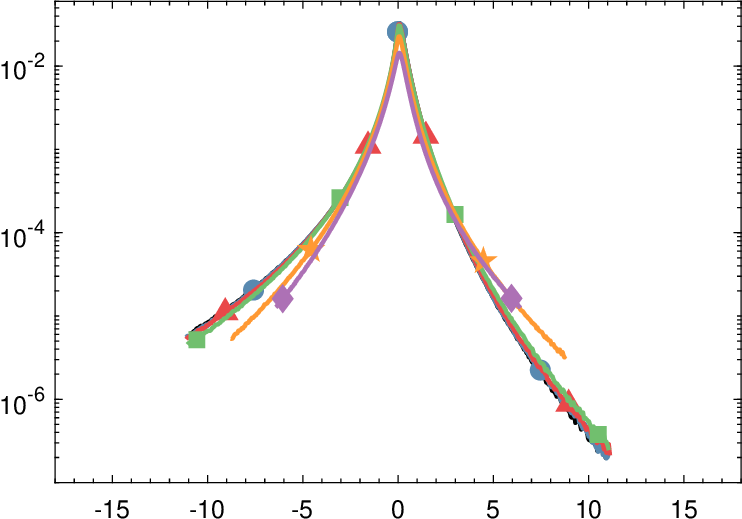} %Eul_Q_Phi_3042.eps
    %\put(20,59.5){{1WC}} \put(42,59.5){$\Phi_2$} \put(69,59.5){$\Phi_4$}
    %\put(21,50){\footnotesize{$\Phi_1$}} \put(42,50){$\Phi_3$} \put(69,50){$\Phi_5$}
    % \put(45,0){$Q$} 
    \put(12,60){(a)}
    \put(-5,25){\rotatebox{90}{$\sigma_\mathcal{Q}\mathcal{P}(\mathcal{Q}/\sigma_\mathcal{Q})$}}
    \put(95,10){
    {\begin{overpic}
        [trim = 10mm 10mm 0mm 0mm,
        scale=0.6,clip,tics=20]{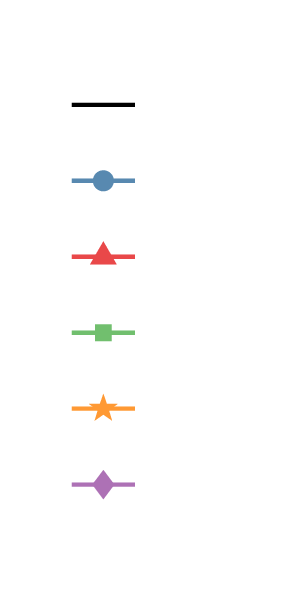}
        \put(18,7){$\Phi_5$ 2WC} 
        \put(18,21.5){$\Phi_4$ 2WC} 
        \put(18,36){$\Phi_3$ 2WC} 
        \put(18,50){$\Phi_2$ 2WC} 
        \put(18,63.5){$\Phi_1$ 2WC}
        \put(18,78){1WC}
    \end{overpic}} 
    }
\end{overpic}} \\
\subfloat
{\begin{overpic}
    [trim = 0mm -5mm -10mm 0mm,
    scale=0.65,clip,tics=20]{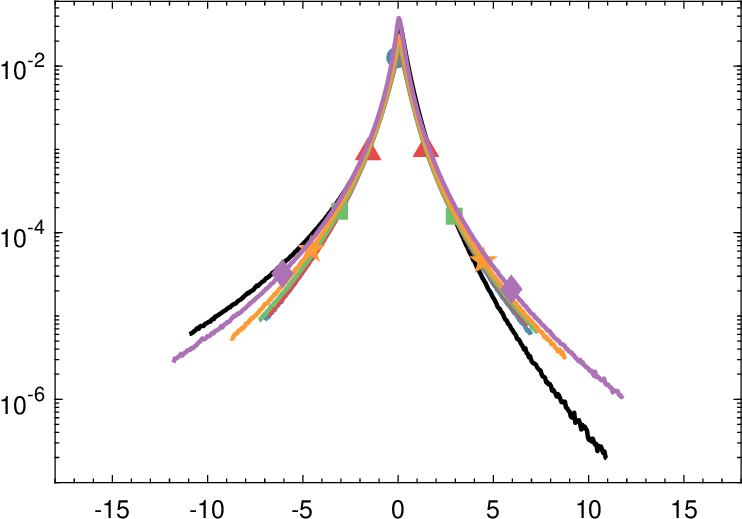} % Eul_Q_St_4143 %Eul_Q_St_4143.eps}
    \put(12,60){(b)}
    % \put(22,63.5){{1WC}} \put(46,63.5){$\rm{St}_2$} \put(69,63.5){$\rm{St}_4$}
    % \put(22,53){\footnotesize{$\rm{St}_1$}} \put(46,53){$\rm{St}_3$} \put(69,53){$\rm{St}_5$}
    \put(45,0){$\mathcal{Q}/\sigma_\mathcal{Q}$}
    \put(-5,25){\rotatebox{90}{$\sigma_\mathcal{Q}\mathcal{P}(\mathcal{Q}/\sigma_\mathcal{Q})$}} 
    \put(95,10){
    {\begin{overpic}
        [trim = 10mm 10mm 0mm 0mm,
        scale=0.6,clip,tics=20]{Figures/leg_Eul_Q.eps}
        \put(18, 7){$St_5$ 2WC} 
        \put(18,21.5){$St_4$ 2WC} 
        \put(18,36){$St_3$ 2WC} 
        \put(18,50){$St_2$ 2WC} 
        \put(18,63.5){$St_1$ 2WC}
        \put(18,78.5){1WC}
    \end{overpic}} 
    }
\end{overpic}} 
\caption{Plot of $\mathcal{P}(\mathcal{Q})$ for 2WC simulations with (a) fixed $St_4=1$, $Fr=0.3$ and varying $\Phi$, and 
(b) fixed $\Phi_4=7\times 10^{-5}$ and $Fr=0.3$ and varying $St$. Black solid lines show $\mathcal{P}(\mathcal{Q})$ for the 1WC simulations, which is the same as for the unladen flow. $\sigma_{\mathcal{Q}}$ denotes the standard deviation of $Q$.}
\label{fig:Eul_PDF_Q}
\end{figure}
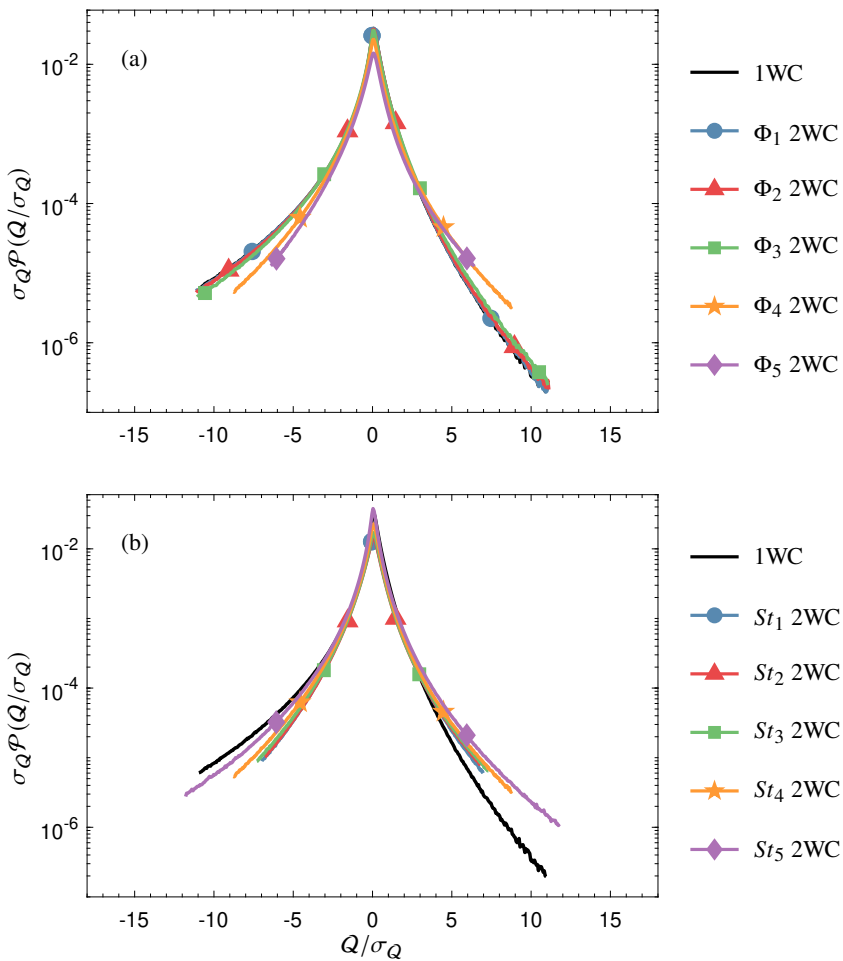

% -------------

Before we explore the parametric dependence of the impact of 2WC on particle settling in turbulence, it is informative to first consider how 2WC affects the global flow statistics. We first consider the probability density function (PDF) $\mathcal{P}(\mathcal{Q})\equiv\langle\delta(Q(\bm{x},t)-\mathcal{Q}\rangle$, where $\bm{x}$ denotes a fixed point in the flow, $Q$ is the second-invariant of the velocity gradient tensor $\bm{A}$
\begin{equation}
    Q \equiv \bm{S:S}-\bm{R:R}; \quad \bm{S}\equiv(\bm{A}+\bm{A}^\top)/2, \quad \bm{R}\equiv(\bm{A}-\bm{A}^\top)/2,
\end{equation}
and $\mathcal{Q}$ is the sample-space coordinate. 
In Fig. \ref{fig:Eul_PDF_Q} we show the results for $\mathcal{P}(\mathcal{Q})$ for the 2WC simulations for varying $St, \Phi$ at $Fr=0.3$, and the corresponding 1WC result (which is the same as for the unladen flow) is also shown for reference. 
Here, $\mathcal{Q}$ has been normalised using the standard-deviation of $Q$ (referred to as $\sigma_{\mathcal{Q}}$) so that the PDFs are in standard form, enabling us to compare the shape of the PDFs for the different cases. 
$\mathcal{P}(\mathcal{Q})$ for the 1WC simulations, which is the same as for the unladen flow, are shown in black solid lines.
The mean value of $\mathcal{Q}$ for 1WC and for all the 2WC simulations is zero, and for the 1WC flows, $\mathcal{P}(\mathcal{Q})$ is negatively skewed due to the vorticity field being more intermittent than the strain-rate field. 
Fig. \ref{fig:Eul_PDF_Q} (a) shows that $\mathcal{P}(\mathcal{Q})$ is only weakly affected by 2WC when $\Phi \leq \Phi_3$, which was also observed in \citet{tom2022} for $\Phi_3$. 
For larger $\Phi$, $\mathcal{P}(\mathcal{Q})$ becomes more symmetric, with the probability of the tail for $\mathcal{Q}<0$ reducing, while it increases for $\mathcal{Q}>0$. 
This means that 2WC is suppressing the probability of regions of strong enstrophy and enhancing  the probability of regions of intense straining motions. 
Figure \ref{fig:Eul_PDF_Q} (b) shows the $St$ dependence of $\mathcal{P}(\mathcal{Q})$ for volume fraction $\Phi_4$. 
We see that changing the particle inertia has in general little effect in altering the skewness of $\mathcal{P}(\mathcal{Q})$, but increases the probability of high-strain and high-rotation regions alike.  
%Beyond this, we see that normalised probabilities become less peaked around the origin. 
%The effect becomes more pronounced with \textcolor{red}{increasing $St$}, and decreasing $Fr$.
%\textcolor{red}{Point N: Is this redundant:} \textit{This is because, in this parameter regime, the tails of $P(Q)$ become heavier as well as the energy dissipation rate $\epsilon$ increases manifold.}
% -------------
\begin{figure}
\centering
\captionsetup{width=\linewidth}
\vspace{0mm}	
% Figure
% St(Fr) vs St, const. Phi
\subfloat
% {\begin{overpic}
%     [trim = 5mm 5mm 0mm 5mm,
%     scale=0.7,clip,tics=20]{Figures/epsilon.eps}
%     \put(5,70){(a)}
%     \put(5,46){(b)}
%     \put(5,24){(c)}
% \end{overpic}} 
{\begin{overpic}
    [trim = 0mm 0mm -10mm 0mm,
    scale=0.5,clip,tics=20]{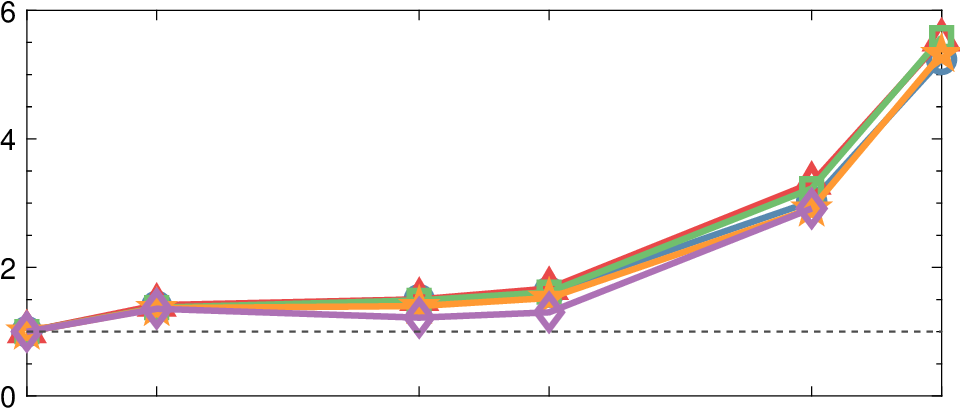} %PQpos-PQneg.eps
    \put(5,32){(a)}
    \put(-4,20){\rotatebox{90}{$\epsilon$}}
\end{overpic}} \\
\subfloat
{\begin{overpic}
    [trim = 0mm 0mm -10mm 0mm,
    scale=0.5,clip,tics=20]{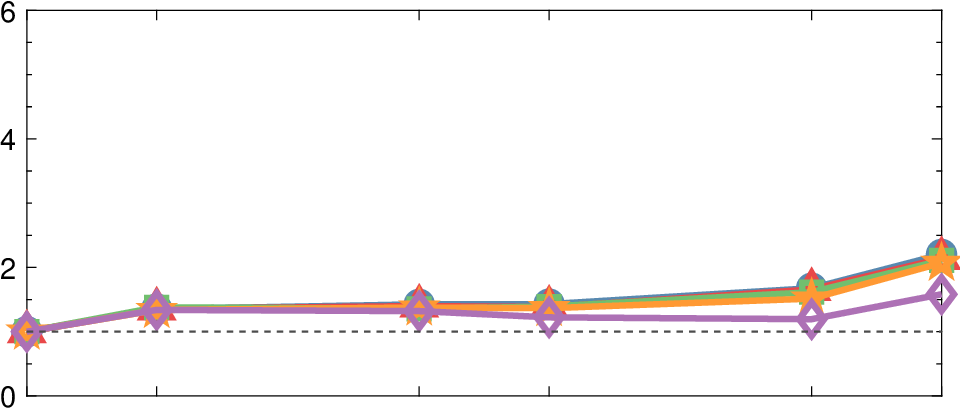}
    \put(5,32){(b)}
    \put(-4,20){\rotatebox{90}{$\epsilon$}}
\end{overpic}} \\
\subfloat
{\begin{overpic}
    [trim = 0mm -4mm -10mm 0mm,
    scale=0.5,clip,tics=20]{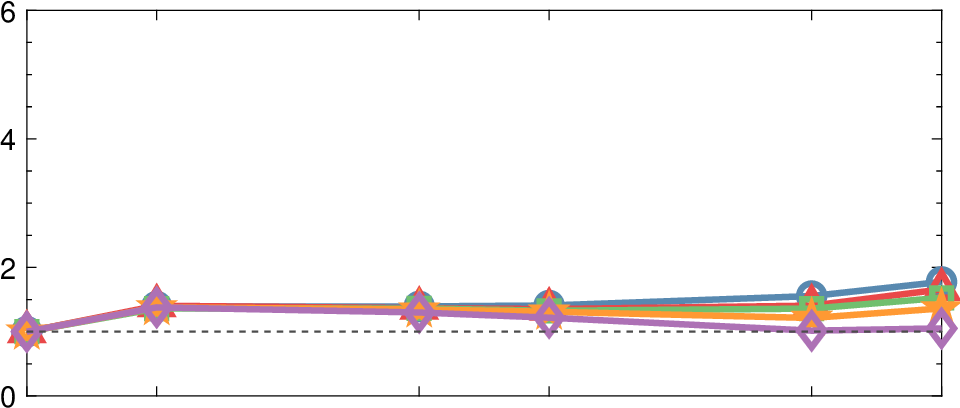}
        \put(5,32){(c)}
        \put(-4,20){\rotatebox{90}{$\epsilon$}}
        \put(2,-0.5){1WC} \put(14.3,-0.5){$\Phi_1$} \put(40,-0.5){$\Phi_2$} \put(52.55,-0.5){$\Phi_3$} \put(78.3,-0.5){$\Phi_4$} \put(91,-0.5){$\Phi_5$}
        \put(96,40){
        {\begin{overpic}
            [trim = 20mm 110mm 60mm 20mm,
            scale=0.6,clip,tics=20]{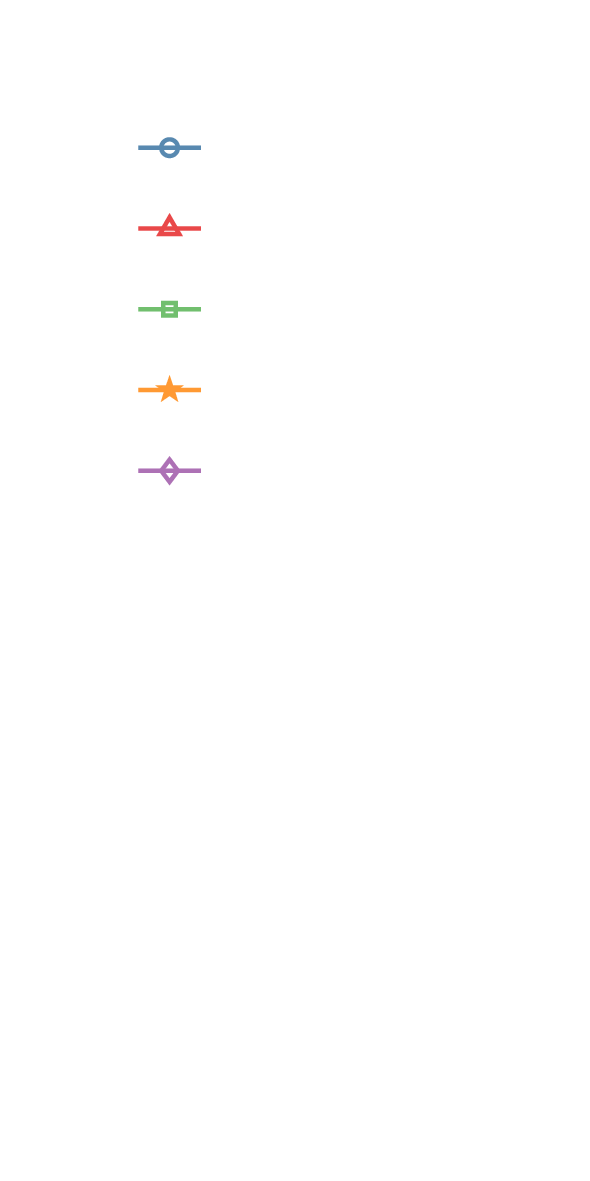}
            \put(21,12.5){$St_5$} 
            \put(21,32.0){$St_4$} 
            \put(21,51.5){$St_2$} 
            \put(21,71.0){$St_2$}
            \put(21,91.0){$St_1$}
        \end{overpic}} 
        }
\end{overpic}}
\caption{Plots showing the turbulent kinetic energy (TKE)  dissipation rate $\epsilon$, normalised by corresponding values from the 1WC runs, for 1WC and 2WC simulations for different volume fractions $\Phi$, and for particles with different $St$. 
Lines in blue, red, green, yellow, and purple correspond to particles with $St_1=0.3$, $St_2=0.5$, $St_3=0.7$, $St_4=1$, and $St_5=2$ respectively. Panels (a), (b) and (c) correspond to simulations with $Fr=0.3$, $1$, and $3$, respectively.
}
\label{fig:Eul_edr}
\end{figure}
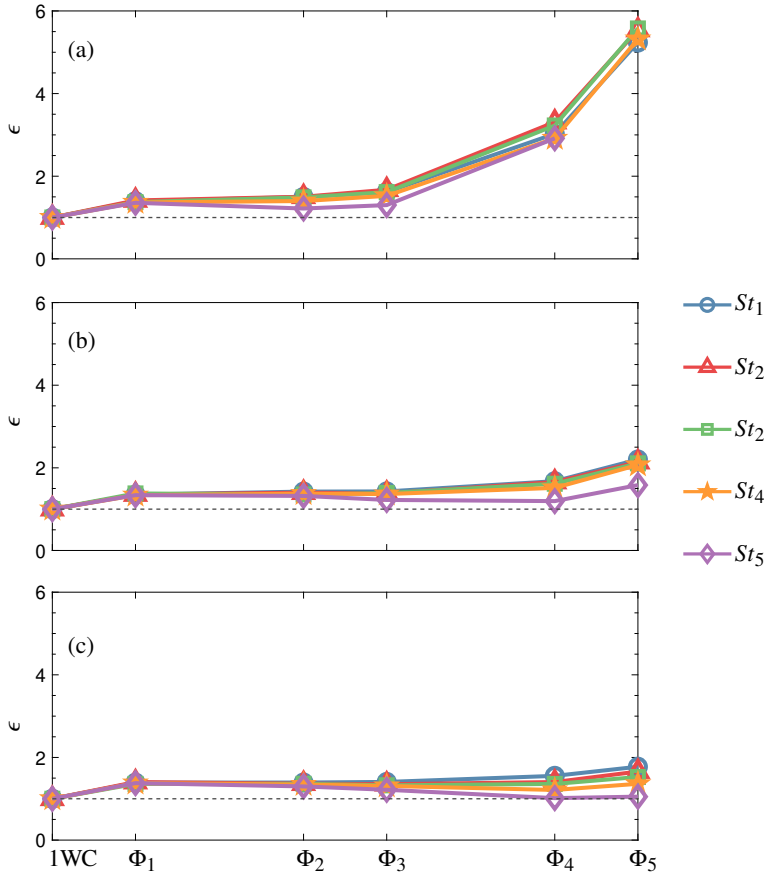
% -------------

Figure \ref{fig:Eul_edr} shows the turbulent kinetic energy (TKE) dissipation rate $\epsilon$ for the different 2WC simulations, normalised by the value of $\epsilon$ from the 1WC runs.
We observe that $\epsilon$ does not change appreciably for small $\Phi$, and shows negligible dependence on the Stokes number $St$. 
However, for volume fractions $\Phi>\Phi_3$, $\epsilon$ generally increases with increasing $\Phi$ (especially in simulations where the Froude number is $Fr=0.3$) and reduces slightly with increasing $St$. 
It is important to note, however, that these results depend on the type of forcing used in the DNS. 
In the present simulations, the flow is forced by keeping the flow TKE constant by re-injecting the TKE that is dissipated back into the flow at the lowest wavenumbers, see \citet{ireland2013}. 
In \citet{carbone2019} where it was the energy injection rate that was controlled, $\epsilon$ was found to decrease with increasing $\Phi$ in 2WC DNS with $Fr=\infty$.

% -------------
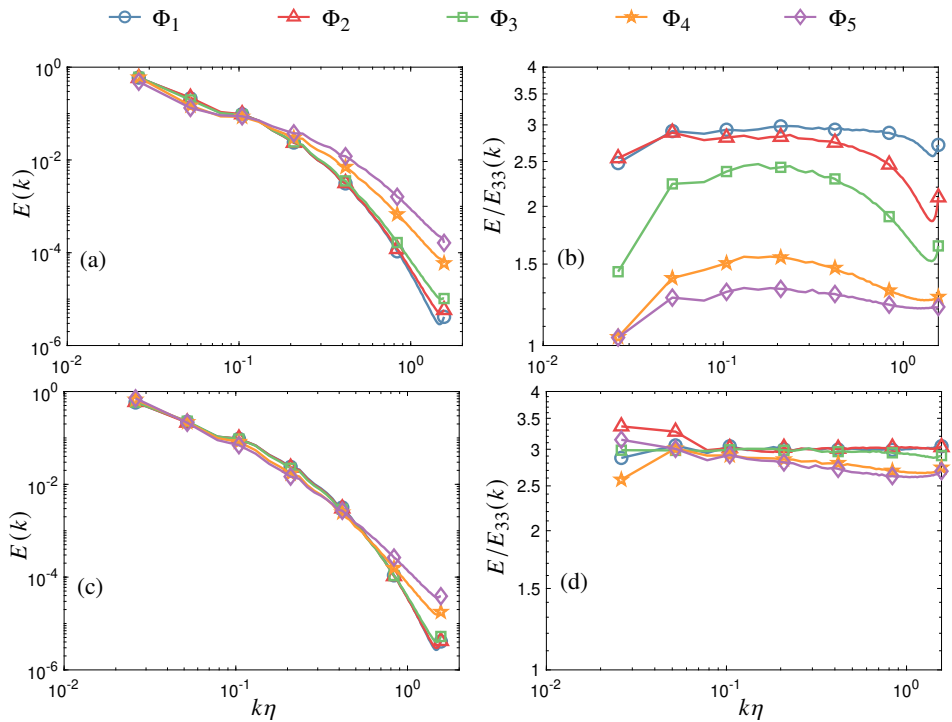
\begin{figure}
\centering
\captionsetup{width=\linewidth}
\vspace{0mm}	
\subfloat
{\begin{overpic}
    [trim = 0mm 0mm -15mm -20mm,
    scale=0.45,clip,tics=20]{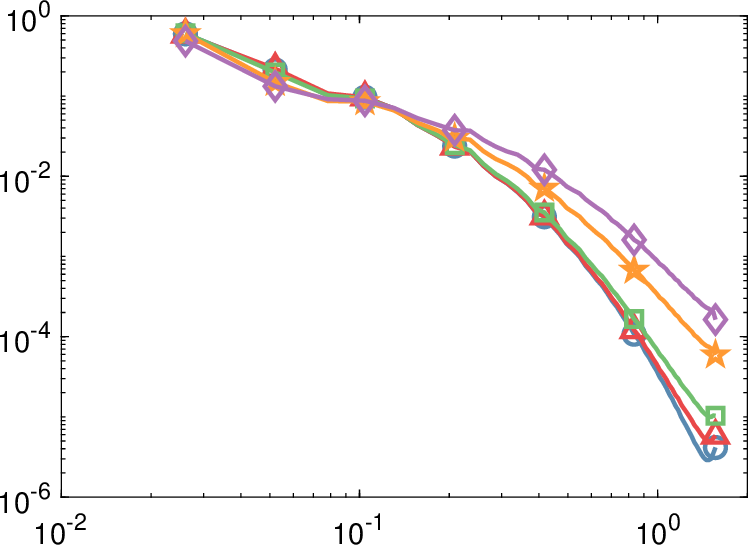}
    \put(10,22){(a)}
    \put(-4,32){\rotatebox{90}{\footnotesize{$E(k)$}}}
    % \put(45,0){$k$} 
    \put(5,70){
        \begin{overpic}
            [trim = 50mm 20mm 0mm 20mm,
            scale=0.4,clip,tics=20]{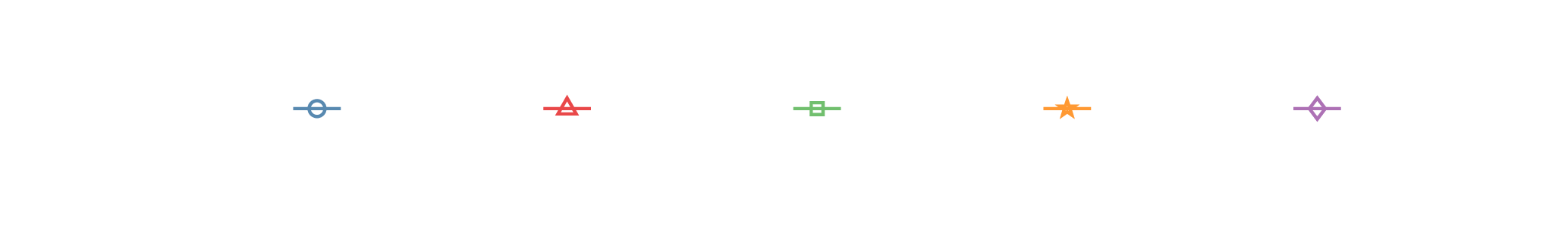}
            \put(10,1){{$\Phi_1$}}
            \put(28.7,1){{$\Phi_2$}}
            \put(47.2,1){{$\Phi_3$}}
            \put(66.2,1){{$\Phi_4$}}
            \put(84.5,1){{$\Phi_5$}}
        \end{overpic}}
\end{overpic}} 
{\begin{overpic}
    [trim = 0mm 0mm 0mm -20mm,
    scale=0.45,clip,tics=20]{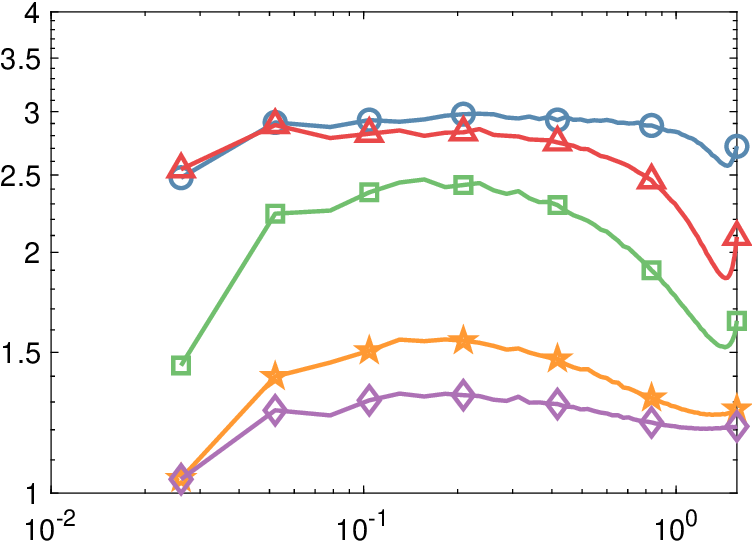}
    \put(10,25){(b)}
    \put(-7,32){\rotatebox{90}{\footnotesize{$E/E_{33}(k)$}}}
    % \put(45,0){$k$} 
\end{overpic}} \\
{\begin{overpic}
    [trim = 0mm -5mm -15mm -2mm,
    scale=0.45,clip,tics=20]{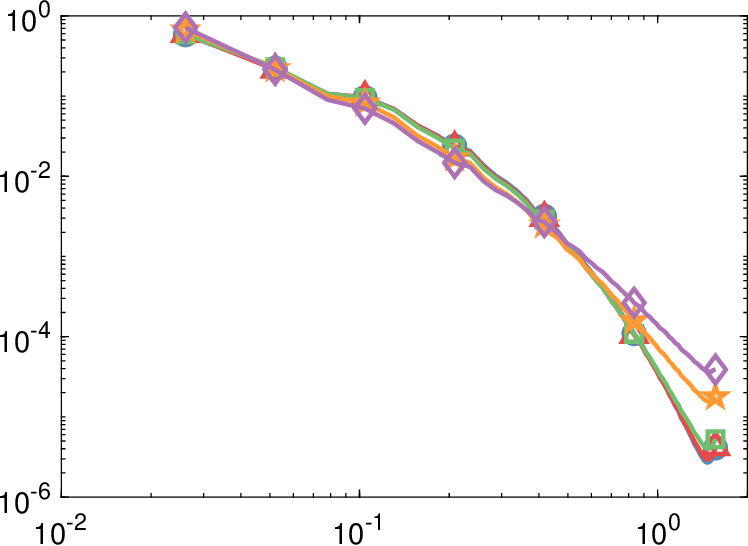}
    \put(10,25){(c)}
    \put(-4,32){\rotatebox{90}{\footnotesize{$E(k)$}}}
    \put(46,0){$k \eta$} 
\end{overpic}} 
{\begin{overpic}
    [trim = 0mm -5mm 0mm -2mm,
    scale=0.45,clip,tics=20]{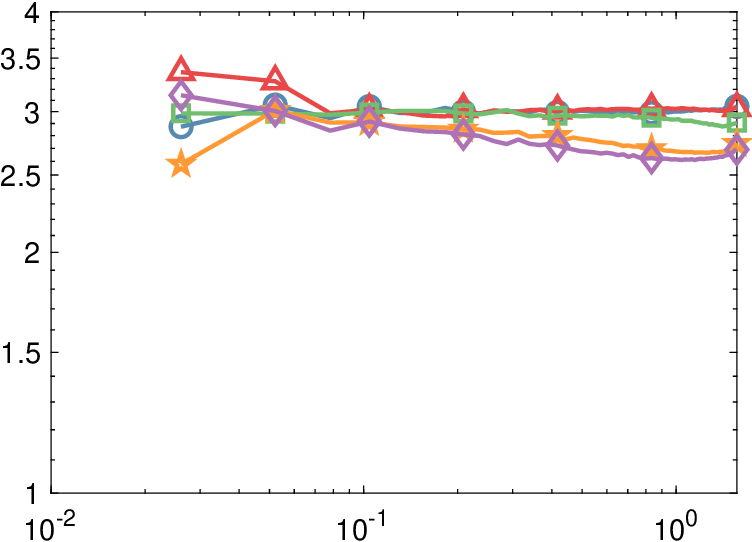}
    \put(10,29){(d)}
    \put(-7,32){\rotatebox{90}{\footnotesize{$E/E_{33}(k)$}}}
    \put(51,0){$k \eta$} 
\end{overpic}} 
\caption{(Turbulent kinetic energy  spectrum $E(k)$ versus the wave-number $k$ for 2WC simulations with $St=1$ particles, for different volume fractions for fixed $Fr$ numbers (a) $Fr=0.3$, (c) $Fr=3.0$.
Ratio of the turbulent kinetic energy spectrum $E(k)$ to the spectrum of the vertical component of the turbulent kinetic energy $E_{33}(k)$ versus the wave-number $k$ for 2WC simulations with $St=1$ particles for different volume fractions for fixed $Fr$ numbers (b) $Fr=0.3$, (d) $Fr=3.0$.}
\label{fig:Eul_spec}
\end{figure}
% -------------

We now consider the energy spectrum in order to consider the impact of 2WC over the entire range of scales in the flow. 
Figure~\ref{fig:Eul_spec} shows (a) $E(k)$ the turbulent kinetic energy spectrum  and (b) $E(k)/E_{33}(k)$ the ratio of the total to the vertical component of the turbulent kinetic energy spectrum versus the wave-number $k$, for 2WC simulations with $St=1$, $Fr=0.3$, and for all volume fractions $\Phi$. 
The bottom panel in Fig.~\ref{fig:Eul_spec} shown corresponding plots of $E(k)$ (c) and $E(k)/E_{33}(k)$ for 2WC simulations with $St=1$, $Fr=3.0$.
It is evident from Fig.~\ref{fig:Eul_spec} (a) that at the highest volume fractions ($\Phi_4$, $\Phi_5$), the energy content at high wavenumbers increases significantly. 
This increase in the high-wavenumber energy content is reflected in the corresponding increase in the turbulent kinetic energy dissipation rate observed in Fig.~\ref{fig:Eul_edr} (a).
There is also a corresponding reduction in the energy at intermediate wavenumbers since in our DNS the TKE is held constant across all the cases due to the forcing scheme used (this behavior at intermediate wavenumbers need not occur for alternative forcing schemes). 
Figure~\ref{fig:Eul_spec} (b) shows that the fraction of the kinetic energy contained in the vertical motions of the flow increases at all wavenumbers, and $E(k)/E_{33}(k)$ approaches one as $\Phi$ increases. 
This occurs due to the transfer of the potential energy of the particles to the vertical component of the kinetic energy, with the amount of potential energy increasing with increasing $\Phi$.
The energy build in the high-wavenumber modes and increase in the share of vertical fluctuations is less pronounced in simulations with $Fr=3.0$; correspondingly, the increase in energy dissipation rate with increasing $\Phi$ is also less pronounced (Fig.~\ref{fig:Eul_edr} (c)).
\subsection{Settling enhancement} \label{subsec:unfilteredsv}

We now consider the settling velocity enhancement, $\langle u_z(\bm{x}^p(t),t)\rangle$, and its dependence on $Fr$, $St$. 
In the limit $Sv\to\infty$ and for a finite Reynolds number flow (so that the range of fluid velocity scales is finite), to leading order, the particles settle in vertical paths and $\langle u_z(\bm{x}^p(t),t)\rangle=0$. 
In the opposite limit $Sv\to0$, $\langle u_z(\bm{x}^p(t),t)\rangle=0$ also occurs because the symmetry breaking effect responsible for generating $\langle u_z(\bm{x}^p(t),t)\rangle\neq0$ (namely gravitation settling) has vanished. 
As a result, $\langle u_z(\bm{x}^p(t),t)\rangle$ is expected to obtain a maximum value for some intermediate value $0<Sv<\infty$, with the value of $Sv$ at which the maximum occurs depending on the other flow parameters. \cite{tom2019} discussed this in detail from the perspective of the multi-scale preferential sweeping mechanism that they developed.

% -------------
% s.v.e (Fr) vs St
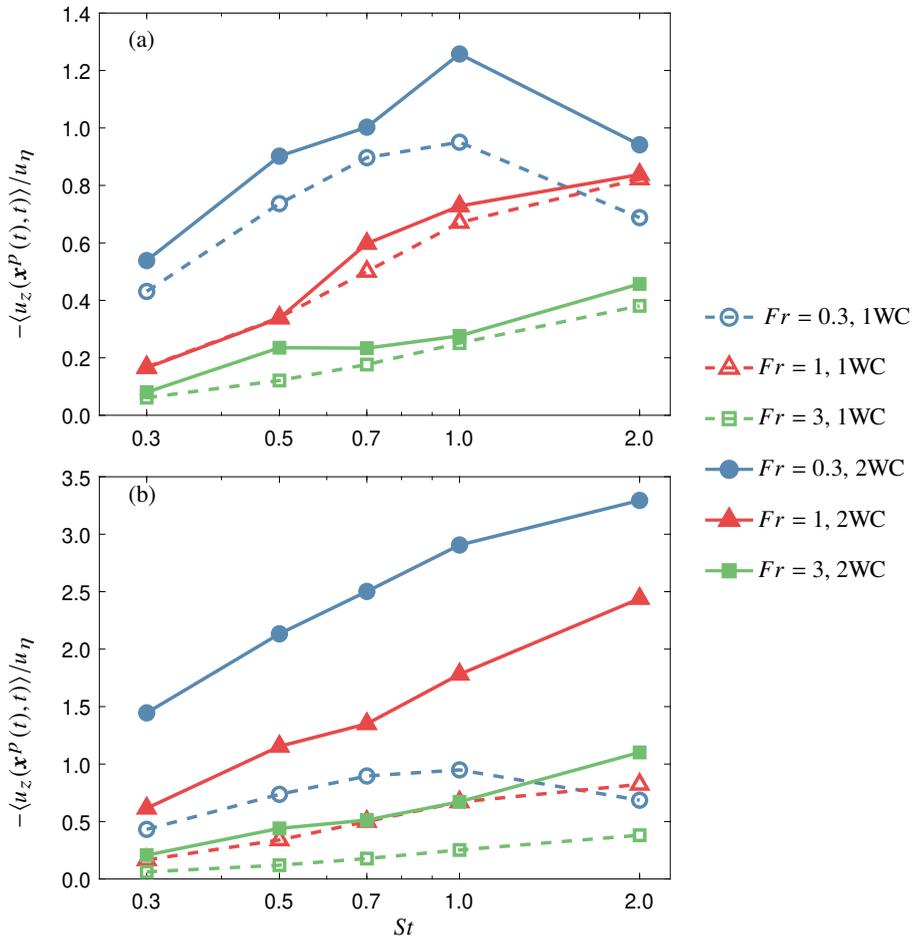
\begin{figure}
\centering
\captionsetup{width=\linewidth}
\vspace{0mm}	
\subfloat	
    {\begin{overpic}
        [trim = 0mm 0mm -11mm -0mm,
        scale=0.65,clip,tics=20]{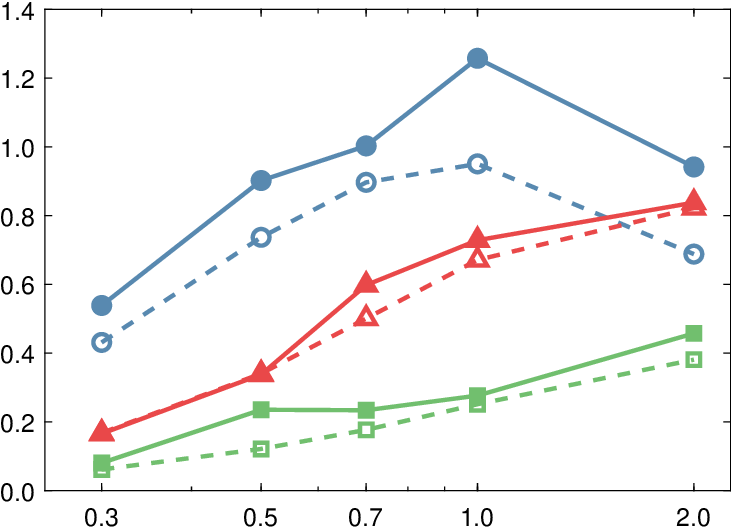}
        % \put(54,0){$St$}
        \put(-8,16){\rotatebox{90}{$-\langle u_z(\bm{x}^p(t),t)\rangle / u_{\eta}$}}
        \put(10,60){(a)} %\put(14.5,49.75){2WC}
    \end{overpic}}
    % \begin{tikzpicture}[overlay]
    %     \node[draw,text width=2.2cm, gray] at (-1.9,0.99) {\small{$\Phi_1=1.5\times 10^{-6}$}};
    % \end{tikzpicture}
    
\subfloat
    {\begin{overpic}
        [trim = 0mm -5mm -11mm 0mm,
        scale=0.65,clip,tics=20]{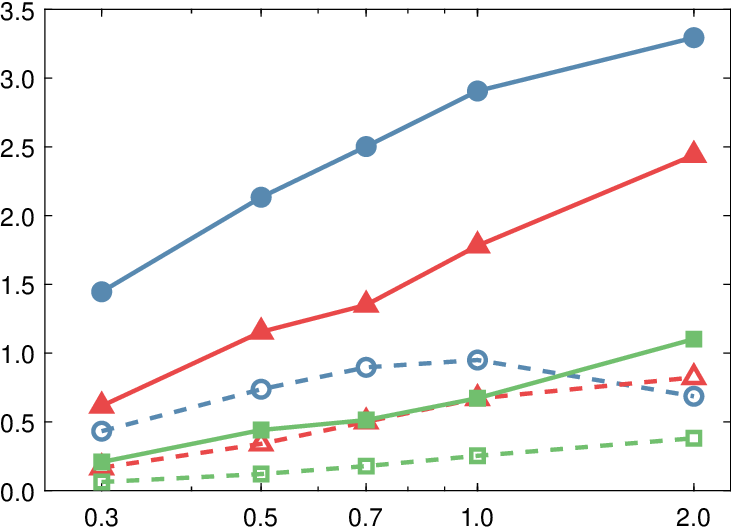}
        \put(-8,16){\rotatebox{90}{$-\langle u_z(\bm{x}^p(t),t)\rangle / u_{\eta}$}}
        \put(50,0){$St$} \put(10,65){(b)}
        \put(95,48){
            {\begin{overpic}
                [trim = 73mm 40mm 0mm 0mm,
                scale=0.6,clip,tics=20]{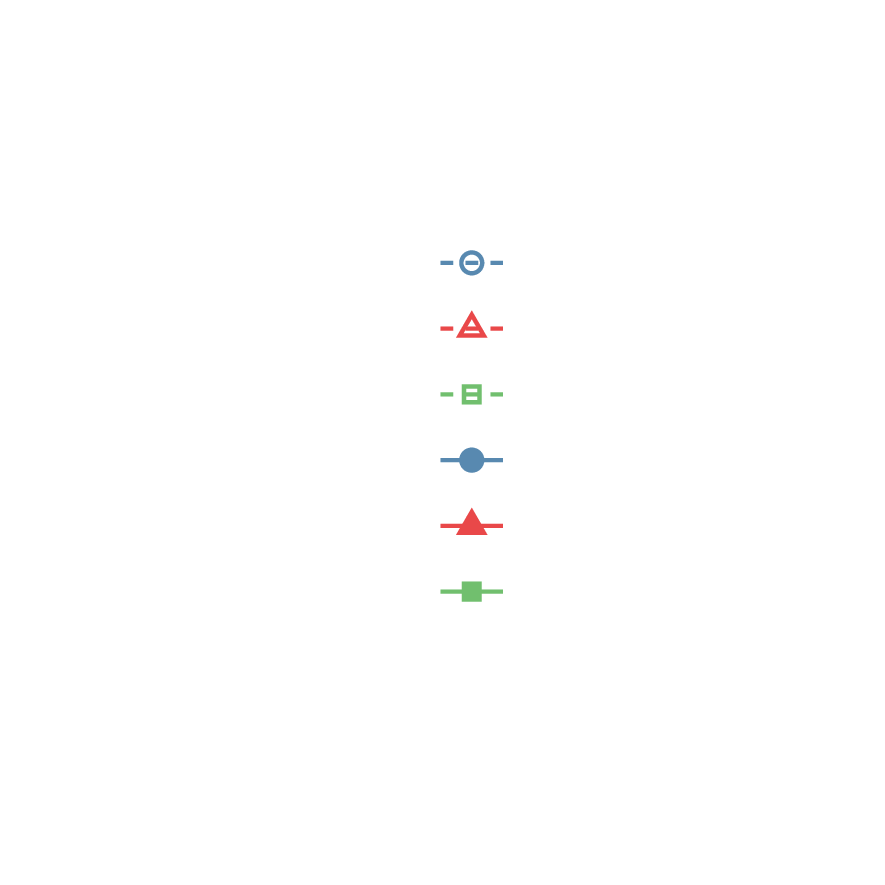} 
                \put(12, 7.4){$Fr=3$,  2WC} 
                \put(12,17.5){$Fr=1$,  2WC}
                \put(12,27.7){$Fr=0.3$, 2WC}
                \put(12,37.9){$Fr=3$,  1WC} 
                \put(12,47.9){$Fr=1$,  1WC} 
                \put(12,58){  $Fr=0.3$, 1WC}
            \end{overpic}} 
            }
    \end{overpic}} 
    % \begin{tikzpicture}[overlay]
        % \draw [help lines] (0,0) grid (-2,6);
        %  \node[draw,text width=2.1cm, gray] at (-1.2,2.5) {\small{\textcolor{red}{$\Phi_3=1.5\times 10^{-5}$}}};
        % \draw[gray, thick] (-1,-1) -- (2,2);
    % \end{tikzpicture}
    \caption{Normalized settling velocity enhancement for the 1WC (dashed) and 2WC (solid) cases for $\Phi=1.5\times 10^{-6}$ (a) and $\Phi=1.5\times 10^{-5}$ (b), versus $St$. See Fig. \ref{fig:App_sve_vs_St} for trends corresponding for other volume fractions.}
    \label{fig:sve_vs_St}
\end{figure}
% -------------

In Fig. \ref{fig:sve_vs_St} we show the normalized settling velocity enhancement, $-\langle{u}_z(\bm{x}^p(t),t)\rangle/u_\eta$, in 1WC (dashed) and the 2WC (solid) simulation regimes, as a function of $St$, for volume-fractions (a) $\Phi_1$ and (b) $\Phi_4$, respectively (plots for other volume fractions are shown in the Appendix). 
Settling enhancements for simulations with $Fr=0.3$, $1$, and $3$ are shown in red, blue, and green respectively. 
For a given $\Phi$ we see that the settling enhancement due to turbulence generally increases with (i) increasing $St$, for a given $Fr$, and (ii) decreasing $Fr$, for a fixed $St$, both of which correspond to increasing $Sv$.
However, for $Fr=0.3$, a decrease in the settling-enhancement is observed when going from $St=1$ to $St=2$ for both the 1WC and 2WC simulations at $\Phi_1$ and for the 1WC simulations at $\Phi_3$.
For the 1WC case (the explanation for its occurrence in the 2WC case is given below), this reduction can be understood by noting that for fixed $Fr$, increasing $St$ corresponds to increasing $Sv$, and as already discussed, $\langle{u}_z(\bm{x}^p(t),t)\rangle$ is expected to exhibit a non-monotonic dependence on $Sv$. 
The values of $St$ and $Fr$ at which $-\langle u_z(\bm{x}^p(t),t)\rangle / u_{\eta}$ is observed to decrease must therefore correspond to values of $Sv$ that exceed the value of $Sv$ at which $-\langle u_z(\bm{x}^p(t),t)\rangle / u_{\eta}$ is maximum.

%% Figure: effect of volume fraction
% 
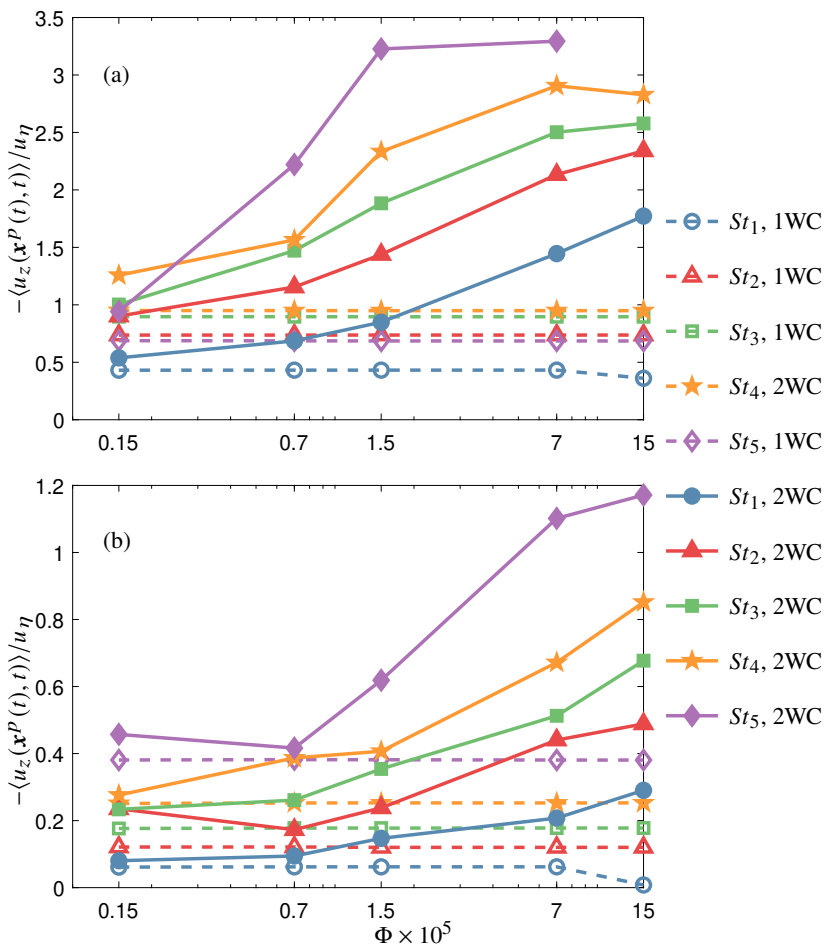
\begin{figure}
\centering
\captionsetup{width=\linewidth}
\vspace{0mm}	
% Figure
% St(Fr) vs St, const. Phi
\subfloat
{\begin{overpic}
    [trim = 0mm 0mm -12mm 0mm,
    scale=0.65,clip,tics=20]{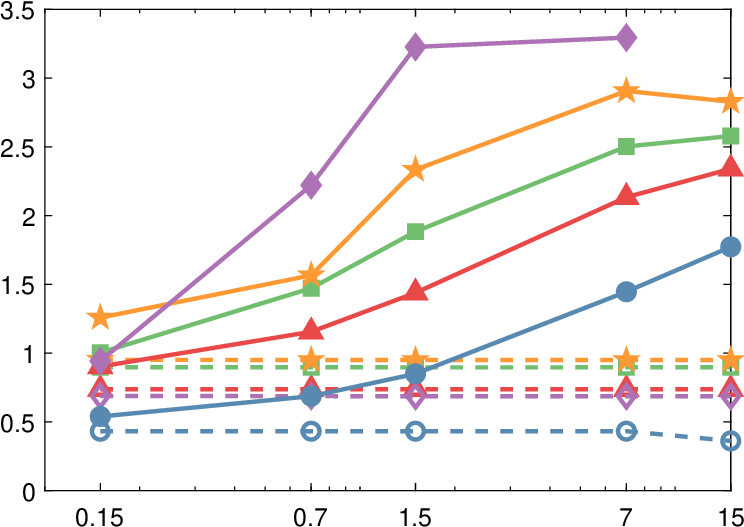}
    \put(-4,20){\rotatebox{90}{$-\langle u_z(\bm{x}^p(t),t)\rangle / u_{\eta}$}}
    \put(10,54){(a)}
    % \put(14.5,57.75){1WC} \put(14.5,49.75){2WC}
\end{overpic}}

\subfloat
{\begin{overpic}
    [trim = 0mm -5mm -12mm 0mm,
    scale=0.65,clip,tics=20]{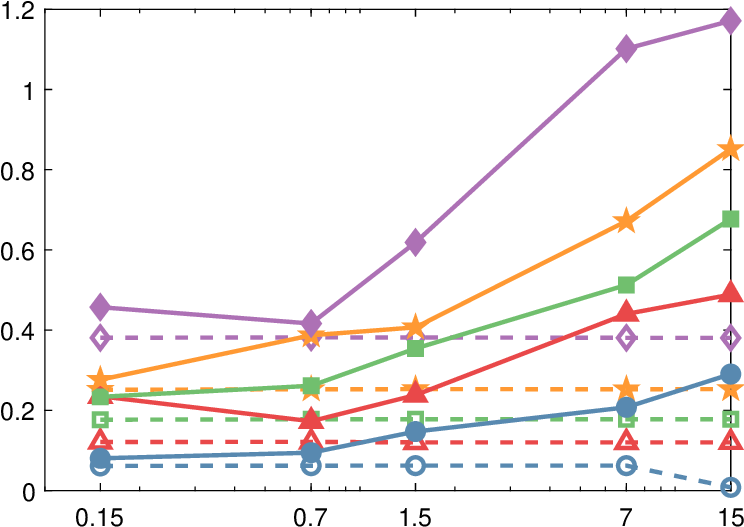}
    \put(-4,20){\rotatebox{90}{$-\langle u_z(\bm{x}^p(t),t)\rangle / u_{\eta}$}}
    \put(50,0){$\Phi \times 10^{5}$}
    \put(10,58){(b)}
    \put(91,30){
    {\begin{overpic}
        [trim = 22mm 25mm 40mm 0mm,
        scale=0.65,clip,tics=20]{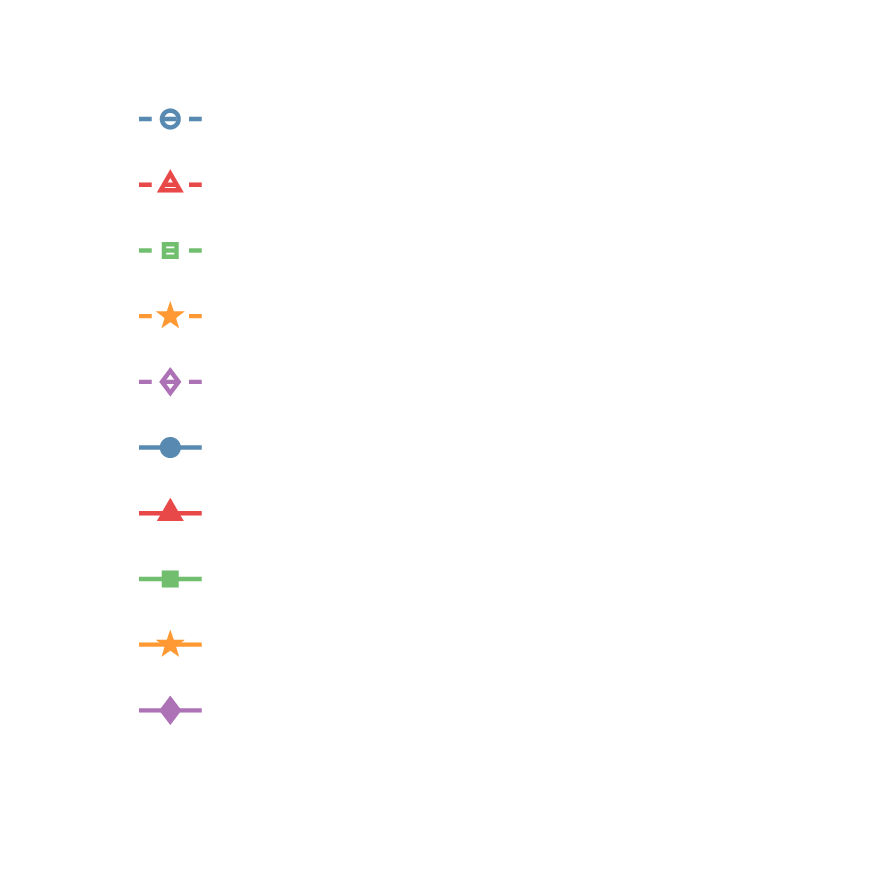} 
        \put(11.5,02.4){$St_5$, 2WC} 
        \put(11.5,11.3){$St_4$, 2WC}
        \put(11.5,20.2){$St_3$, 2WC}
        \put(11.5,29.0){$St_2$, 2WC} 
        \put(11.5,38.0){$St_1$, 2WC} 
        \put(11.5,47.0){$St_5$, 1WC}
        \put(11.5,55.8){$St_4$, 2WC}
        \put(11.5,64.6){$St_3$, 1WC} 
        \put(11.5,73.7){$St_2$, 1WC} 
        \put(11.5,82.6){$St_1$, 1WC}
    \end{overpic}} 
    }
\end{overpic}}
\caption{Normalized settling velocity enhancement versus volume fraction $\Phi$, for the 1WC (dashed) and 2WC (solid) cases, for fixed $Fr$ numbers: (a) $Fr=0.3$, (b) $Fr=3$. See Fig. \ref{fig:App_sve_vs_Phi} for simulations with $Fr=1$.}
\label{fig:sve_vs_Phi}
\end{figure}

Figure \ref{fig:sve_vs_Phi} shows the volume fraction $\Phi$ dependence of the normalized settling enhancement for simulations with (a) $Fr=0.3$, and (b) $Fr=3$.
Dashed and solid lines denote the results for the 1WC and 2WC simulations, respectively, whereas different line colors denote different $St$. 
The results show that even for the smallest volume fraction considered ($\Phi_1$), the momentum-coupling results in a finite, additional settling enhancement compared to the 1WC case, indicating that the particles are modifying the flow field in their vicinity. 
As shown in \S\ref{subsec:fluid_stats}, for $\Phi_1$ the global fluid statistics are almost identical to those of the unladen flow and this is because for low $\Phi$, modifications to the flow in the vicinity of the particles makes a negligible contribution to the global fluid behavior. 
As $\Phi$ is increased, we observe increasingly large differences between the 1WC and 2WC cases, as expected (note that physically, the 1WC results should be independent of $\Phi$. 
Slight variations of the 1WC results when varying $\Phi$ must therefore be due to statistical convergence effects due to the varying numbers of particles in the flow in each case). 
In \citet{monchaux2017} it was argued that the enhancement of particle settling speeds in the presence of 2WC compared with the 1WC case is because for a 2WC system, the settling inertial particles drag the fluid down with them which reduces the drag force on the particles and enables them to settle faster than in the 1WC case. 
As $\Phi$ is increased, this dragging effect on the fluid by the settling particles becomes stronger because of the increased mass-loading of the particles, and hence $-\langle u_z(\bm{x}^p(t),t)\rangle / u_{\eta}$ becomes larger. 
It was argued in \citet{monchaux2017} that this fluid dragging effect takes over from the preferential sweeping mechanism as being the dominant effect leading to $-\langle u_z(\bm{x}^p(t),t)\rangle / u_{\eta}>0$. 
However, in \citet{tom2022} it was demonstrated that this is not the case, and that at least for the parameters they considered, the fluid drag effect actually works together with the preferential sweeping mechanism to further enhance $-\langle u_z(\bm{x}^p(t),t)\rangle / u_{\eta}$ compared to its value for the 1WC system. This will be investigated further later in this paper.

The effect of 2WC on $-\langle u_z(\bm{x}^p(t),t)\rangle / u_{\eta}$ is seen to become stronger for increasing $St$ (at fixed $Fr$) and/or decreasing $Fr$ (at fixed $St$), i.e. for increasing $Sv$. 
This occurs because an increase in $Sv$ leads to greater potential energy of the particles, so that there is more energy available to be converted into fluid turbulent kinetic energy through the fluid dragging effect, which can lead to an increase for $-\langle u_z(\bm{x}^p(t),t)\rangle / u_{\eta}$. 
An implication of this is that whereas $-\langle u_z(\bm{x}^p(t),t)\rangle / u_{\eta}$ is expected to exhibit a non-monotonic dependence on $Sv$ in a 1WC system, in the presence of 2WC it may monotonically increase with increasing $Sv$ when $\Phi$ is sufficiently large for the effects of 2WC to be strong. 
The data in Fig. \ref{fig:sve_vs_Phi} is consistent with this, except for $\Phi_1$ and $Fr=0.3$ where a decrease of $-\langle u_z(\bm{x}^p(t),t)\rangle / u_{\eta}$ is observed in going from $St=1$ to $St=2$. 
However, this is likely because for $\Phi_1$, the effects of 2WC are relatively weak and so the non-monotonic dependence of $-\langle u_z(\bm{x}^p(t),t)\rangle / u_{\eta}$ on $Sv$ that is expected for a 1WC system (and which is also observed in Fig. \ref{fig:sve_vs_Phi}(a)) would also be expected for the 2WC case. 
This also explains the decrease in $-\langle u_z(\bm{x}^p(t),t)\rangle / u_{\eta}$ observed in Fig. \ref{fig:sve_vs_St} (a) for $\Phi_1$ and $Fr=0.3$ for the 2WC case when going from $St=1$ to $St=2$, which is not observed for the 2WC at the higher volume fraction $\Phi_3$ in Fig. \ref{fig:sve_vs_St} (b).

\subsection{Preferential sampling: $St$, $Fr$, $\Phi$ dependence\label{subsec:PrefConc}}
% -------------
\begin{figure}
\centering
\captionsetup{width=\linewidth}
\subfloat
{\begin{overpic}
    [trim = 0mm -5mm -10mm 0mm,
    scale=0.65,clip,tics=20]{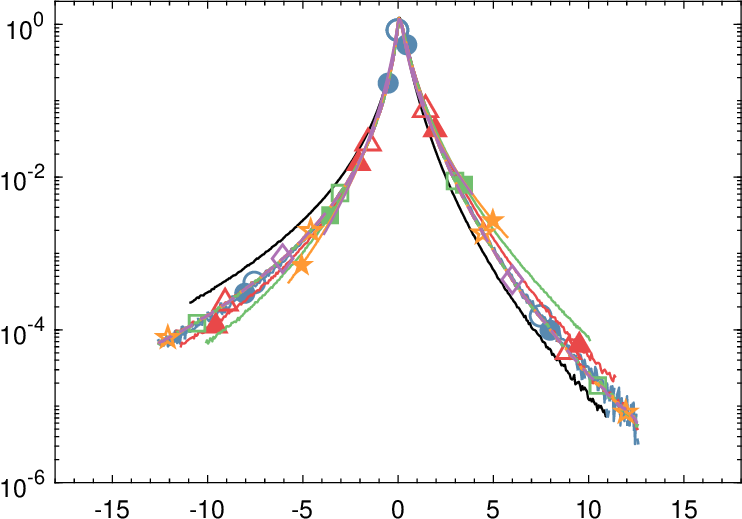} % PDF_Qp_303
    % \put(48,0){$\mathcal{Q}^p$} 
    \put(-5,25){\rotatebox{90}{$\sigma_{\mathcal{Q}}\mathscr{P}(\mathcal{Q}/\sigma_{\mathcal{Q}})$}}
    \put(10,60){(a)}
    \put(95.5,40){
    {\begin{overpic}
        [trim = 60mm 15mm 75mm 60mm,
        scale=0.5,clip,tics=20]{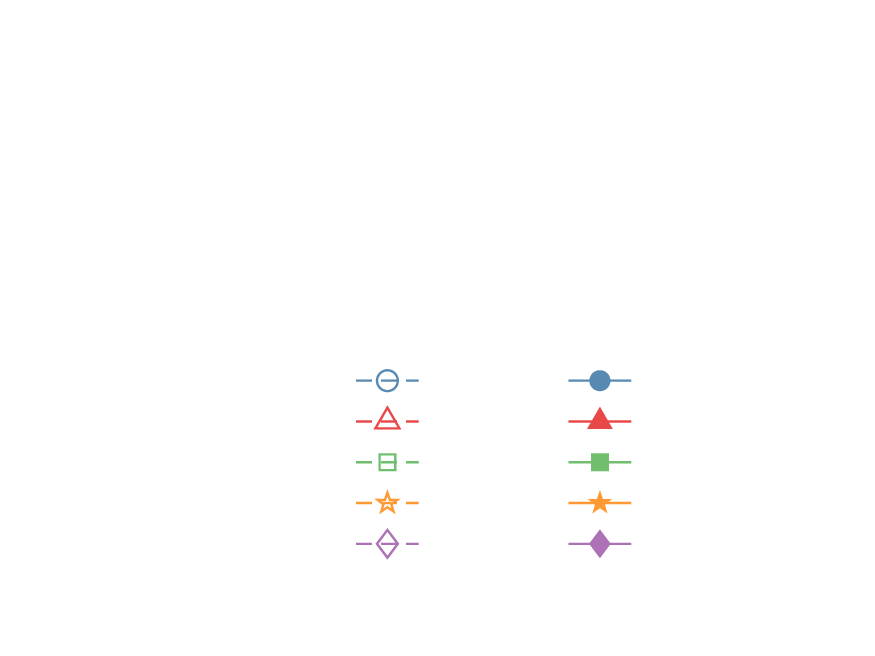}
        \put(33.5,82){$\Phi_1$ 1WC}
        \put(33.5,62){$\Phi_2$ 1WC} 
        \put(33.5,45){$\Phi_3$ 1WC} 
        \put(33.5,25){$\Phi_4$ 1WC} 
        \put(33.5, 6){$\Phi_5$ 1WC} 
    \end{overpic}} 
    }
    \put(95,10){
    {\begin{overpic}
        [trim = 95mm 15mm 35mm 57mm,
        scale=0.5,clip,tics=20]{Figures/leg_PDF_Q.eps}
        \put(33,75){$\Phi_1$ 2WC}
        \put(33,58){$\Phi_2$ 2WC}
        \put(33,40){$\Phi_3$ 2WC}
        \put(33,22){$\Phi_4$ 2WC}
        \put(33, 6){$\Phi_5$ 2WC}
    \end{overpic}} 
    }
\end{overpic}} 

\subfloat
{\begin{overpic}
    [trim = 0mm -5mm -10mm 0mm,
    scale=0.65,clip,tics=20]{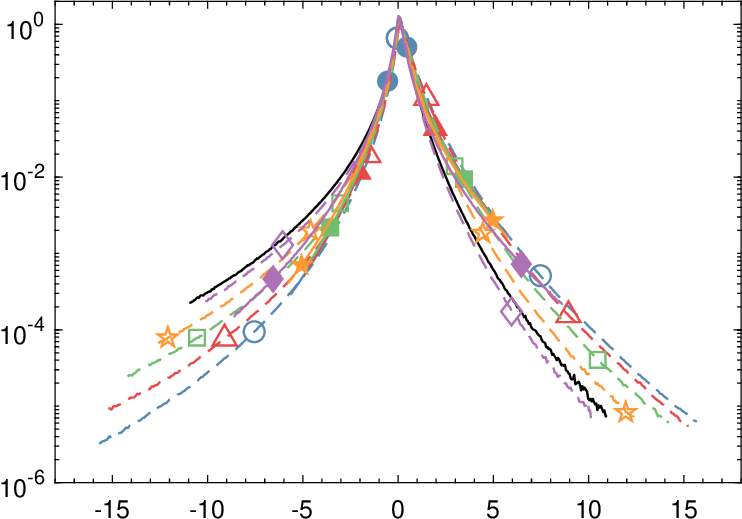} % PDF_Q310.eps
    \put(10,60){(b)}
    \put(52,0){$\mathcal{Q}/\sigma_{\mathcal{Q}}$} 
    \put(-5,25){\rotatebox{90}{$\sigma_{\mathcal{Q}}\mathscr{P}(\mathcal{Q}/\sigma_{\mathcal{Q}})$}}
    \put(95.5,40){
    {\begin{overpic}
        [trim = 60mm 15mm 75mm 60mm,
        scale=0.5,clip,tics=20]{Figures/leg_PDF_Q.eps}
        \put(33.5,82){$St_1$ 1WC}
        \put(33.5,62){$St_2$ 1WC} 
        \put(33.5,45){$St_3$ 1WC} 
        \put(33.5,25){$St_4$ 1WC} 
        \put(33.5, 6){$St_5$ 1WC} 
    \end{overpic}} 
    }
    \put(95,10){
    {\begin{overpic}
        [trim = 95mm 15mm 35mm 57mm,
        scale=0.5,clip,tics=20]{Figures/leg_PDF_Q.eps}
        \put(33,75){$St_1$ 2WC}
        \put(33,57){$St_2$ 2WC} 
        \put(33,40){$St_3$ 2WC} 
        \put(33,22){$St_4$ 2WC} 
        \put(33, 6){$St_5$ 2WC} 
    \end{overpic}} 
    }
\end{overpic}} 
\caption{Results for $\mathscr{P}(\mathcal{Q})$ in 1WC and 2WC simulations, along particle trajectories (a) for $\rm{St}_1=1$, $Fr=0.3$ showing $\Phi$ dependence, and 
(b) at a constant volume fraction $\Phi_4=7\times 10^{-5}$, and $Fr=0.3$ showing $St$ dependence. 
Dashed and solid lines mark the trends for 1WC and 2WC simulations respectively. $\sigma_{\mathcal{Q}}$ denotes the standard deviation of $Q^p$.}
\label{fig:Lag_PDF_Q}
\end{figure}
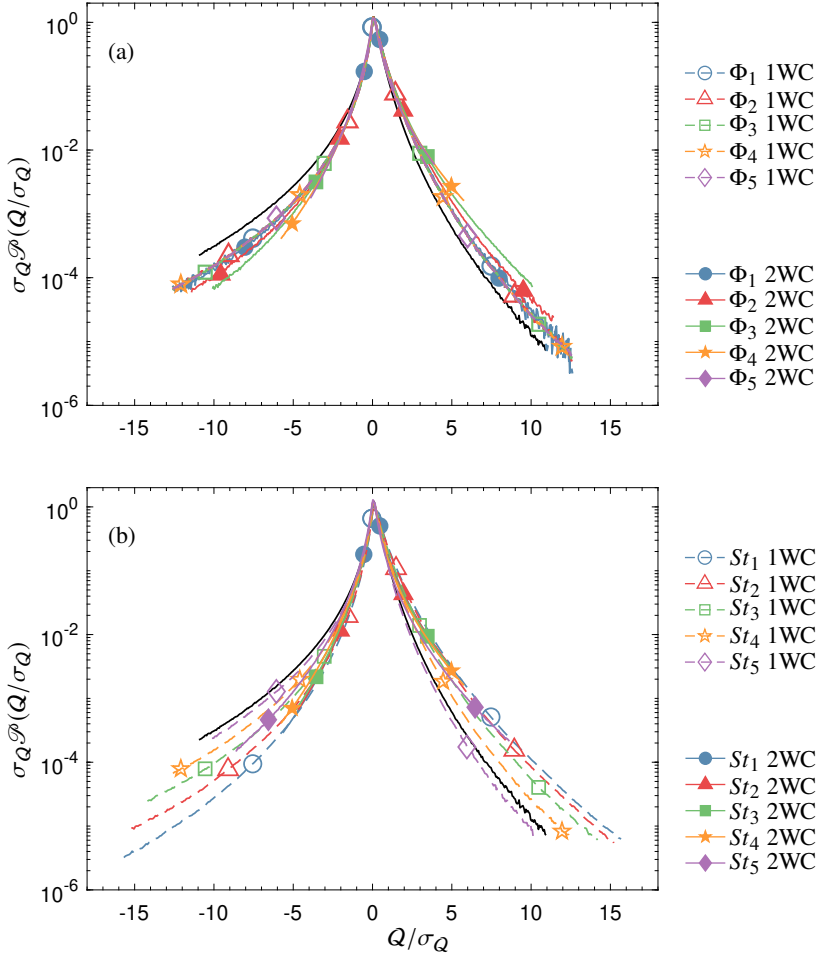

Having considered the effect of the parameters $St, Fr, \Phi$ on the enhancement of the particle settling velocity, we now turn to consider the mechanism underlying this enhancement. 
In \citet{monchaux2017} it had been argued that in the presence of 2WC, the enhanced settling velocities due to turbulence is no longer due to the preferential sweeping mechanism but due to the fluid dragging effect, i.e. the particles drag the fluid down with them which reduces the drag force acting on them leading to enhanced settling speeds. 
However, in a later study by \citet{tom2022}, some aspects of this argument was brought into question and the authors shed light on a nuanced perspective. 
It was shown that 
% \citet{tom2022} challenged this argument and showed that what is going on is more subtle; 
the preferential sweeping mechanism still operates in the presence of 2WC, but the particles drag the fluid with them as they are swept down, so that preferential sweeping and the fluid dragging effect act together to enhance the particle settling velocities. 
We want to explore whether this argument remains true as the parameters are varied, especially when $\Phi$ is larger.

The preferential sweeping mechanism is associated with the preference for inertial particles to move in downward moving strain-dominated regions of the flow. 
To explore the role this mechanism is playing in governing the settling enhancement, we therefore first consider statistical measures that can quantify the preferential sampling of the flow field. 
We do this by analyzing the PDF $\mathscr{P}(\mathcal{Q})\equiv\langle\delta(Q^p(t)-\mathcal{Q}\rangle$, where $Q^p(t)\equiv Q(\bm{x}^p(t),t)$ is the velocity gradient invariant $Q$ measured along the inertial particle trajectory. 
Figure \ref{fig:Lag_PDF_Q} (a) shows $\mathscr{P}(\mathcal{Q})$ normalized using the standard deviation of $Q^p$ (referred to as $\sigma_{\mathcal{Q}}$), for 1WC (dashed) and 2WC (solid) simulations with $St=1$ particles and with $Fr=0.3$, for all volume fractions $\Phi$. 
For small $\Phi$, the 2WC results are quite similar to the 1WC results, as was also observed in \cite{tom2022} for the case $\Phi_3$ with $Fr=1$. 
As $\Phi$ is increased, the effect of 2WC on $\mathscr{P}(\mathcal{Q})$ becomes apparent, with the primary impact being on the left tail, corresponding to enstrophy dominated regions of the flow. 
It is important to note that these differences between the 1WC and 2WC results are not only due to differences in the way that the particles are sampling the flow, but also because the statistics of the flow field itself changes due to 2WC, as was previously shown in Fig.~\ref{fig:Eul_PDF_Q} for the global statistics of $Q(\bm{x},t)$. 

In Fig. \ref{fig:Lag_PDF_Q} (b), we show the normalised PDF $\mathscr{P}(\mathcal{Q})$ at a constant volume fraction $\Phi_4$, and $Fr=0.3$ for different $St$. 
Here, solid lines correspond to 2WC simulations, whereas dashed lines correspond to 1WC simulations. 
For the 1WC simulations, the results show that $\mathscr{P}(\mathcal{Q})$ depends strongly on $St$, indicating that the way that the particles sample the flow changes significantly as $St$ is varied. 
Interestingly, both the left and right tails of the PDF change significantly as $St$ is varied, whereas the results in \citet{tom2022} for $Fr=1$ show a much weaker dependence on $St$ of the right tail of the PDF than the left tail. 
Since the results in Fig.~\ref{fig:Lag_PDF_Q} (b) are for $Fr=0.3$, this suggests that in some regimes of $Sv$, the particles not only move away from regions of strong enstrophy, but also migrate significantly towards regions of stronger strain-rate.

% ------------------
\begin{figure}
\centering
\captionsetup{width=\linewidth}
\vspace{0mm}	
% Figure
% St(Fr) vs St, const. Phi
\subfloat
{\begin{overpic}
    [trim = 0mm 0mm -10mm 0mm,
    scale=0.5,clip,tics=20]{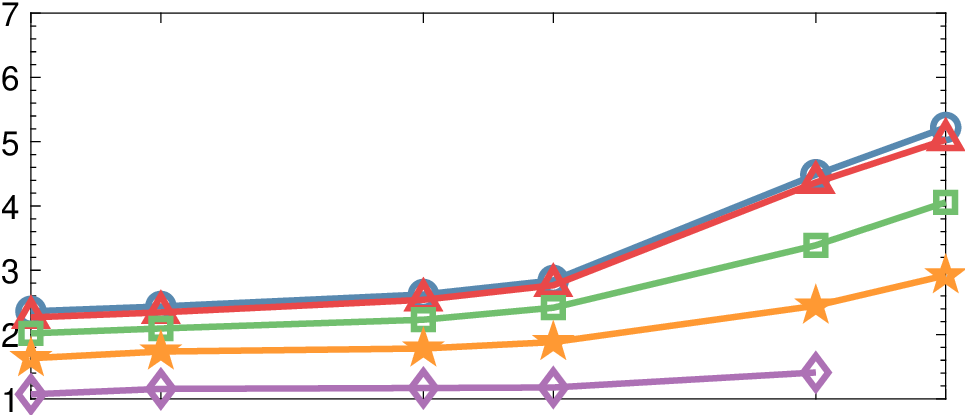} %PQpos-PQneg.eps
    \put(5,32){(a)}
    \put(-4,20){\rotatebox{90}{$\varphi$}}
\end{overpic}} \\
\subfloat
{\begin{overpic}
    [trim = 0mm 0mm -10mm 0mm,
    scale=0.5,clip,tics=20]{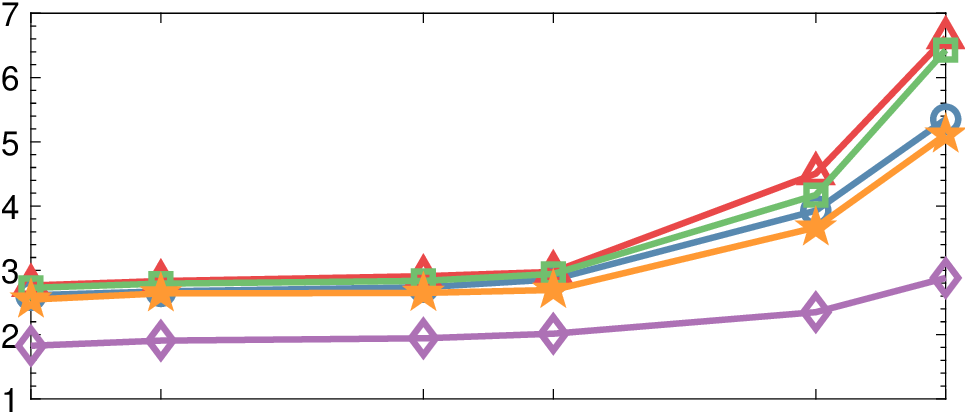}
    \put(5,32){(b)}
    \put(-4,20){\rotatebox{90}{$\varphi$}}
\end{overpic}} \\
\subfloat
{\begin{overpic}
    [trim = 0mm -4mm -10mm 0mm,
    scale=0.5,clip,tics=20]{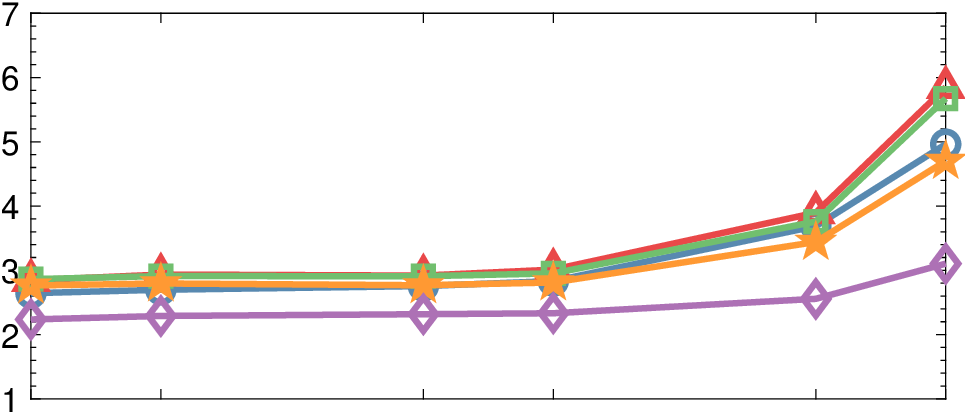}
        \put(5,32){(c)}
        \put(-4,20){\rotatebox{90}{$\varphi$}}
        \put(2,-0.5){1WC} \put(14.3,-0.5){$\Phi_1$} \put(40,-0.5){$\Phi_2$} \put(52.55,-0.5){$\Phi_3$} \put(78.3,-0.5){$\Phi_4$} \put(91,-0.5){$\Phi_5$}
        \put(96,40){
        {\begin{overpic}
            [trim = 20mm 110mm 60mm 20mm,
            scale=0.6,clip,tics=20]{Figures/leg_LagEul_rat_PQd.eps}
            \put(21,12.5){$St_5$} 
            \put(21,32.0){$St_4$} 
            \put(21,51.5){$St_2$} 
            \put(21,71.0){$St_2$}
            \put(21,91.0){$St_1$}
        \end{overpic}} 
        }
\end{overpic}} 
\caption{Plots of $\varphi$ (Eq.~\ref{eq:def_varphi}) versus volume fraction $\Phi$ for simulations with (a) $Fr=0.3$, (b) $Fr=1$, and (c) $Fr=3$. When $\varphi> 1$ this indicates that the inertial particles exhibit a preference to move in strain-dominated regions of the flow.}
\label{fig:PQp-PQn}
\end{figure}
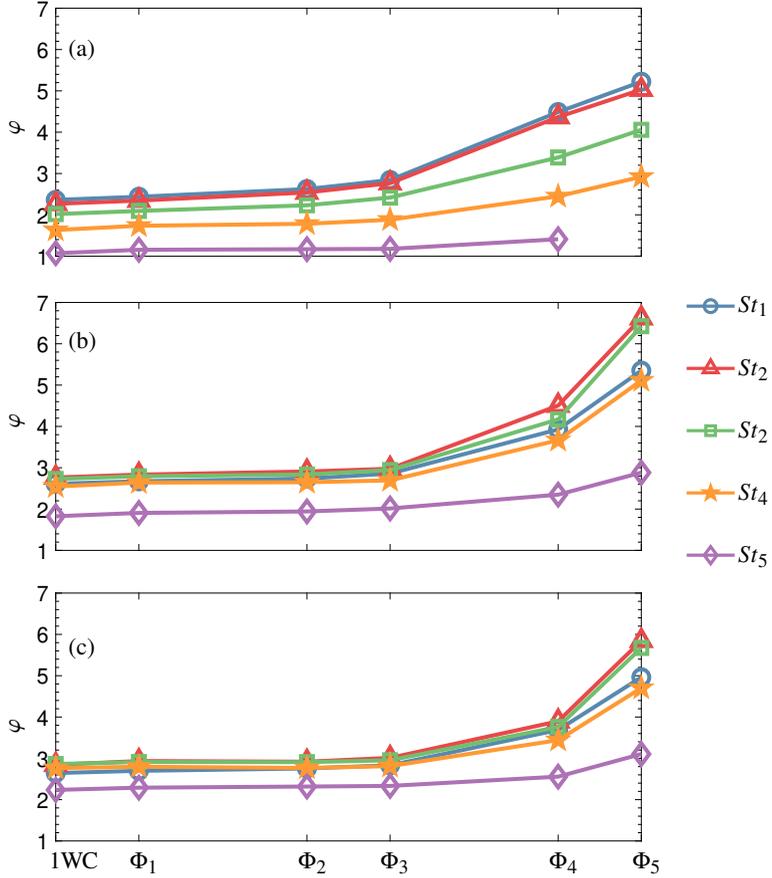

% ------------------

To provide clearer quantitative insight into the degree to which the inertial particles are preferentially sampling the flow we can consider the the difference between the probability that the particles are in a strain-dominated or vorticity dominated regions, denoted by $\mathbb{P}(\mathcal{Q}>0)~\equiv \int_0^\infty \mathscr{P}(\mathcal{Q})\,d \mathcal{Q}$ and $\mathbb{P}(\mathcal{Q}<0)\equiv \int_{-\infty}^0  \mathscr{P}(\mathcal{Q})\,d \mathcal{Q}$, respectively. However, even for fluid particles (which sample the flow uniformly), $\mathbb{P}(\mathcal{Q}>0)-\mathbb{P}(\mathcal{Q}<0)\neq 0$, and therefore simply considering $\mathbb{P}(\mathcal{Q}>0)-\mathbb{P}(\mathcal{Q}<0)$ will not provide a direct measure of the degree to which the inertial particles are preferentially sampling the flow. Therefore, the measure of preferential sampling that we will use is
\begin{align}\label{eq:def_varphi}
\varphi\equiv\frac{\mathbb{P}(\mathcal{Q}>0)-\mathbb{P}(\mathcal{Q}<0)}{[\mathbb{P}(\mathcal{Q}>0)-\mathbb{P}(\mathcal{Q}<0)]\vert_{St=0}},
\end{align}
where $[\mathbb{P}(\mathcal{Q}>0)-\mathbb{P}(\mathcal{Q}<0)]\vert_{St=0}$ denotes $\mathbb{P}(\mathcal{Q}>0)-\mathbb{P}(\mathcal{Q}<0)$ evaluated along the trajectories of fluid particles (and since the flow is incompressible, these single-time statistics are the same as those based on $Q$ measured at fixed locations $\bm{x}$ in the flow). When $\varphi> 1$ this indicates that the inertial particles exhibit a preference to move in strain-dominated regions of the flow.

In Fig.~\ref{fig:PQp-PQn} we plot $\varphi$ for different parameter choices. 
Most strikingly, the results show that the preferential sampling of strain-dominated regions of the flow becomes stronger as $\Phi$ is increased. 
This is contrary to the argument of \citet{monchaux2017} that 2WC weakens the preferential sampling of the flow and the associated centrifuge mechanism. 
A potential reason for this discrepancy is that \citet{monchaux2017} based their conclusion on results for the joint PDF of $\bm{S:S}$ and $\bm{R:R}$ (see \S\ref{subsec:fluid_stats} for notation) measured along the inertial particle trajectories, and compared results for different $\Phi$. 
However, as $\Phi$ is varied, this PDF is affected both by changes in the particle motion and also changes in the fluid velocity field due to 2WC. 
As a result, this joint PDF of $\bm{S:S}$ and $\bm{R:R}$ measured along the inertial particle trajectories does not give a direct measure of the preferential sampling of the flow by the inertial particles. 
To test for preferential sampling of the fluid-velocity-gradient field, one must compare the properties of the fluid velocity gradients measured along the inertial particle trajectories to those measured along fluid particle trajectories. In the Appendix, Figs. \ref{fig:App_PQp_pos-PQp_neg} and \ref{fig:App_PQe_pos-PQe_neg} show separately the quantities $\mathbb{P}(\mathcal{Q}>0)-\mathbb{P}(\mathcal{Q}<0)$ and $[\mathbb{P}(\mathcal{Q}>0)-\mathbb{P}(\mathcal{Q}<0)]\vert_{St=0}$, respectively. 
The results for $\mathbb{P}(\mathcal{Q}>0)-\mathbb{P}(\mathcal{Q}<0)$ show that this quantity reduces as $\Phi$ is increased, and this is consistent with the results from \citep{monchaux2017} which show that the joint PDF of $\bm{S:S}$ and $\bm{R:R}$ measured along the inertial particle trajectories becomes more symmetric about the line $\bm{S:S}=\bm{R:R}$ as $\Phi$ is increased, i.e. that the enhanced probability to be in strain-dominated regions reduces as $\Phi$ is increased. 
However, the results also show that $[\mathbb{P}(\mathcal{Q}>0)-\mathbb{P}(\mathcal{Q}<0)]\vert_{St=0}$ decreases as $\Phi$ is increased, and this is why $\varphi$ increases as $\Phi$ increases, even though $\mathbb{P}(\mathcal{Q}>0)-\mathbb{P}(\mathcal{Q}<0)$ decreases as $\Phi$ increases.

\subsection{Preferential sweeping: $St$, $Fr$, $\Phi$ dependence\label{subsec:PrefSweep}}

The preferential sweeping mechanism not only argues that inertial particles preferentially sample strain dominated regions of the flow (which we have just demonstrated is indeed the case), but also that due to gravity, they will preferentially sample strain-dominated regions of the flow where the vertical fluid velocity points down. 
\citet{tom2022} tested this by computing the quantities
\begin{align}
    A &\equiv\int_{-\infty}^{0}\Big\langle u_z(\bm{x}^p(t),t)\Big\rangle_{\mathcal{Q}}\,\mathscr{P}(\mathcal{Q})\, d\mathcal{Q},\label{eq:def_A}\\
    B &\equiv\int_{0}^{\infty}\Big\langle u_z(\bm{x}^p(t),t)\Big\rangle_{\mathcal{Q}}\,\mathscr{P}(\mathcal{Q})\, d\mathcal{Q},\label{eq:def_B}
\end{align}
As discussed in \citet{tom2022}, the preferential sweeping predicts that $B<0$ and $|B|>|A|$, such that the settling enhancement arises primarily due to contributions from particles in
strain-dominated regions of the flow where fluid velocity is moving downward.\\

% -------------
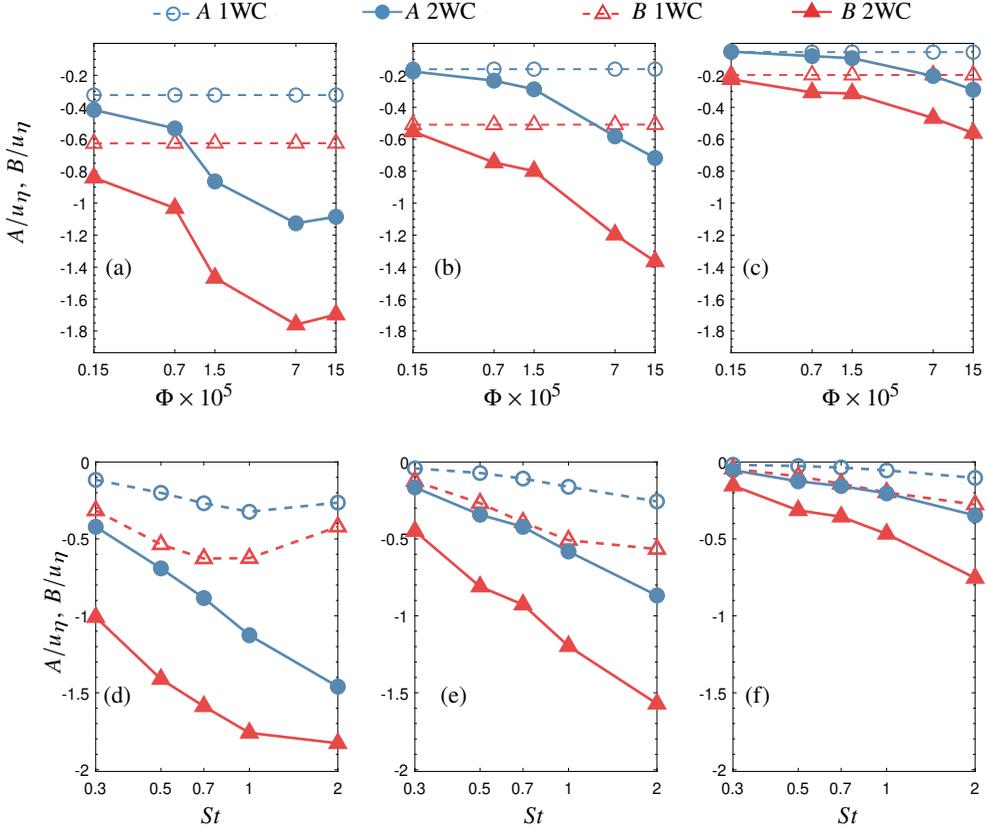
\begin{figure}
\centering
\captionsetup{width=\linewidth}
\subfloat
{\begin{overpic}
    [trim = -9mm -5mm -10mm -15mm,
    scale=0.5,clip,tics=20]{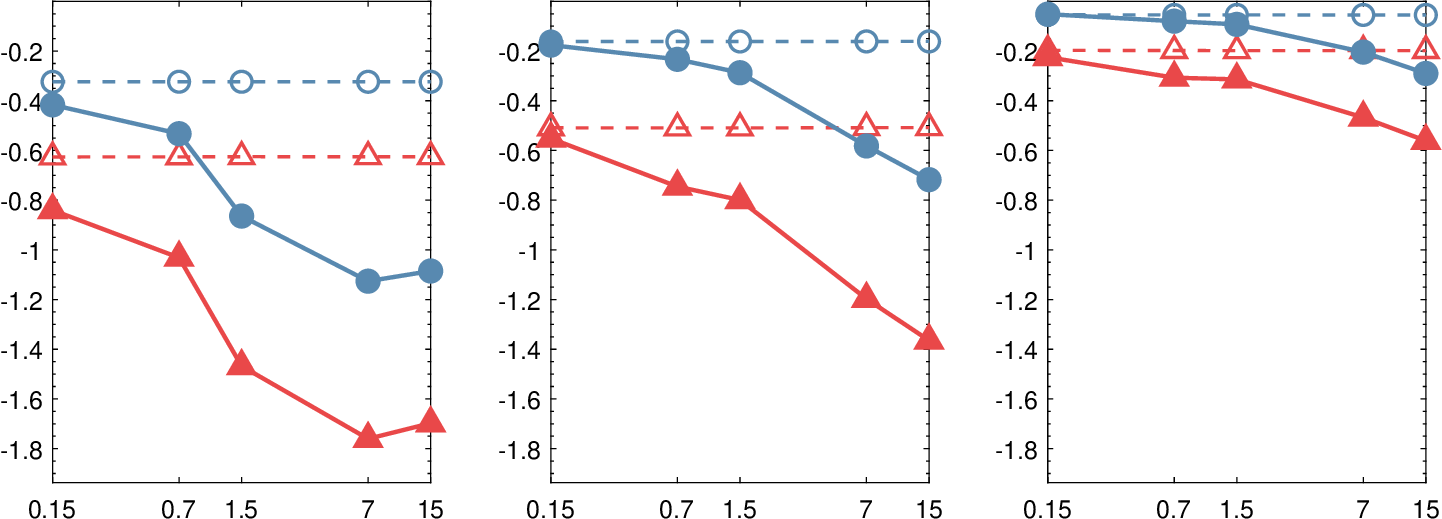}
    \put(8,12){(a)} \put(41,12){(b)} \put(72,12){(c)}
    % \put(82, 7){B 2WC}  % \put(82,10){B 1WC} % \put(82,13){A 2WC}% \put(82,16){A 1WC}
    \put(13,-1){$\Phi\times 10^5$} \put(46,-1){$\Phi\times 10^5$} \put(80,-1){$\Phi\times 10^5$} 
    \put(-2,15){\rotatebox{90}{$A/u_\eta,\, B/u_\eta$}}
    % 1st legend
    \put(54,37){
        \begin{overpic}
        [trim = 1mm 27mm 240mm 0mm,
        scale=0.15,clip,tics=20]{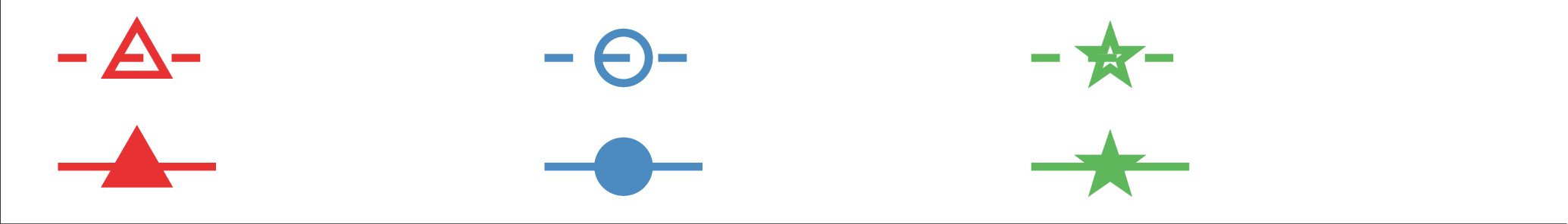}
        \put(50, 5){{$B$ 1WC}}
    \end{overpic}}
    % 2nd legend
    \put(75,37){
        \begin{overpic}
        [trim = 1mm 1mm 240mm 25mm,
        scale=0.15,clip,tics=20]{Figures/sve_Stdep_Leg.eps}
        \put(50, 5){{$B$ 2WC}}
    \end{overpic}}  
    % 3rd legend
    \put(12,37){
        \begin{overpic}
        [trim = 120mm 27mm 120mm 0mm,
        scale=0.15,clip,tics=20]{Figures/sve_Stdep_Leg.eps}
        \put(35,5){{$A$ 1WC}}
    \end{overpic}}  
    % 4th legend
    \put(33,37){
        \begin{overpic}
        [trim = 120mm 1mm 120mm 25mm,
        scale=0.15,clip,tics=20]{Figures/sve_Stdep_Leg.eps}
        \put(35, 5){{$A$ 2WC}}
    \end{overpic}}  
\end{overpic}} 

\subfloat
{\begin{overpic}
    [trim = 0mm -5mm 0mm -10mm,
    scale=0.5,clip,tics=20]{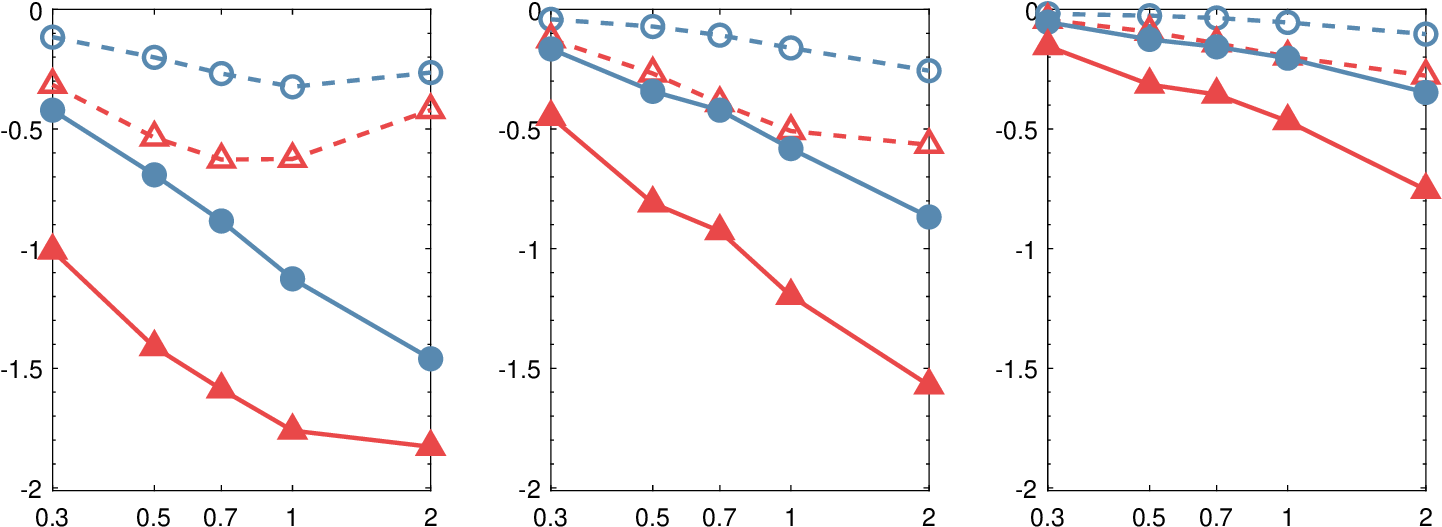}
    \put(-2,15){\rotatebox{90}{$A/u_\eta,\, B/u_\eta$}}
    \put(4.5,12){(d)} \put(41,12){(e)} \put(74,12){(f)}
    \put(15,-1){$St$} \put(50,-1){$St$} \put(84,-1){$St$} 
    % \put(82, 7){B 2WC}  % \put(82,10){B 1WC} % \put(82,13){A 2WC}% \put(82,16){A 1WC}
\end{overpic}} 
\caption{Plots of $A$ (Eq.~\ref{eq:def_A}), in blue, and $B$ (Eq.~\ref{eq:def_B}), in red, for 1WC (dashed) and 2WC (solid) flows for $St=1$ particles as a function of $\Phi$ with (a) $Fr=0.3$, (b) $Fr=1$, and (c) $Fr=3$.
Plot of $A$ (blue) and $B$ (red) at a fixed volume fraction $\Phi_4=7\times 10^{-5}$ for 1WC (dashed) and 2WC (solid) simulations, with (e) $Fr=0.3$, (f) $Fr=1$, (g) $Fr=3$ showing the $St$ dependence.
See Figs.~\ref{fig:App_AB_plot_vs_Phi} and ~\ref{fig:App_AB_plot_vs_St} for $\Phi$-, $St$-dependence for other simulations.
}
\label{fig:AB_plot}
\end{figure}
% -------------

The top row of Fig. \ref{fig:AB_plot} shows results for $A$ (blue) and $B$ (red), for 1WC (dashed) and 2WC (solid) cases as a function of $\Phi$ at $St=1$ and for (a) $Fr=0.3$, (b) $Fr=1$, (c) $Fr=3$. 
For the 1WC results, $B$ is seen to become more negative as $Fr$ is reduced, indicating that the preferential sweeping is becoming stronger as $Sv$ is reduced over the range of $Sv$ considered. 
The results also show that $A<0$ is also negative, implying that there is also a contribution to the settling enhancement arising from particles in vorticity dominated regions of the flow. 
While this may seem surprising, it was argued in \citet{tom2022} that this is not inconsistent with the preferential sweeping mechanism (the reader is referred to that paper for the explanation why). 
For $Fr=1$ and $Fr=3$ the values of $A$ and $B$ are almost identical for the 1WC and 2WC cases at the lowest $\Phi$, but for $Fr=0.3$ there is an enhancement in the magnitudes of both $A$ and $B$ due to 2WC even for the lowest $\Phi$. 
For each $Fr$, however, both $A$ and $B$ become increasingly negative as $\Phi$ is increased, while always preserving $|B|>|A|$. 
As argued in \citet{tom2022}, this implies that as $\Phi$ is increased and 2WC becomes increasingly important, the preferential sweeping mechanism remains active and is the mechanism responsible for the turbulent enhancement of the particle settling speeds. 
The difference 2WC makes is that when the particles are swept into downward moving, strain-dominated regions of the flow, they are not only swept downwards but also drag the fluid down with them, and this is why $B$ is more negative for the 2WC cases than the 1WC cases at all $St, Fr, \Phi$ combinations considered.

The bottom row of Fig. \ref{fig:AB_plot} illustrates the $St$ dependence of $A$ and $B$ at a fixed volume fraction $\Phi_4=7\times 10^{-5}$ and for (e) $Fr=0.3$, (f) $Fr=1$, (g) $Fr=3$. 
The results show that the increased negativity of $B$ (as well as $A$) occurs at all $St$ considered. The results also indicate that the difference between the 1WC and 2WC values for $A$ and $B$ becomes larger as $St$ is increased, for a fixed $\Phi$ and a given $Fr$. This is again because for a given $Fr$, increasing $St$ corresponds to increased potential energy for the particles, and hence a greater ability to modify the fluid velocity field.

% -------------
\begin{figure}
\centering
\captionsetup{width=\linewidth}
\subfloat
{\begin{overpic}
    [trim = -9mm -5mm -10mm -5mm,
    scale=0.65,clip,tics=20]{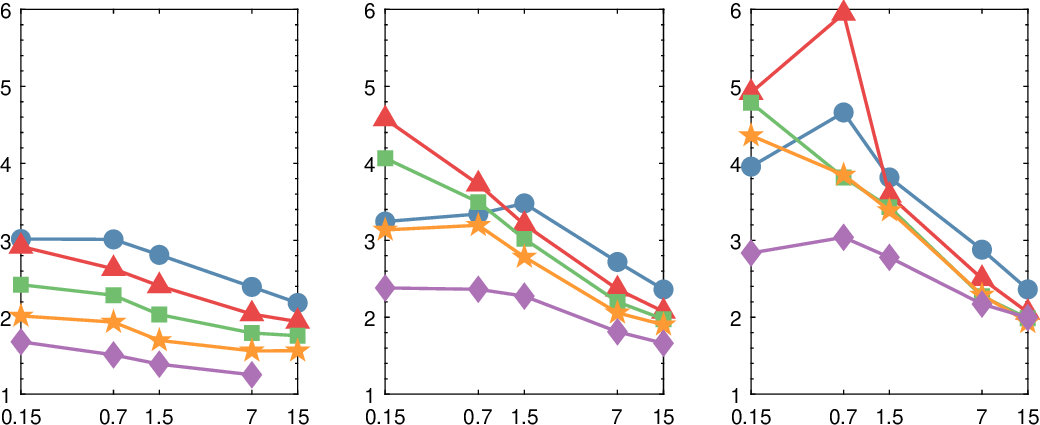}
    \put(8,29){(a)} \put(41,29){(b)} \put(72,29){(c)}
    % \put(82, 7){B 2WC}  % \put(82,10){B 1WC} % \put(82,13){A 2WC}% \put(82,16){A 1WC}
    \put(13,-1){$\Phi\times 10^5$} \put(46,-1){$\Phi\times 10^5$} \put(80,-1){$\Phi\times 10^5$} 
    \put(-1,20){\rotatebox{90}{$B/A$}}
    % legend
    \put(7,41){
        \begin{overpic}
        [trim = 150mm 24mm 40mm 21mm,
        scale=0.65,clip,tics=20]{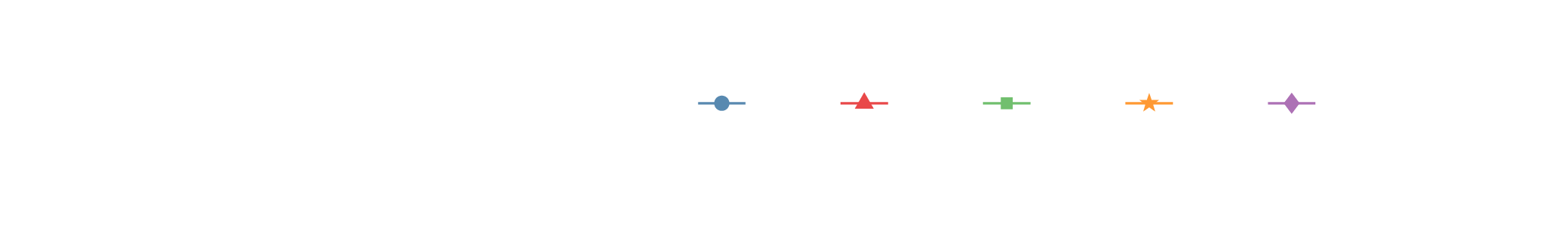}
        \put(11,1.0){{$St_1$}}
        \put(31,1.0){{$St_2$}}
        \put(51,1.0){{$St_3$}}
        \put(71,1.0){{$St_4$}}
        \put(91,1.0){{$St_5$}}
    \end{overpic}}
\end{overpic}} 
\caption{Plots of the ratio $B/A$ (Eqs.~\ref{eq:def_A}, \ref{eq:def_B}) for 2WC flows versus $\Phi$ with (a) $Fr=0.3$, (b) $Fr=1$, and (c) $Fr=3$.}
\label{fig:ABratio_plot}
\end{figure}
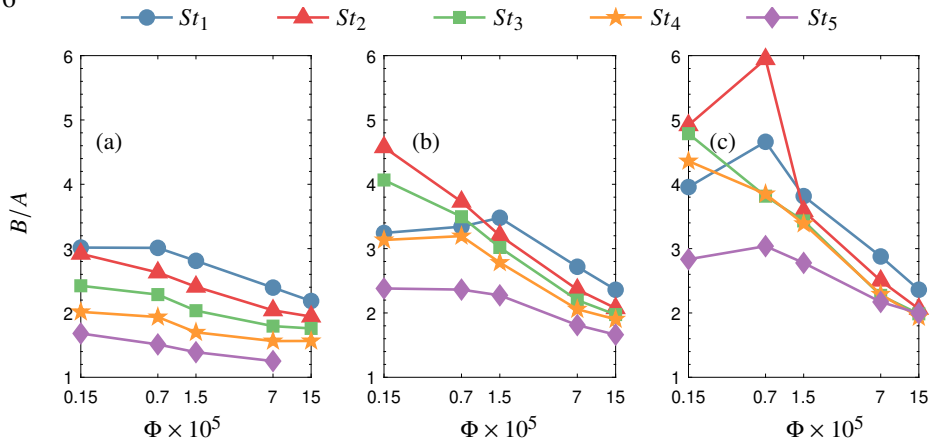
% -------------
To examine more carefully the relative contributions of both $A$ and $B$ and their dependence on $\Phi$, Fig.~\ref{fig:ABratio_plot} shows the ratio $B/A$ for 2WC cases as a function of $\Phi$ for different $St$ with (a) $Fr=0.3$, (b) $Fr=1$, and (c) $Fr=3$. 
Large values of $B/A$ indicate cases where preferential sweeping plays a significant role, whereas $B/A=1$ corresponds to the limit where the contribution to $\langle u_z(\bm{x}^p(t),t)\rangle$ from strain- and rotation-dominated regions are equal, and therefore preferential sweeping is playing no role.
The results show that the ratio decreases (except for a few cases where it initially increases, but then decreases) with increasing $\Phi$, showing that the role of preferential sweeping becomes less important as $\Phi$ increases, in partial agreement with \cite{monchaux2017}. However, it is still important for all of the cases; the ratio is greater than 1 even at the largest $\Phi$, and for many of the cases it is closer to 2 at this $\Phi$. An interesting and important question is whether preferential sweeping will continue to play an important role as $\Phi$ is increased even further. The results for $Fr=0.3$ seem to indicate that $B/A$ becomes independent of $\Phi$ as $\Phi$ is increased, which could suggest that preferential sweeping remains important even at much larger $\Phi$. The results for $Fr=1,3$ do not show evidence of a regime where $B/A$ becomes independent of $\Phi$, but this could be simply because this asymptotic regime occurs at larger $\Phi$ than we have simulated.

Taken together, these results provide strong evidence that even in regimes where the mass loading is high enough for the particles to substantially modify the global fluid statistics, the preferential sweeping mechanism remains important for governing how turbulence enhances particle settling speeds. The difference compared with the 1WC case is that in the 2WC case, the particles are not merely swept around the downward moving side of vortices in the flow but they also (if the mass loading is sufficient) drag the fluid down with them in these regions as they fall. Our results also suggest that preferential sweeping may continue to be important at mass loadings much greater than those considered here, but the evidence is not conclusive.
\section{Conclusions}\label{sec:C}
We have conducted 1WC and 2WC DNS of settling, sub-Kolmogorov scale, heavy, inertial particles in homogeneous turbulence over a wide range of Froude numbers $Fr$, Stokes numbers $St$, and volume fractions $\Phi$ (the particle density was held constant, so that the mass loading $\Phi_m$ is proportional to $\Phi$). For each combination of $St, Fr$, five different values of $\Phi$ were considered that span two orders of magnitude. 
We first considered the impact of 2WC on various global fluid statistics, including the properties of the strain-rate and vorticity fields and TKE dissipation rates, as well as the energy spectrum to consider the impact of 2WC on different scales in the flow. 
At the smallest $\Phi$, the PDF of $Q$ (the second invariant of the velocity gradient tensor) retains the characteristic skewness that is seen for the unladen flow which is associated with the vorticity field being more intermittent than the strain-rate field. 
However, the PDF becomes increasingly symmetric as $\Phi$ is increased, indicating that the momentum feedback from the particles suppresses regions of high vorticity while amplifying regions of high strain-rate. 
This effect of 2WC depends much more strongly on $Fr$ than on $St$ for the parameter regimes considered. 
For the energy spectrum we observe that as $\Phi$ is increased there is an accumulation of energy in the high wave-number modes, which also corresponds to an increase in the TKE dissipation rate. 
Moreover, as $\Phi$ is increased, the fraction of TKE associated with the vertical velocity fluctuations also increases, and this is due to the conversion of particle potential energy to the vertical component of TKE through the effect of the particles dragging the surrounding fluid down with them as they settle, and thereby doing work on the surrounding flow.

We then turned to consider how the settling velocity of the particles depends on $St, Fr$ and $\Phi$. 
We find that even for the smallest $\Phi$ considered ($\Phi=1.5\times 10^{-6}$, corresponding to a mass loading of $7.5\times 10^{-3}$), the mean settling speed of the particles can be appreciably larger for the 2WC case than the 1WC case for some choices of $St$ and $Fr$. 
For both 1WC flows and 2WC flows at low $\Phi$, the largest enhancement of the mean settling velocity compared with the Stokes settling velocity occurs for $St=1$ and $Fr=0.3$, and the enhancement exhibits a non-monotonic dependence on $Sv$ which is expected since the settling enhancement must go to zero in the two limits $Sv \to \infty$ and $Sv \to 0$. 
However, for 2WC flows with higher $\Phi$ this non-monotonic dependence does not occur since as $Sv$ is increased, the particles are increasingly effective in dragging the fluid down with them which reduces the drag force on the particle and enhances their settling velocity. 
Due to this, the difference between the particle settling speeds in the 1WC and 2WC flows increases with increasing $Sv$.

The next step was to understand how the mechanisms governing the enhanced particle settling depends on $\Phi$. 
In particular, it was to understand whether preferential sweeping remains relevant in 2WC flows when $\Phi$ becomes large enough for the global fluid statistics to be strongly affected by the particles. 
The preferential sweeping mechanism is based on the argument that the following two things should occur, 1) the particles should preferentially sample strain-dominated regions of the flow (which they would do even in the absence of gravity), and 2) the particles should preferentially sample strain-dominated regions where the vertical fluid velocity points down (which they would not do in the absence of gravity, if the flow is isotropic). 
To test the first part, we computed the PDF of $Q$ measured along the trajectories of the settling inertial particles and compared them to results based on $Q$ measured along fluid particle trajectories. 
We also compared the probability of particles being in $Q>0$ regions compared with $Q<0$ regions, normalized by the corresponding probability for fluid particles. 
The results showed that not only is preferential sampling still active in the 2WC flows, but that it actually becomes stronger as $\Phi$ is increased. 
To test the second part, we computed the settling velocity enhancement conditioned on the particles being in $Q>0$ or $Q<0$ regions. 
The results showed that at all $St, Fr, \Phi$ combinations considered, the strongest contribution to the enhanced particle settling velocities comes from particles in strain dominated regions of the flow, consistent with the preferential sweeping mechanism. 
The results do indicate that the imbalance in the contribution from $Q>0$ and $Q<0$ regions reduces as $\Phi$ is increased, for a given $St, Fr$. 
However, the imbalance does not vanish and remains significant at the highest $\Phi$ considered. 
The results also indicate that at least for low $Fr$, the dominance of the contribution from strain dominated regions might persist to higher $\Phi$ than we have considered.

In conclusion, our results strongly support the idea that preferential sweeping remains the key mechanism governing the enhanced settling velocity of inertial particles in turbulent flows with 2WC, even though it was originally presented as a mechanism in the context of 1WC flows by \citet{maxey1987}. 
This was demonstrated in \citet{tom2022} for low values of $\Phi$ where the global fluid statistics are weakly affected by the particles, and only the flow in the vicinity of the particles is strongly affected. 
We have now demonstrated that this conclusion also holds in 2WC flows where $\Phi$ is large enough for the particles to significantly affect the global fluid statistics.

In our DNS, a relatively small Taylor Reynolds number $Re_\lambda=O(10^2)$ was considered, a restriction imposed due to the computational expense of exploring a significant portion of the $St, Fr, \Phi$ parameter space. 
It is important in future work to consider 2WC flows with much larger $Re_\lambda$ where there is a large range of scales in the flow. 
The results could then be analyzed from the perspective of the multiscale preferential sweeping mechanism that was presented in \cite{tom2019}, which would provide a way to understand how flow scales of different sizes contribute to the enhanced settling velocity of inertial particles in turbulent flows. 
Such an attempt was done in \citet{tom2022}, however, that study also used $Re_\lambda=O(10^2)$ and so the range of scales in the flow was quite small. 
New DNS of 2WC flows with high $Re_\lambda$ would provide insight into how 2WC modifies the sweeping contribution from different scales in the flow which is important for atmospheric contexts, for example, where typically $Re_\lambda\geq O(10^4)$.

\backsection[Acknowledgements]
{ S.B. gratefully acknowledges Sarma L. Rani for their invaluable feedback and the insightful discussion on numerical techniques.
}

\backsection[Funding]
{
This work was supported by the National Aeronautics and Space Administration Weather and Atmospheric Dynamics program (grant number NASA 80NSSC20K0912). The computational resources used were provided by the Extreme Science and Engineering Discovery Environment (XSEDE) under allocation CTS170009, which is supported by National Science Foundation (NSF) grant number ACI-1548562 \citep{xsede}. Specifically, Stampede2 cluster operated by Texas Advanced Computing Center (TACC) and the Expanse cluster operated by San Diego Supercomputer Center (SDSC) were used to obtain the results in this work. The Duke Computing Cluster (DCC) operated by Duke University Research Computing was also used to obtain some of the preliminary results for this study.}

\backsection[Declaration of interests]
{ 
The authors report no conflict of interest.
}

\backsection[Author ORCID]{Soumak Bhattacharjee: https://orcid.org/0000-0002-3123-8973, Josin Tom: https://orcid.org/0000-0002-2717-089X, Maurizio Carbone: https://orcid.org/0000-0003-0409-6946, Andrew D.~Bragg: https://orcid.org/0000-0001-7068-8048.}

%\appendix
%\subfile{Appendicies/A01_timeaverage.tex}

\bibliographystyle{jfm}
\bibliography{main}
%Use of the above commands will create a bibliography using the .bib file. Shown below is a bibliography built from individual items.

%\bibliographystyle{jfm}
%\bibliography{jfm2esam}

\clearpage
\section*{Appendix}\label{sec:A}

% \begin{table}
% \centering
% \captionsetup{width=\linewidth}
% \begin{tabular}{c c c c c c c c c c c c} %\hline
% \begin{longtable}{ p{.14\textwidth}  p{.05\textwidth}  p{.05\textwidth}  p{.065\textwidth}  p{.065\textwidth}  p{.065\textwidth}  p{.065\textwidth}  p{.065\textwidth}  p{.065\textwidth}  p{.065\textwidth}  p{.065\textwidth}  p{.065\textwidth} } \hline \\
% \centering
\begin{longtblr}[
caption = {Flow parameters in DNS of 2WC simulations for different volume fraction $\Phi$, Stokes number $St$ and Froude number $Fr$. The Stokes number and Froude number are defined with respect to the unladen DNS statistics.},
  label = {tab:test},
]{
  colspec = {X[3]XXXXXXXXXXX},
  rowhead = 1 %, hlines
}
\hline
$\Phi$     & $St$       & $Fr$     & $\langle\epsilon\rangle$  & $u_{\eta}$    & $\eta$   &  $u'$&  $u'_x$&  $u'_y$&  $u'_z$   & $\tau_L$  & $Re_{\lambda}$ \\ \hline \\ %\hline
$1.5 \times 10^{-6}$ & $0.3$  & $0.3$   & $0.265$   & $0.190$   & $0.026$   & $0.916$   & $0.934$   & $0.866$   & $0.946$   & $3.243$   & $ 89.6$\\
$1.5 \times 10^{-6}$ & $0.3$  & $1.0$   & $0.256$   & $0.189$   & $0.026$   & $0.916$   & $0.940$   & $0.926$   & $0.881$   & $3.285$   & $ 91.2$\\
$1.5 \times 10^{-6}$ & $0.3$  & $3.0$   & $0.264$   & $0.190$   & $0.026$   & $0.916$   & $0.914$   & $0.888$   & $0.945$   & $3.230$   & $ 89.7$\\
$1.5 \times 10^{-6}$ & $0.5$  & $0.3$   & $0.268$   & $0.191$   & $0.026$   & $0.916$   & $0.917$   & $0.942$   & $0.888$   & $3.210$   & $ 88.9$\\
$1.5 \times 10^{-6}$ & $0.5$  & $1.0$   & $0.257$   & $0.189$   & $0.026$   & $0.916$   & $0.916$   & $0.914$   & $0.919$   & $3.283$   & $ 90.9$\\
$1.5 \times 10^{-6}$ & $0.5$  & $3.0$   & $0.267$   & $0.191$   & $0.026$   & $0.916$   & $1.018$   & $0.847$   & $0.874$   & $3.210$   & $ 89.2$\\
$1.5 \times 10^{-6}$ & $0.7$  & $0.3$   & $0.261$   & $0.190$   & $0.026$   & $0.916$   & $0.861$   & $0.960$   & $0.925$   & $3.267$   & $ 90.3$\\
$1.5 \times 10^{-6}$ & $0.7$  & $1.0$   & $0.262$   & $0.190$   & $0.026$   & $0.916$   & $0.883$   & $0.900$   & $0.963$   & $3.251$   & $ 90.0$\\
$1.5 \times 10^{-6}$ & $0.7$  & $3.0$   & $0.258$   & $0.189$   & $0.026$   & $0.916$   & $0.873$   & $0.932$   & $0.942$   & $3.270$   & $ 90.6$\\
$1.5 \times 10^{-6}$ & $1.0$  & $0.3$   & $0.259$   & $0.190$   & $0.026$   & $0.916$   & $0.966$   & $0.892$   & $0.888$   & $3.254$   & $ 90.5$\\
$1.5 \times 10^{-6}$ & $1.0$  & $1.0$   & $0.254$   & $0.188$   & $0.027$   & $0.916$   & $0.891$   & $0.937$   & $0.919$   & $3.287$   & $ 91.5$\\
$1.5 \times 10^{-6}$ & $1.0$  & $3.0$   & $0.262$   & $0.190$   & $0.026$   & $0.916$   & $0.923$   & $0.927$   & $0.897$   & $3.228$   & $ 90.0$\\
$1.5 \times 10^{-6}$ & $2.0$  & $0.3$   & $0.258$   & $0.189$   & $0.026$   & $0.916$   & $0.920$   & $0.897$   & $0.931$   & $3.224$   & $ 90.8$\\
$1.5 \times 10^{-6}$ & $2.0$  & $1.0$   & $0.254$   & $0.189$   & $0.027$   & $0.916$   & $0.906$   & $0.949$   & $0.892$   & $3.278$   & $ 91.4$\\
$1.5 \times 10^{-6}$ & $2.0$  & $3.0$   & $0.262$   & $0.190$   & $0.026$   & $0.916$   & $0.898$   & $0.903$   & $0.946$   & $3.218$   & $ 90.0$\\ \hline 
$7.0 \times 10^{-6}$ & $0.3$  & $0.3$   & $0.284$   & $0.194$   & $0.026$   & $0.916$   & $0.901$   & $0.891$   & $0.955$   & $3.188$   & $ 86.4$\\
$7.0 \times 10^{-6}$ & $0.3$  & $1.0$   & $0.270$   & $0.192$   & $0.026$   & $0.916$   & $0.885$   & $0.903$   & $0.959$   & $3.216$   & $ 88.6$\\
$7.0 \times 10^{-6}$ & $0.3$  & $3.0$   & $0.264$   & $0.190$   & $0.026$   & $0.916$   & $0.929$   & $0.909$   & $0.911$   & $3.234$   & $ 89.6$\\
$7.0 \times 10^{-6}$ & $0.5$  & $0.3$   & $0.285$   & $0.194$   & $0.026$   & $0.916$   & $0.930$   & $0.872$   & $0.944$   & $3.191$   & $ 86.1$\\
$7.0 \times 10^{-6}$ & $0.5$  & $1.0$   & $0.266$   & $0.191$   & $0.026$   & $0.916$   & $0.920$   & $0.898$   & $0.930$   & $3.222$   & $ 89.3$\\
$7.0 \times 10^{-6}$ & $0.5$  & $3.0$   & $0.259$   & $0.189$   & $0.026$   & $0.916$   & $0.889$   & $0.956$   & $0.902$   & $3.249$   & $ 90.6$\\
$7.0 \times 10^{-6}$ & $0.7$  & $0.3$   & $0.282$   & $0.194$   & $0.026$   & $0.916$   & $0.965$   & $0.878$   & $0.903$   & $3.180$   & $ 86.6$\\
$7.0 \times 10^{-6}$ & $0.7$  & $1.0$   & $0.257$   & $0.189$   & $0.026$   & $0.916$   & $0.937$   & $0.893$   & $0.918$   & $3.265$   & $ 90.9$\\
$7.0 \times 10^{-6}$ & $0.7$  & $3.0$   & $0.259$   & $0.189$   & $0.026$   & $0.916$   & $0.920$   & $0.918$   & $0.911$   & $3.265$   & $ 90.7$\\
$7.0 \times 10^{-6}$ & $1.0$  & $0.3$   & $0.266$   & $0.191$   & $0.026$   & $0.916$   & $0.968$   & $0.923$   & $0.854$   & $3.227$   & $ 89.4$\\
$7.0 \times 10^{-6}$ & $1.0$  & $1.0$   & $0.261$   & $0.190$   & $0.026$   & $0.916$   & $0.976$   & $0.892$   & $0.878$   & $3.213$   & $ 90.1$\\
$7.0 \times 10^{-6}$ & $1.0$  & $3.0$   & $0.255$   & $0.189$   & $0.026$   & $0.916$   & $0.883$   & $0.916$   & $0.948$   & $3.257$   & $ 91.1$\\
$7.0 \times 10^{-6}$ & $2.0$  & $0.3$   & $0.231$   & $0.184$   & $0.027$   & $0.916$   & $1.133$   & $0.784$   & $0.788$   & $3.375$   & $ 95.9$\\
$7.0 \times 10^{-6}$ & $2.0$  & $1.0$   & $0.251$   & $0.188$   & $0.027$   & $0.916$   & $0.923$   & $0.894$   & $0.931$   & $3.232$   & $ 92.0$\\
$7.0 \times 10^{-6}$ & $2.0$  & $3.0$   & $0.247$   & $0.187$   & $0.027$   & $0.916$   & $0.913$   & $0.908$   & $0.927$   & $3.258$   & $ 92.7$\\ \\ \hline \\ 
$1.5 \times 10^{-5}$ & $0.3$  & $0.3$   & $0.307$   & $0.198$   & $0.025$   & $0.916$   & $0.917$   & $0.957$   & $0.872$   & $3.196$   & $ 83.1$\\
$1.5 \times 10^{-5}$ & $0.3$  & $1.0$   & $0.271$   & $0.192$   & $0.026$   & $0.916$   & $0.932$   & $0.902$   & $0.914$   & $3.269$   & $ 88.5$\\
$1.5 \times 10^{-5}$ & $0.3$  & $3.0$   & $0.267$   & $0.191$   & $0.026$   & $0.916$   & $0.907$   & $0.931$   & $0.910$   & $3.250$   & $ 89.1$\\
$1.5 \times 10^{-5}$ & $0.5$  & $0.3$   & $0.317$   & $0.199$   & $0.025$   & $0.916$   & $1.010$   & $0.855$   & $0.875$   & $3.159$   & $ 81.7$\\
$1.5 \times 10^{-5}$ & $0.5$  & $1.0$   & $0.262$   & $0.190$   & $0.026$   & $0.916$   & $0.892$   & $0.931$   & $0.926$   & $3.276$   & $ 89.9$\\
$1.5 \times 10^{-5}$ & $0.5$  & $3.0$   & $0.257$   & $0.189$   & $0.026$   & $0.916$   & $0.937$   & $0.905$   & $0.906$   & $3.266$   & $ 90.9$\\
$1.5 \times 10^{-5}$ & $0.7$  & $0.3$   & $0.307$   & $0.198$   & $0.025$   & $0.916$   & $1.036$   & $0.839$   & $0.861$   & $3.197$   & $ 83.2$\\
$1.5 \times 10^{-5}$ & $0.7$  & $1.0$   & $0.264$   & $0.190$   & $0.026$   & $0.916$   & $0.914$   & $0.889$   & $0.945$   & $3.248$   & $ 89.6$\\
$1.5 \times 10^{-5}$ & $0.7$  & $3.0$   & $0.252$   & $0.188$   & $0.027$   & $0.916$   & $0.962$   & $0.914$   & $0.870$   & $3.289$   & $ 91.8$\\
$1.5 \times 10^{-5}$ & $1.0$  & $0.3$   & $0.289$   & $0.195$   & $0.026$   & $0.916$   & $1.186$   & $0.742$   & $0.749$   & $3.265$   & $ 85.6$\\
$1.5 \times 10^{-5}$ & $1.0$  & $1.0$   & $0.259$   & $0.190$   & $0.026$   & $0.916$   & $0.932$   & $0.915$   & $0.901$   & $3.243$   & $ 90.4$\\
$1.5 \times 10^{-5}$ & $1.0$  & $3.0$   & $0.249$   & $0.188$   & $0.027$   & $0.916$   & $0.919$   & $0.936$   & $0.893$   & $3.264$   & $ 92.3$\\
$1.5 \times 10^{-5}$ & $2.0$  & $0.3$   & $0.247$   & $0.187$   & $0.027$   & $0.916$   & $1.319$   & $0.625$   & $0.622$   & $3.416$   & $ 92.9$\\
$1.5 \times 10^{-5}$ & $2.0$  & $1.0$   & $0.232$   & $0.184$   & $0.027$   & $0.916$   & $0.963$   & $0.866$   & $0.917$   & $3.308$   & $ 95.7$\\
$1.5 \times 10^{-5}$ & $2.0$  & $3.0$   & $0.231$   & $0.184$   & $0.027$   & $0.916$   & $0.902$   & $0.931$   & $0.915$   & $3.321$   & $ 95.7$\\ \\ \hline \\ 
$7.0 \times 10^{-5}$ & $0.3$  & $0.3$   & $0.574$   & $0.231$   & $0.022$   & $0.916$   & $1.116$   & $0.812$   & $0.783$   & $3.065$   & $ 60.7$\\
$7.0 \times 10^{-5}$ & $0.3$  & $1.0$   & $0.318$   & $0.200$   & $0.025$   & $0.916$   & $0.948$   & $0.901$   & $0.898$   & $3.340$   & $ 81.6$\\
$7.0 \times 10^{-5}$ & $0.3$  & $3.0$   & $0.295$   & $0.196$   & $0.026$   & $0.916$   & $0.929$   & $0.851$   & $0.964$   & $3.319$   & $ 84.8$\\
$7.0 \times 10^{-5}$ & $0.5$  & $0.3$   & $0.628$   & $0.237$   & $0.021$   & $0.916$   & $1.199$   & $0.720$   & $0.750$   & $3.010$   & $ 58.0$\\
$7.0 \times 10^{-5}$ & $0.5$  & $1.0$   & $0.315$   & $0.199$   & $0.025$   & $0.916$   & $1.009$   & $0.824$   & $0.905$   & $3.363$   & $ 82.0$\\
$7.0 \times 10^{-5}$ & $0.5$  & $3.0$   & $0.267$   & $0.191$   & $0.026$   & $0.916$   & $0.928$   & $0.912$   & $0.908$   & $3.406$   & $ 89.1$\\
$7.0 \times 10^{-5}$ & $0.7$  & $0.3$   & $0.613$   & $0.235$   & $0.021$   & $0.916$   & $1.320$   & $0.628$   & $0.618$   & $3.007$   & $ 58.8$\\
$7.0 \times 10^{-5}$ & $0.7$  & $1.0$   & $0.307$   & $0.198$   & $0.025$   & $0.916$   & $1.022$   & $0.808$   & $0.905$   & $3.371$   & $ 83.1$\\
$7.0 \times 10^{-5}$ & $0.7$  & $3.0$   & $0.258$   & $0.189$   & $0.026$   & $0.916$   & $0.871$   & $0.925$   & $0.951$   & $3.409$   & $ 90.6$\\
$7.0 \times 10^{-5}$ & $1.0$  & $0.3$   & $0.553$   & $0.229$   & $0.022$   & $0.916$   & $1.429$   & $0.486$   & $0.491$   & $3.132$   & $ 61.9$\\
$7.0 \times 10^{-5}$ & $1.0$  & $1.0$   & $0.288$   & $0.195$   & $0.026$   & $0.916$   & $1.142$   & $0.775$   & $0.782$   & $3.407$   & $ 85.8$\\
$7.0 \times 10^{-5}$ & $1.0$  & $3.0$   & $0.230$   & $0.184$   & $0.027$   & $0.916$   & $0.960$   & $0.878$   & $0.908$   & $3.511$   & $ 95.9$\\
$7.0 \times 10^{-5}$ & $2.0$  & $0.3$   & $0.554$   & $0.229$   & $0.022$   & $0.916$   & $1.502$   & $0.363$   & $0.361$   & $2.417$   & $ 61.9$\\
$7.0 \times 10^{-5}$ & $2.0$  & $1.0$   & $0.227$   & $0.183$   & $0.027$   & $0.916$   & $1.228$   & $0.685$   & $0.735$   & $3.536$   & $ 96.7$\\
$7.0 \times 10^{-5}$ & $2.0$  & $3.0$   & $0.193$   & $0.176$   & $0.028$   & $0.916$   & $0.999$   & $0.899$   & $0.843$   & $3.529$   & $104.8$\\ \\ \hline \\ 
$1.5 \times 10^{-4}$ & $0.3$  & $0.3$   & $0.993$   & $0.265$   & $0.019$   & $0.916$   & $1.177$   & $0.755$   & $0.751$   & $2.846$   & $ 46.1$\\
$1.5 \times 10^{-4}$ & $0.3$  & $1.0$   & $0.418$   & $0.214$   & $0.023$   & $0.916$   & $0.897$   & $0.930$   & $0.921$   & $3.339$   & $ 71.2$\\
$1.5 \times 10^{-4}$ & $0.3$  & $3.0$   & $0.337$   & $0.202$   & $0.025$   & $0.916$   & $0.909$   & $0.919$   & $0.920$   & $3.434$   & $ 79.3$\\
$1.5 \times 10^{-4}$ & $0.5$  & $0.3$   & $1.052$   & $0.269$   & $0.019$   & $0.916$   & $1.370$   & $0.556$   & $0.576$   & $2.915$   & $ 44.8$\\
$1.5 \times 10^{-4}$ & $0.5$  & $1.0$   & $0.406$   & $0.212$   & $0.024$   & $0.916$   & $1.009$   & $0.847$   & $0.885$   & $3.466$   & $ 72.2$\\
$1.5 \times 10^{-4}$ & $0.5$  & $3.0$   & $0.315$   & $0.199$   & $0.025$   & $0.916$   & $0.881$   & $0.909$   & $0.955$   & $3.499$   & $ 82.0$\\
$1.5 \times 10^{-4}$ & $0.7$  & $0.3$   & $1.059$   & $0.270$   & $0.019$   & $0.916$   & $1.398$   & $0.529$   & $0.534$   & $2.796$   & $ 44.7$\\
$1.5 \times 10^{-4}$ & $0.7$  & $1.0$   & $0.400$   & $0.211$   & $0.024$   & $0.916$   & $1.135$   & $0.748$   & $0.818$   & $3.475$   & $ 72.7$\\
$1.5 \times 10^{-4}$ & $0.7$  & $3.0$   & $0.290$   & $0.195$   & $0.026$   & $0.916$   & $0.859$   & $0.957$   & $0.929$   & $3.555$   & $ 85.5$\\
$1.5 \times 10^{-4}$ & $1.0$  & $0.3$   & $1.010$   & $0.267$   & $0.019$   & $0.916$   & $1.460$   & $0.440$   & $0.440$   & $2.683$   & $ 45.8$\\
$1.5 \times 10^{-4}$ & $1.0$  & $1.0$   & $0.393$   & $0.210$   & $0.024$   & $0.916$   & $1.130$   & $0.756$   & $0.818$   & $3.376$   & $ 73.4$\\
$1.5 \times 10^{-4}$ & $1.0$  & $3.0$   & $0.259$   & $0.190$   & $0.026$   & $0.916$   & $0.909$   & $0.909$   & $0.930$   & $3.637$   & $ 90.4$\\
$1.5 \times 10^{-4}$ & $2.0$  & $1.0$   & $0.300$   & $0.197$   & $0.025$   & $0.916$   & $1.434$   & $0.484$   & $0.477$   & $3.589$   & $ 84.0$\\
$1.5 \times 10^{-4}$ & $2.0$  & $3.0$   & $0.200$   & $0.178$   & $0.028$   & $0.916$   & $0.916$   & $0.972$   & $0.856$   & $3.699$   & $103.0$ \\ \hline
\end{longtblr}

% -------------
\begin{figure}
\centering
\captionsetup{width=\linewidth}
\vspace{0mm}	
    % Figure
    % St(Fr) vs St, const. Phi
\subfloat	
    {\begin{overpic}
        [trim = 0mm 0mm -15mm 0mm,
        scale=0.65,clip,tics=20]{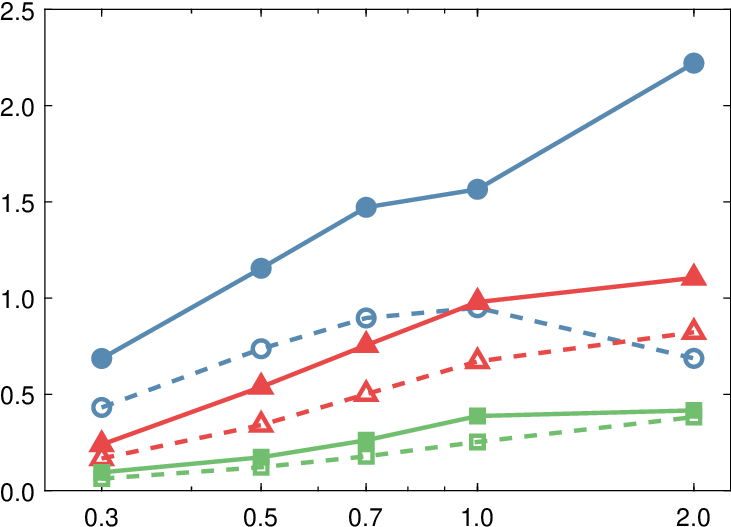}
        % \put(54,0){$St$}
        \put(-8,16){\rotatebox{90}{$-\langle u_z(\bm{x}^p(t),t)\rangle / u_{\eta}$}}
        \put(12,50){(a)} %\put(14.5,49.75){2WC}
    \end{overpic}}\\

\subfloat
    {\begin{overpic}
        [trim = 0mm 0mm -15mm 0mm,
        scale=0.65,clip,tics=20]{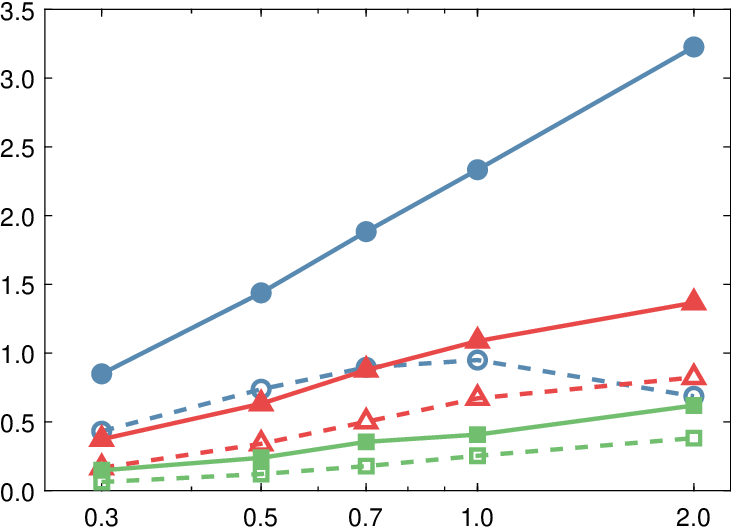}
        \put(-8,16){\rotatebox{90}{$-\langle u_z(\bm{x}^p(t),t)\rangle / u_{\eta}$}}
        \put(54,0){$St$}
        \put(12,50){(b)}
        \put(92,8){
            {\begin{overpic}
                [trim = 73mm 40mm 0mm 0mm,
                scale=0.6,clip,tics=20]{Figures/leg_sve_St_Fr.eps} 
                \put(12, 7.4){$Fr=3$,  2WC}
                \put(12,17.5){$Fr=1$,  2WC}
                \put(12,27.7){$Fr=0.3$, 2WC}
                \put(12,37.9){$Fr=3$,  1WC}
                \put(12,47.9){$Fr=1$,  1WC}
                \put(12,58){  $Fr=0.3$, 1WC}
            \end{overpic}} 
            }
    \end{overpic}} \\
\subfloat
    {\begin{overpic}
        [trim = 0mm -5mm -15mm 0mm,
        scale=0.65,clip,tics=20]{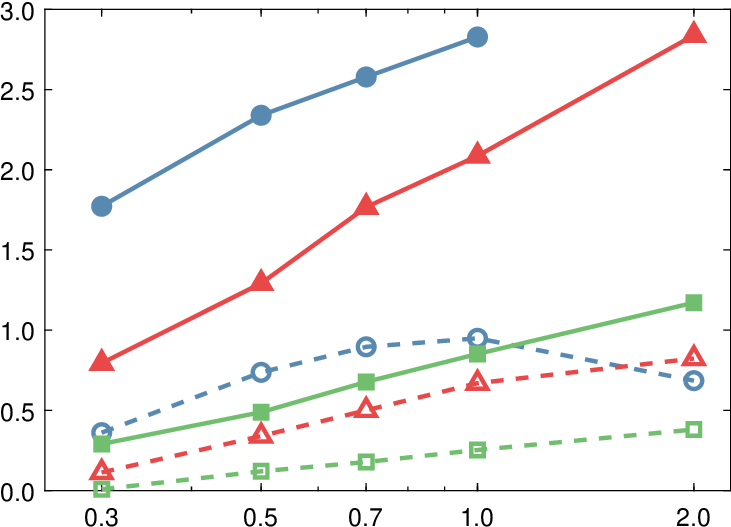}
        \put(-8,16){\rotatebox{90}{$-\langle u_z(\bm{x}^p(t),t)\rangle / u_{\eta}$}}
        \put(54,0){$St$}
        \put(12,50){(c)}
    \end{overpic}} 
    \caption{Normalized settling velocity enhancement for the 1WC (dashed) and 2WC (solid) cases for (a) $\Phi_2=7\times 10^{-6}$, (b) $\Phi_3=1.5\times 10^{-5}$ and (c) $\Phi_5=1.5\times 10^{-4}$, versus $St$. Figure \ref{fig:sve_vs_St} in the text shows the results for other volume fractions.}
    \label{fig:App_sve_vs_St}
\end{figure}
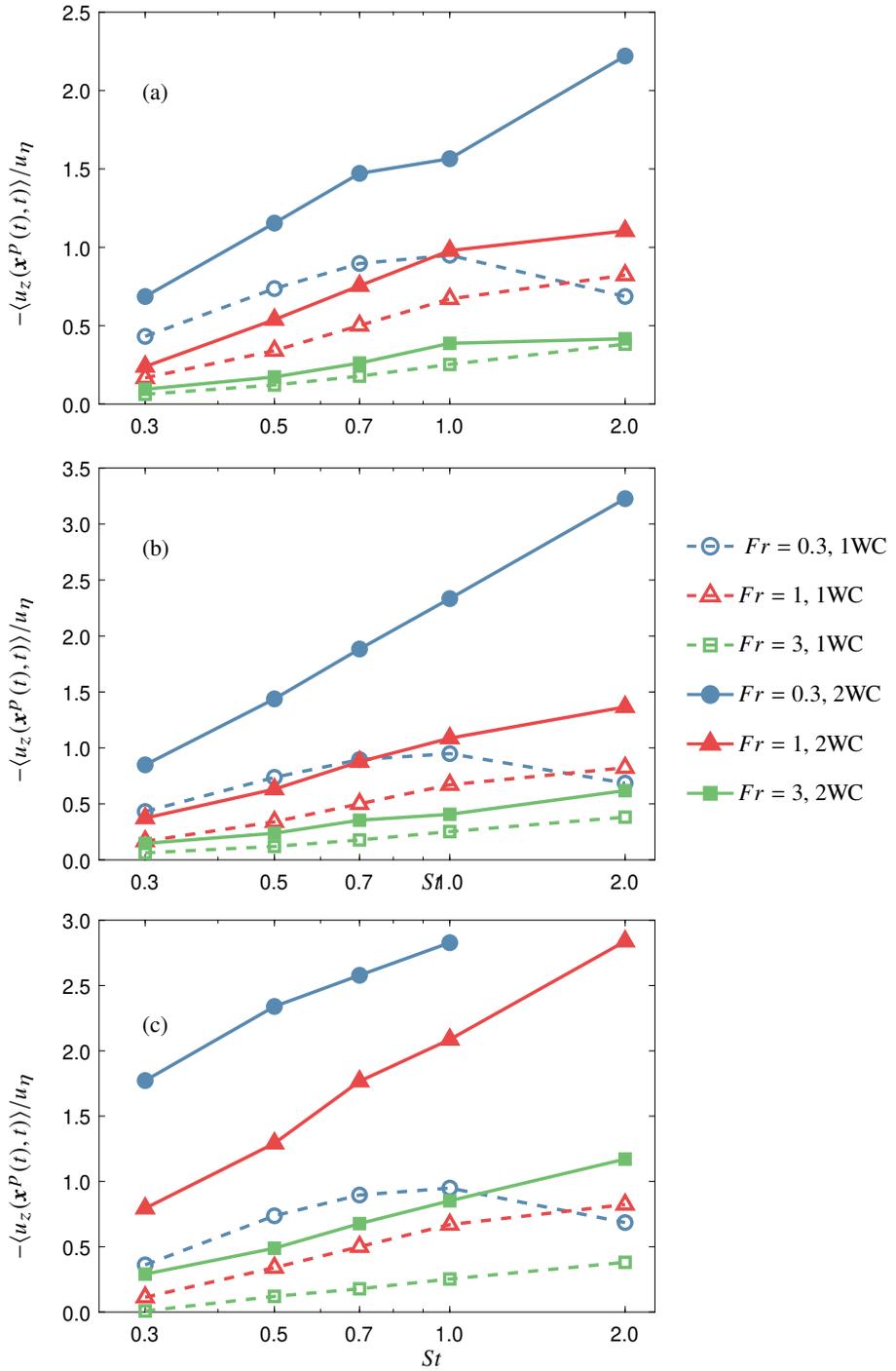
% -------------

%% Figure: effect of volume fraction
% 
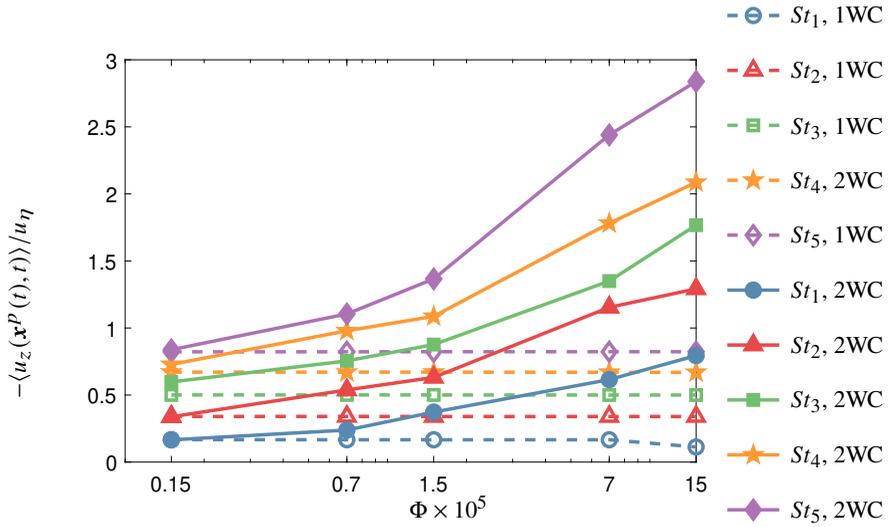
\begin{figure}
\centering
\captionsetup{width=\linewidth}
\vspace{0mm}	
% Figure
% St(Fr) vs St, const. Phi
\subfloat
{\begin{overpic}
    [trim = -10mm -5mm -15mm 0mm,
    scale=0.65,clip,tics=20]{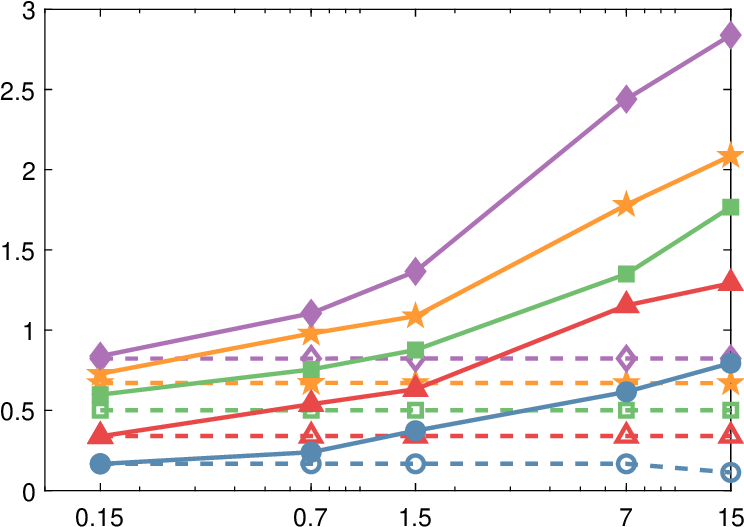}
    \put(50,0){$\Phi \times 10^{5}$}
    \put(-4,16){\rotatebox{90}{$-\langle u_z(\bm{x}^p(t),t)\rangle / u_{\eta}$}}
    % \put(80,58){(a)}
    % \put(14.5,57.75){1WC} \put(14.5,49.75){2WC}
        \put(91,-2){
    {\begin{overpic}
        [trim = 22mm 25mm 40mm 0mm,
        scale=0.65,clip,tics=20]{Figures/leg_sve_Phi_St.eps} 
        \put(11.5,02.4){$St_5$, 2WC} 
        \put(11.5,11.3){$St_4$, 2WC}
        \put(11.5,20.2){$St_3$, 2WC}
        \put(11.5,29){$St_2$, 2WC}
        \put(11.5,38){$St_1$, 2WC}
        \put(11.5,47){$St_5$, 1WC}
        \put(11.5,55.8){$St_4$, 2WC}
        \put(11.5,64.6){$St_3$, 1WC} 
        \put(11.5,73.7){$St_2$, 1WC} 
        \put(11.5,82.6){$St_1$, 1WC}
    \end{overpic}} 
    }
\end{overpic}}
\caption{Normalized settling velocity enhancement versus volume fraction $\Phi$, for the 1WC (dashed) and 2WC (solid) cases for $Fr=1$. Figure~\ref{fig:sve_vs_Phi} in the text shows the results for $Fr=0.3$ and $Fr=3$. }
\label{fig:App_sve_vs_Phi}
\end{figure}
% -------------

% ------------------------------------------------------------------------
% Lagrangian statistics P(Qp>0)-P(Qp<0)

% ------------------
% \begin{figure}
% \centering
% \captionsetup{width=\linewidth}
% \vspace{0mm}	
% % Figure
% % St(Fr) vs St, const. Phi
% \subfloat
% {\begin{overpic}
%     [trim = 0mm -5mm 0mm 50mm,
%     scale=0.7,clip,tics=20]{Figures/PQpos-PQneg.eps}
%     \put(-3,50){(a)}
%     \put(-3,22){(b)}
% \end{overpic}} 
% \caption{Bar graph showing $P(Q^p>0)-P(Q^p<0)$ along Lagrangian trajectories of particles in 1WC and 2WC simulations, for different Stokes number $St$ and volume fraction $\Phi$. Panels (a) and (b) correspond to simulations with $Fr=1.0$ and $Fr=3.0$, respectively. Data for $St=0.3$, $0.5$, $0.7$, $1.0$, and $2.0$ are shown in blue, orange, yellow, purple and green, respectively. Corresponding plot for $Fr=0.3$ is shown in Fig. \ref{fig:PQp-PQn} (a). }
% \label{fig:App_PQp-PQn}
% \end{figure}
% ------------------
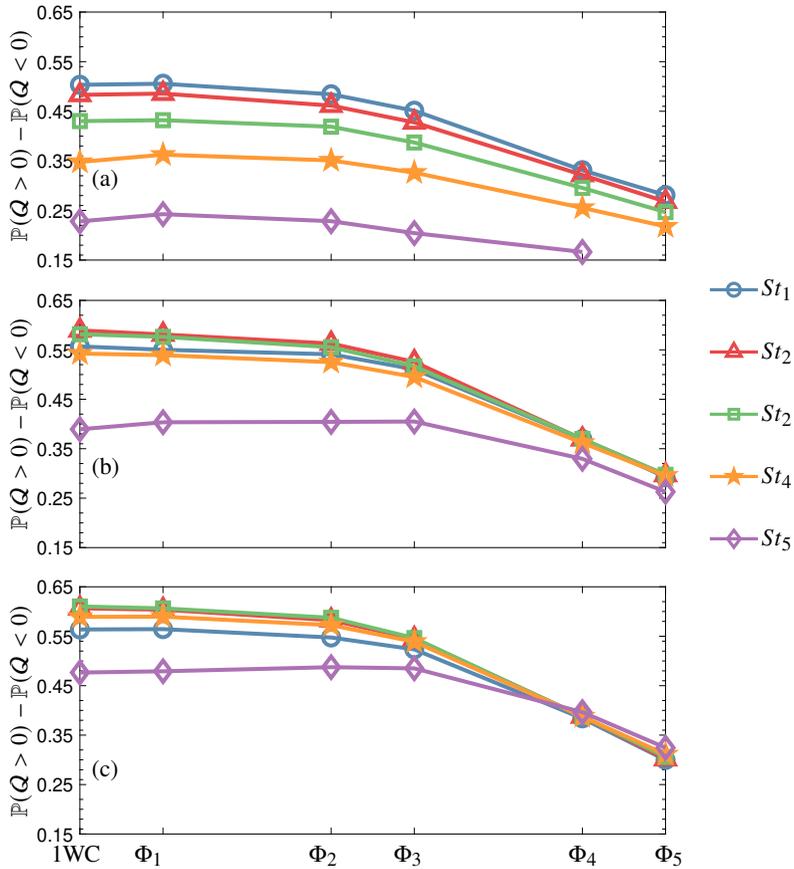
\begin{figure}
\centering
\captionsetup{width=\linewidth}
\vspace{0mm}	
% Figure
% St(Fr) vs St, const. Phi
\subfloat
{\begin{overpic}
    [trim = 0mm 0mm -10mm 0mm,
    scale=0.5,clip,tics=20]{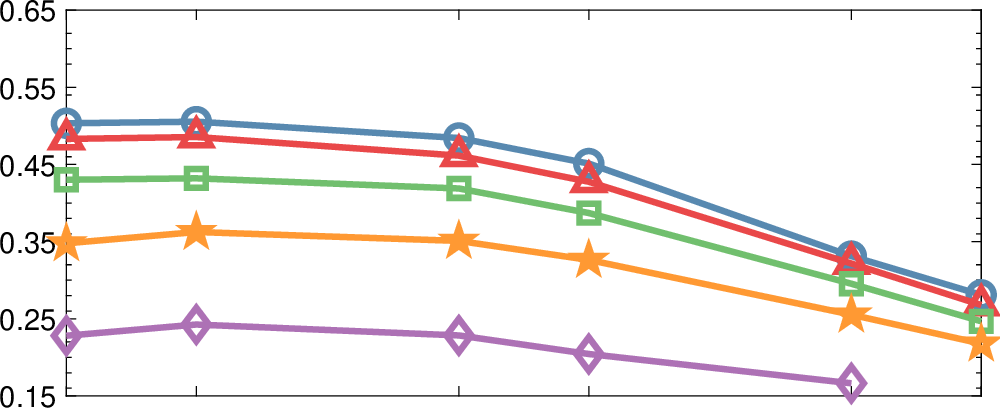} %PQpos-PQneg.eps
    \put(8,12){(a)}
    \put(-4,4){\rotatebox{90}{$\mathbb{P}(\mathcal{Q}>0)-\mathbb{P}(\mathcal{Q}<0)$}}
\end{overpic}} \\
\subfloat
{\begin{overpic}
    [trim = 0mm 0mm -10mm 0mm,
    scale=0.5,clip,tics=20]{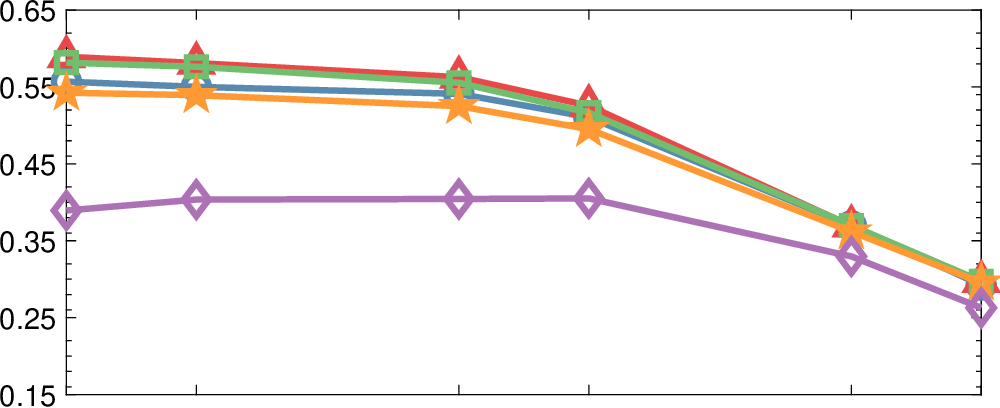}
    \put(8,12){(b)}
    \put(-4,4){\rotatebox{90}{$\mathbb{P}(\mathcal{Q}>0)-\mathbb{P}(\mathcal{Q}<0)$}}
\end{overpic}} \\
\subfloat
{\begin{overpic}
    [trim = 0mm -4mm -10mm 0mm,
    scale=0.5,clip,tics=20]{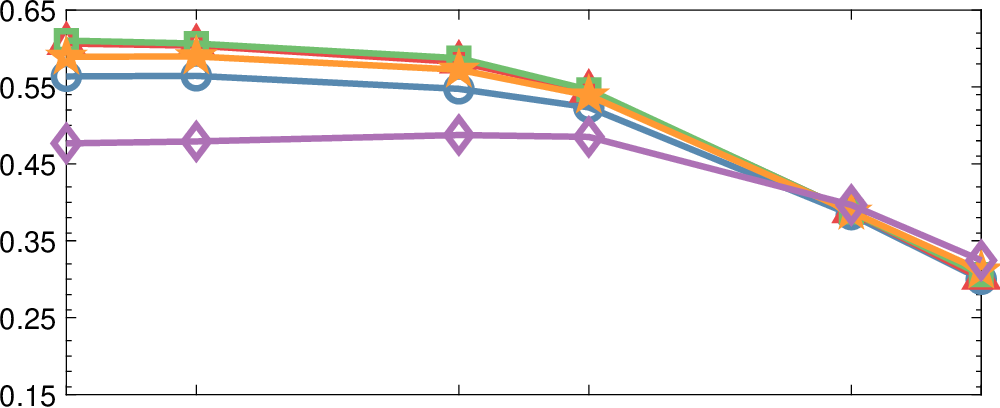}
        \put(8,12){(c)}
        \put(-4,6){\rotatebox{90}{$\mathbb{P}(\mathcal{Q}>0)-\mathbb{P}(\mathcal{Q}<0)$}}
        \put(2,-0.5){1WC} \put(14.3,-0.5){$\Phi_1$} \put(40,-0.5){$\Phi_2$} \put(52.55,-0.5){$\Phi_3$} \put(78.3,-0.5){$\Phi_4$} \put(91,-0.5){$\Phi_5$}
        \put(96,40){
        {\begin{overpic}
            [trim = 20mm 110mm 60mm 20mm,
            scale=0.6,clip,tics=20]{Figures/leg_LagEul_rat_PQd.eps}
            \put(21,12.5){$St_5$} 
            \put(21,32.0){$St_4$} 
            \put(21,51.5){$St_2$} 
            \put(21,71.0){$St_2$}
            \put(21,91.0){$St_1$}
        \end{overpic}} 
        }
\end{overpic}} 
\caption{Plots showing $\mathbb{P}(\mathcal{Q}>0)-\mathbb{P}(\mathcal{Q}<0)$ versus volume fraction $\Phi$ for particles with different Stokes number $St$, for simulations with (a) $Fr=0.3$, (b) $Fr=1$, and (c) $Fr=3$.
The corresponding values for the 1WC simulations are also shown for reference.}
\label{fig:App_PQp_pos-PQp_neg}
\end{figure}

% ------------------------------------------------------------------------

% Eulerian statistics P(Q>0)-P(Q<0)
% -------------
% \begin{figure}
% \centering
% \captionsetup{width=\linewidth}
% \vspace{0mm}	
% % Figure
% % St(Fr) vs St, const. Phi
% \subfloat
% {\begin{overpic}
%     [trim = 0mm -5mm 0mm 50mm,
%     scale=0.7,clip,tics=20]{Figures/EQp_Qn_diff.eps}
%     \put(-3,50){(a)}
%     \put(-3,22){(b)}
% \end{overpic}} 
% \caption{Bar graph showing $P(Q>0)-P(Q<0)$ (where $P(Q)$ denotes the Eulerian probability) in 1WC and 2WC simulations loaded with particles of different Stokes number $St$ and volume fraction $\Phi$. Panels (a) and (b) and (c) correspond to simulations with $Fr=1.0$ and $Fr=3.0$, respectively. Data for $St=0.3$, $0.5$, $0.7$, $1.0$, and $2.0$ are shown in blue, orange, yellow, purple and green, respectively. Corresponding plot for $Fr=0.3$ is shown in Fig. \ref{fig:PQp-PQn} (b).}
% \label{fig:App_EQp_Qn_diff}
% \end{figure}
% -------------
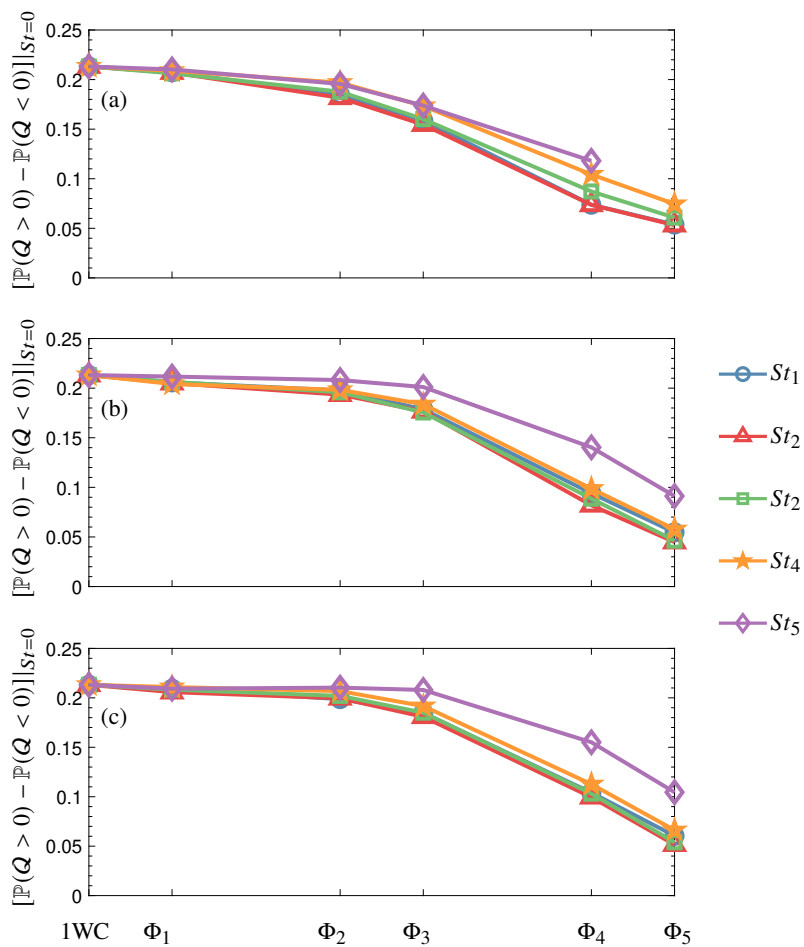
\begin{figure}
\centering
\captionsetup{width=\linewidth}
% Figure
% St(Fr) vs St, const. Phi
\subfloat
{\begin{overpic}
    [trim = 0mm -5mm -10mm 0mm,
    scale=0.5,clip,tics=20]{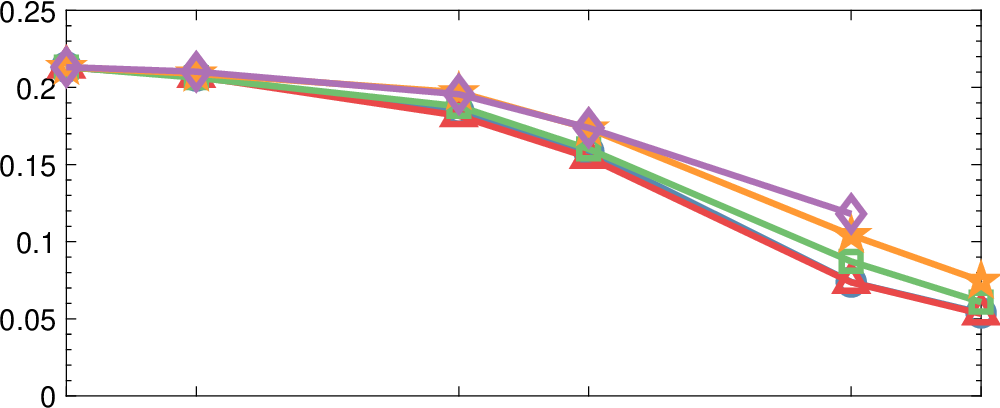} %PQpos-PQneg.eps
    \put(8,29){(a)}
    \put(-5,2){\rotatebox{90}{$[\mathbb{P}(\mathcal{Q}>0)-\mathbb{P}(\mathcal{Q}<0)]\vert_{St=0}$}}
\end{overpic}} \\
\subfloat
{\begin{overpic}
    [trim = 0mm -5mm -10mm 0mm,
    scale=0.5,clip,tics=20]{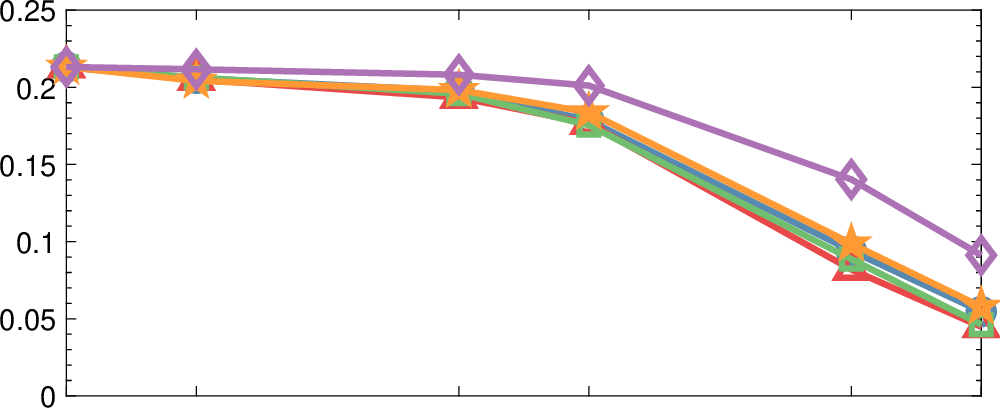}
    \put(8,29){(b)}
    \put(-5,2){\rotatebox{90}{\small{$[\mathbb{P}(\mathcal{Q}>0)-\mathbb{P}(\mathcal{Q}<0)]\vert_{St=0}$}}}
\end{overpic}} \\
\subfloat
{\begin{overpic}
    [trim = 0mm -10mm -10mm 0mm,
    scale=0.5,clip,tics=20]{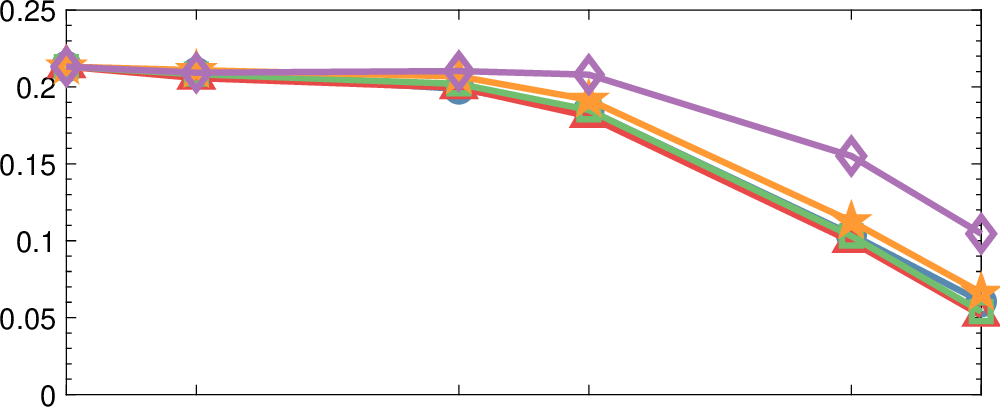}
        \put(8,32){(c)}
        \put(-5,5){\rotatebox{90}{$[\mathbb{P}(\mathcal{Q}>0)-\mathbb{P}(\mathcal{Q}<0)]\vert_{St=0}$}}
        \put(2,0.5){1WC} \put(14.3,0.5){$\Phi_1$} \put(40,0.5){$\Phi_2$} \put(52.55,0.5){$\Phi_3$} \put(78.3,0.5){$\Phi_4$} \put(91,0.5){$\Phi_5$}
        \put(96,40){
        {\begin{overpic}
            [trim = 20mm 110mm 60mm 20mm,
            scale=0.6,clip,tics=20]{Figures/leg_LagEul_rat_PQd.eps}
            \put(21,12.5){$St_5$} 
            \put(21,32.0){$St_4$} 
            \put(21,51.5){$St_2$} 
            \put(21,71.0){$St_2$}
            \put(21,91.0){$St_1$}
        \end{overpic}} 
        }
\end{overpic}} 
\caption{Plots showing $[\mathbb{P}(\mathcal{Q}>0)-\mathbb{P}(\mathcal{Q}<0)]\vert_{St=0}$ versus volume fraction $\Phi$ for simulations with (a) $Fr=0.3$, (b) $Fr=1$, and (c) $Fr=3$.
The corresponding values for the 1WC simulations are also shown for reference.}
\label{fig:App_PQe_pos-PQe_neg}
\end{figure}
% ------------------------------------------------------------------------

% -------------
% A,B vs Phi
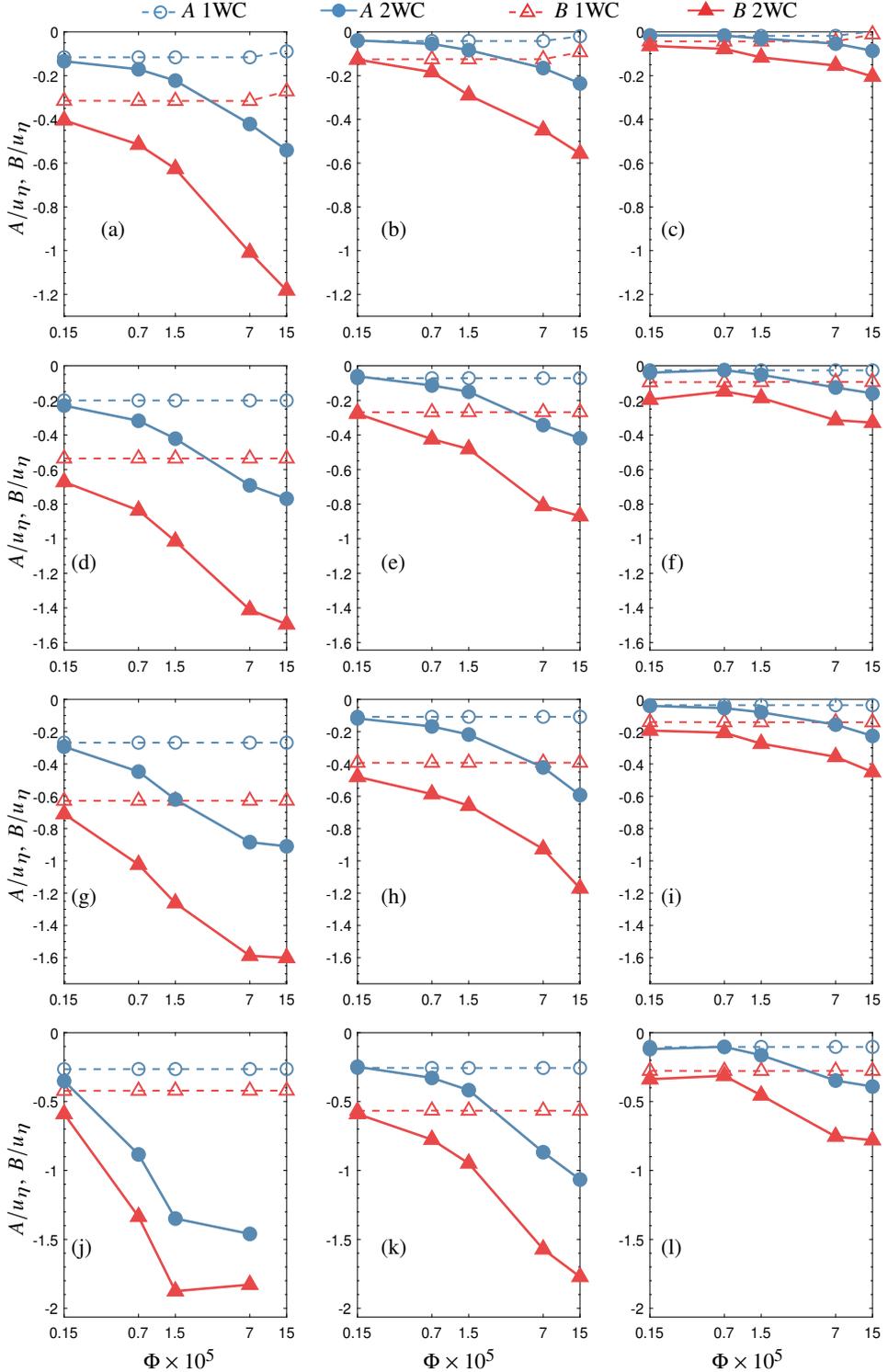
\begin{figure}
\centering
\captionsetup{width=\linewidth}
\subfloat
{\begin{overpic}
    [trim = 0mm 0mm 0mm -15mm,
    scale=0.5,clip,tics=20]{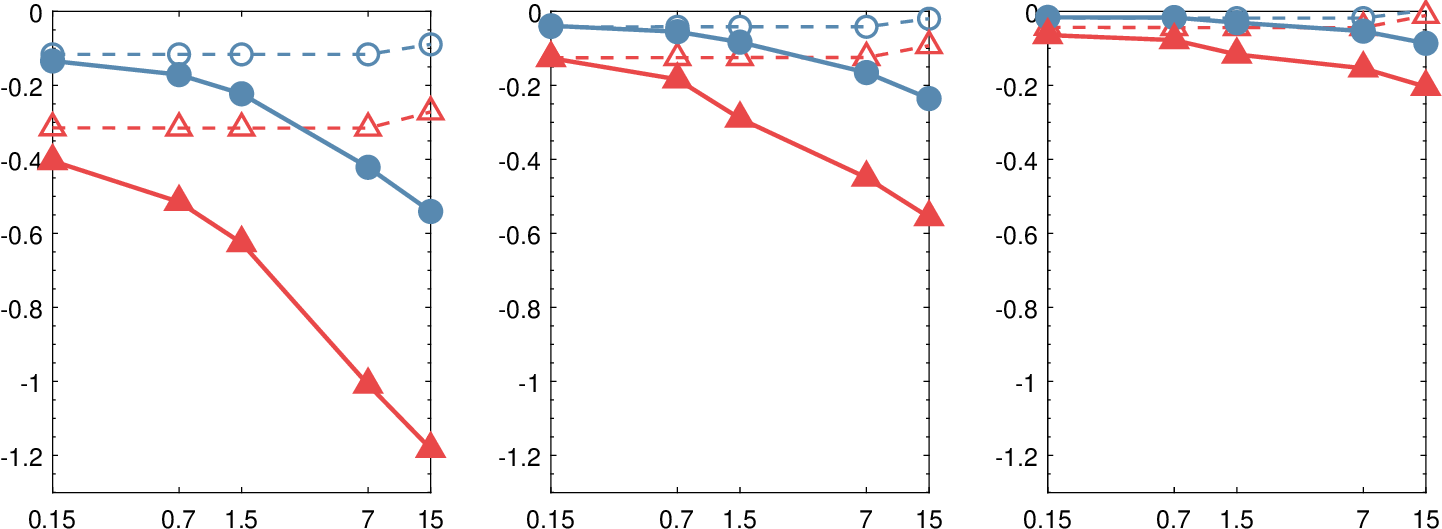}
    \put(8,12){(a)} \put(41,12){(b)} \put(74,12){(c)}
    % \put(82, 7){B 2WC}  % \put(82,10){B 1WC} % \put(82,13){A 2WC}% \put(82,16){A 1WC}
    \put(-3,12){\rotatebox{90}{$A/u_\eta,\, B/u_\eta$}}
    % 1st legend
        \put(54,37){
            \begin{overpic}
            [trim = 1mm 27mm 240mm 0mm,
            scale=0.15,clip,tics=20]{Figures/sve_Stdep_Leg.eps}
            \put(50, 5){{$B$ 1WC}}
        \end{overpic}}
    % 2nd legend
        \put(75,37){
            \begin{overpic}
            [trim = 1mm 1mm 240mm 25mm,
            scale=0.15,clip,tics=20]{Figures/sve_Stdep_Leg.eps}
            \put(50, 5){{$B$ 2WC}}
        \end{overpic}}  
    % 3rd legend
        \put(12,37){
            \begin{overpic}
            [trim = 120mm 27mm 120mm 0mm,
            scale=0.15,clip,tics=20]{Figures/sve_Stdep_Leg.eps}
            \put(35,5){{$A$ 1WC}}
        \end{overpic}}  
    % 4th legend
        \put(33,37){
            \begin{overpic}
            [trim = 120mm 1mm 120mm 25mm,
            scale=0.15,clip,tics=20]{Figures/sve_Stdep_Leg.eps}
            \put(35, 5){{$A$ 2WC}}
        \end{overpic}}  
\end{overpic}} 

\subfloat
{\begin{overpic}
    [trim = 0mm 0mm 0mm 0mm,
    scale=0.5,clip,tics=20]{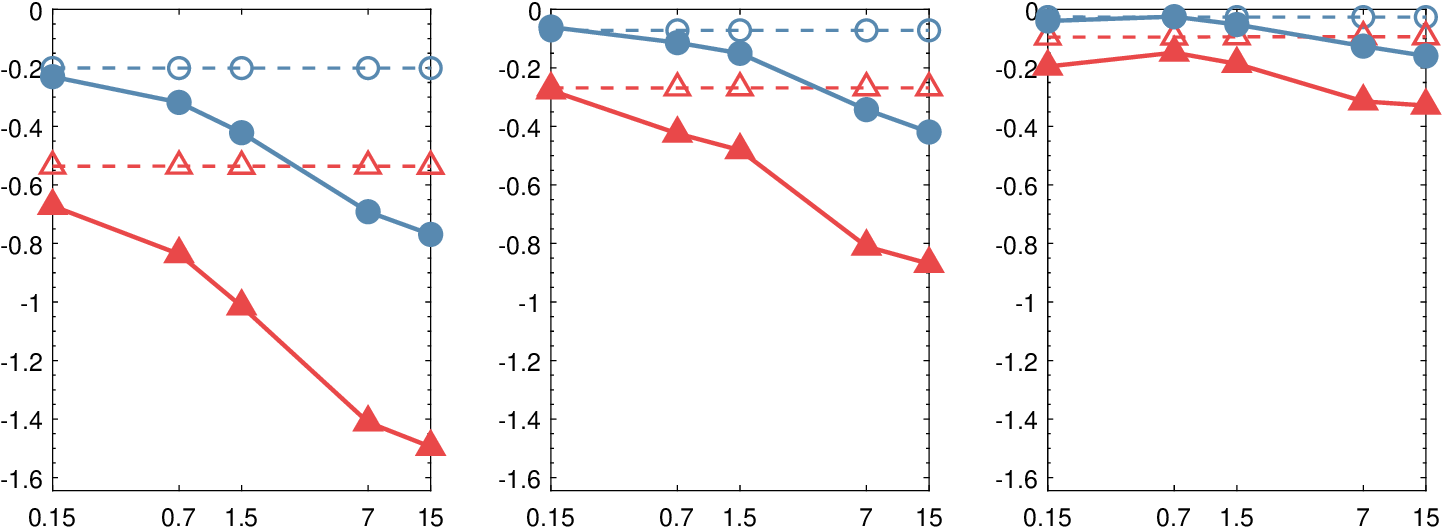}
    \put(-3,12){\rotatebox{90}{$A/u_\eta,\, B/u_\eta$}}
    \put(4.5,12){(d)} \put(41,12){(e)} \put(74,12){(f)}
    % \put(82, 7){B 2WC}  % \put(82,10){B 1WC} % \put(82,13){A 2WC}% \put(82,16){A 1WC}
\end{overpic}} 

\subfloat
{\begin{overpic}
    [trim = 0mm 0mm 0mm 0mm,
    scale=0.5,clip,tics=20]{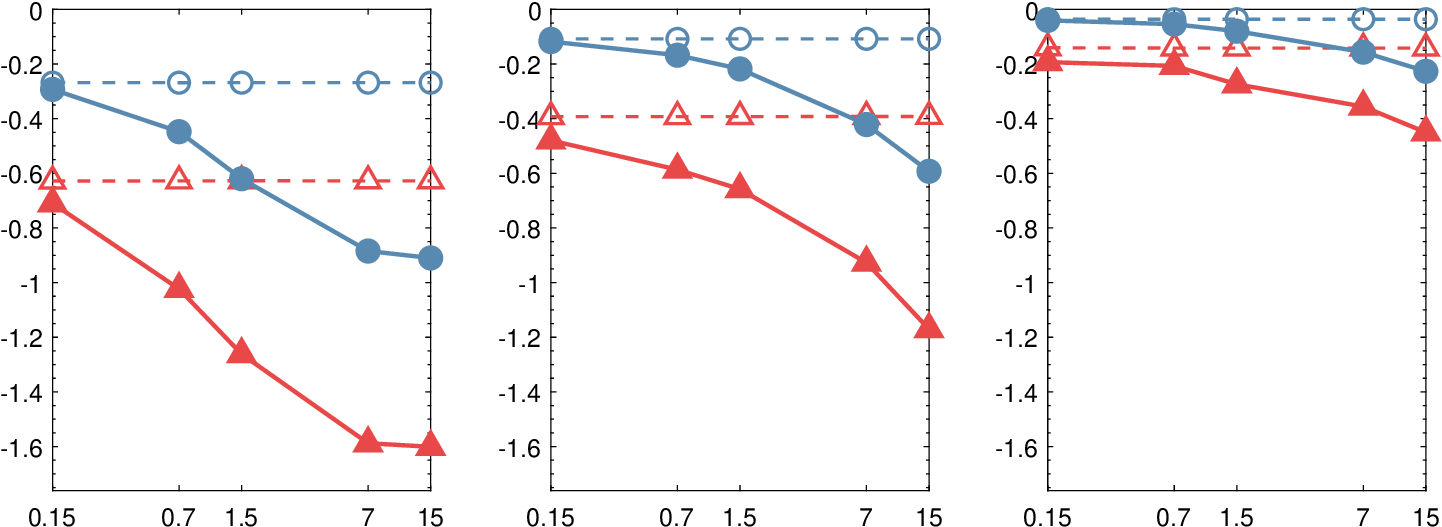}
    \put(-3,12){\rotatebox{90}{$A/u_\eta,\, B/u_\eta$}}
    \put(4.5,12){(g)} \put(41,12){(h)} \put(74,12){(i)}
    % \put(82, 7){B 2WC}  % \put(82,10){B 1WC} % \put(82,13){A 2WC}% \put(82,16){A 1WC}
\end{overpic}} 

\subfloat
{\begin{overpic}
    [trim = 0mm -5mm 0mm 0mm,
    scale=0.5,clip,tics=20]{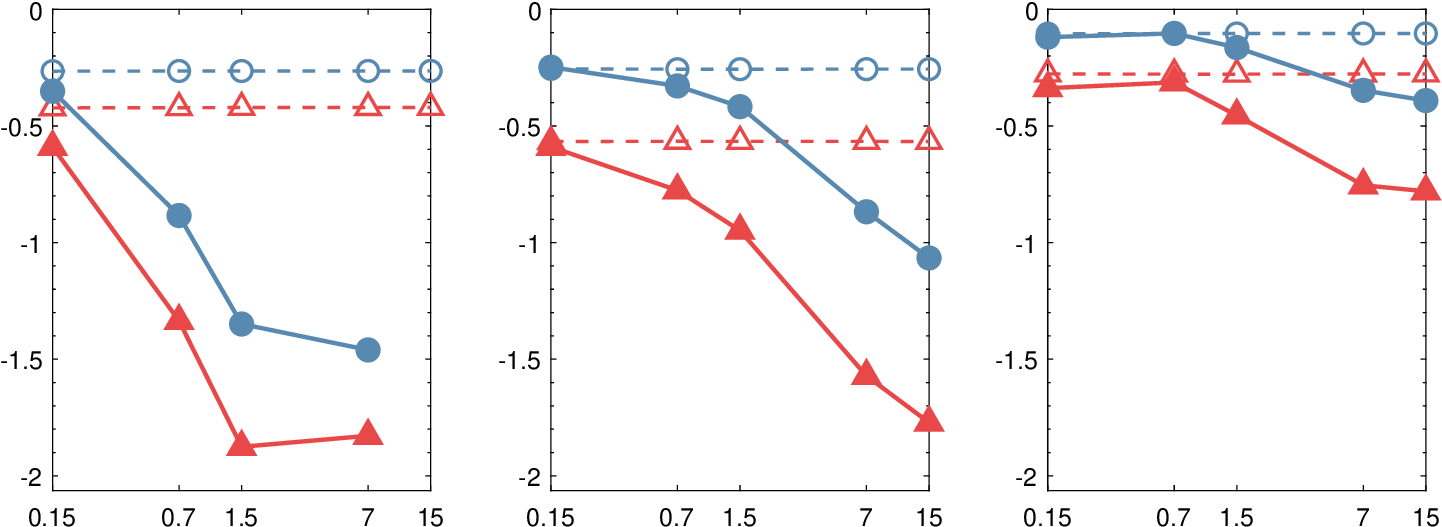}
    \put(-3,14){\rotatebox{90}{$A/u_\eta,\, B/u_\eta$}}
    \put(4.5,12){(j)} \put(41,12){(k)} \put(74,12){(l)}
    \put(13,-1.5){$\Phi\times 10^5$} \put(46,-1.5){$\Phi\times 10^5$} \put(80,-1.5){$\Phi\times 10^5$} 
    % \put(82, 7){B 2WC}  % \put(82,10){B 1WC} % \put(82,13){A 2WC}% \put(82,16){A 1WC}
\end{overpic}} 

\caption{Plots of $A$ (blue) and $B$ (red) versus volume fraction $\Phi$, for 1WC (dashed) and 2WC (solid) flows for particles with (a, b, c) $\rm{St}_1=0.3$, (d, e, f) $\rm{St}_2=0.5$, (g, h, i) $\rm{St}_3=0.7$, and (j, k, l) $\rm{St}_5=2$.
Subplots (a), (d), (g) and (j) correspond to simulations with $Fr=0.3$; 
(b), (e), (h) and (k) correspond to simulations with $Fr=1$;
(c), (f), (i) and (j) correspond to simulations with $Fr=3$.
See Fig.~\ref{fig:AB_plot} for the volume fraction dependence in simulations with $\rm{St}_4$ particles.
\label{fig:App_AB_plot_vs_Phi}}
\end{figure}

% -------------

% -------------
% A,B vs St
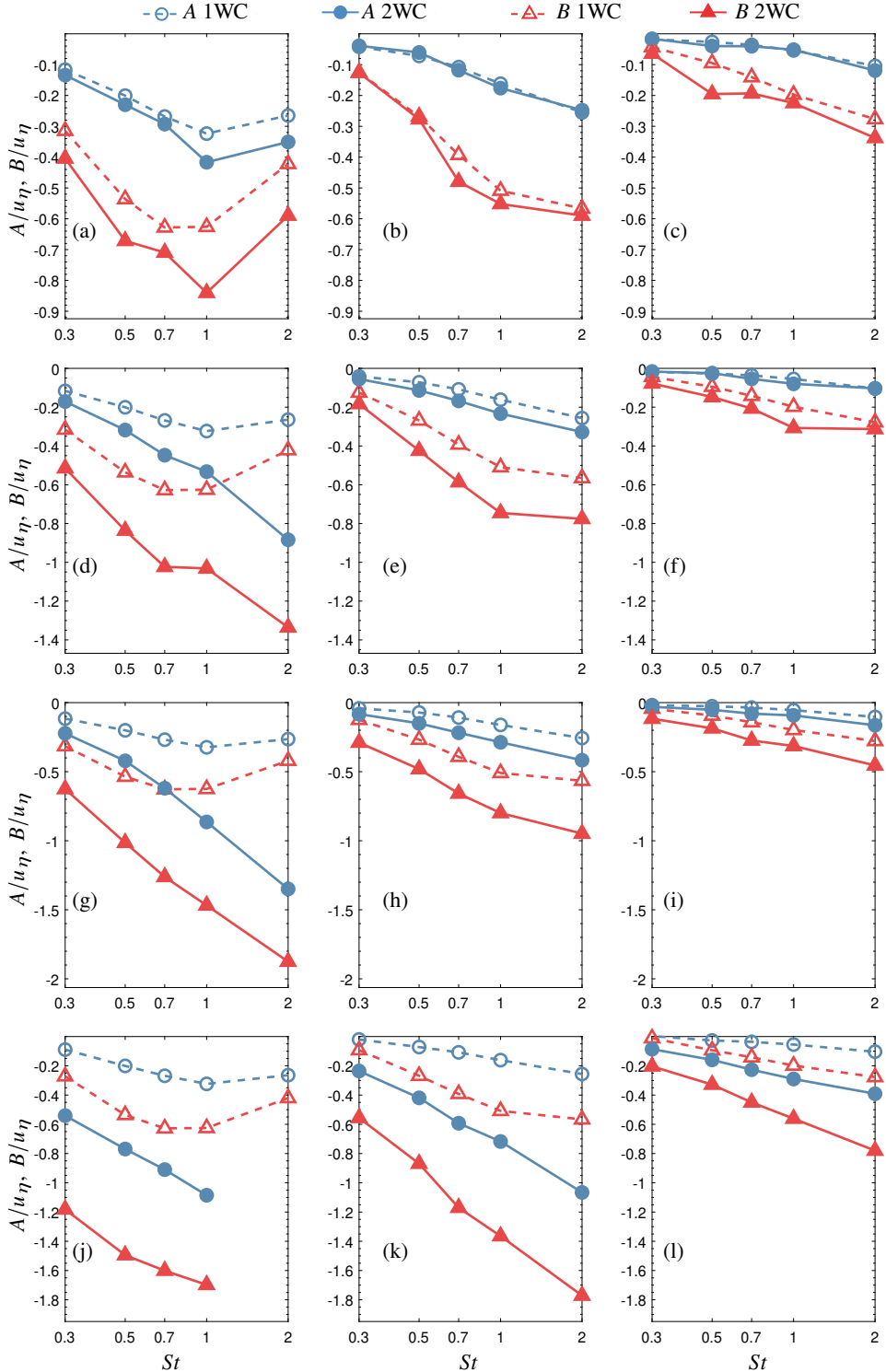
\begin{figure}
\centering
\captionsetup{width=\linewidth}
\subfloat
{\begin{overpic}
    [trim = 0mm 0mm 0mm -15mm,
    scale=0.5,clip,tics=20]{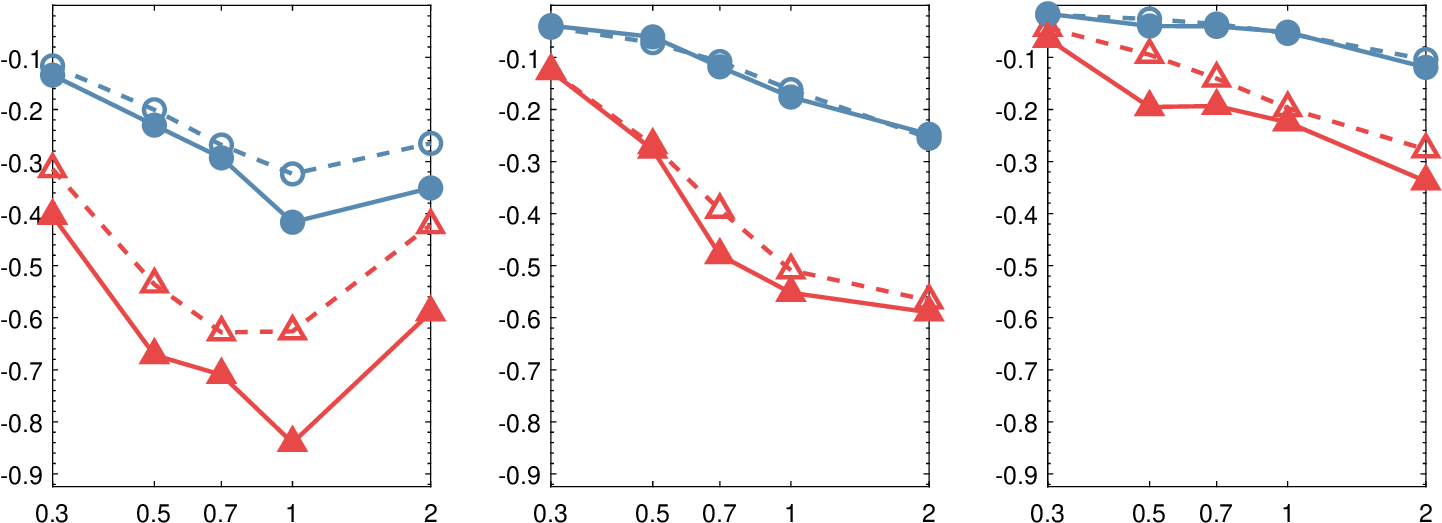}
    \put(4.5,12){(a)} \put(41,12){(b)} \put(74,12){(c)}
    % \put(82, 7){B 2WC}  % \put(82,10){B 1WC} % \put(82,13){A 2WC}% \put(82,16){A 1WC}
    \put(-3,12){\rotatebox{90}{$A/u_\eta,\, B/u_\eta$}}
    % 1st legend
    \put(54,37){
        \begin{overpic}
        [trim = 1mm 27mm 240mm 0mm,
        scale=0.15,clip,tics=20]{Figures/sve_Stdep_Leg.eps}
        \put(50, 5){{$B$ 1WC}}
    \end{overpic}}
    % 2nd legend
    \put(75,37){
        \begin{overpic}
        [trim = 1mm 1mm 240mm 25mm,
        scale=0.15,clip,tics=20]{Figures/sve_Stdep_Leg.eps}
        \put(50, 5){{$B$ 2WC}}
    \end{overpic}}  
    % 3rd legend
    \put(12,37){
        \begin{overpic}
        [trim = 120mm 27mm 120mm 0mm,
        scale=0.15,clip,tics=20]{Figures/sve_Stdep_Leg.eps}
        \put(35,5){{$A$ 1WC}}
    \end{overpic}}  
    % 4th legend
    \put(33,37){
        \begin{overpic}
        [trim = 120mm 1mm 120mm 25mm,
        scale=0.15,clip,tics=20]{Figures/sve_Stdep_Leg.eps}
        \put(35, 5){{$A$ 2WC}}
    \end{overpic}}  
\end{overpic}} 

\subfloat
{\begin{overpic}
    [trim = 0mm 0mm 0mm 0mm,
    scale=0.5,clip,tics=20]{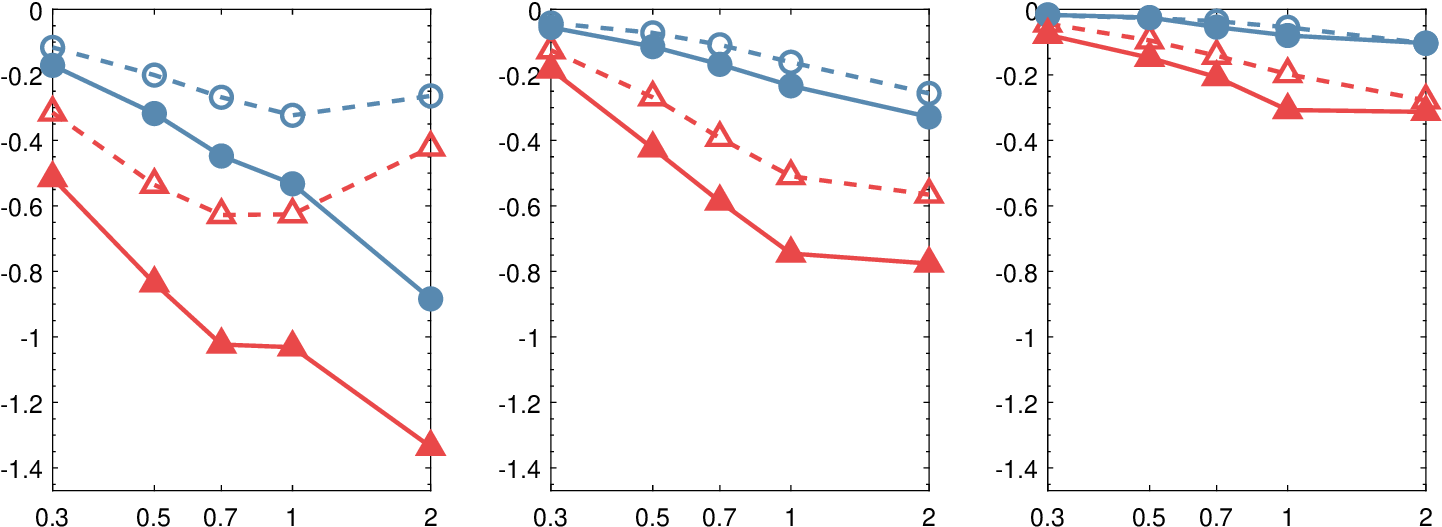}
    \put(-3,12){\rotatebox{90}{$A/u_\eta,\, B/u_\eta$}}
    \put(4.5,12){(d)} \put(41,12){(e)} \put(74,12){(f)}
    % \put(82, 7){B 2WC}  % \put(82,10){B 1WC} % \put(82,13){A 2WC}% \put(82,16){A 1WC}
\end{overpic}} 

\subfloat
{\begin{overpic}
    [trim = 0mm 0mm 0mm 0mm,
    scale=0.5,clip,tics=20]{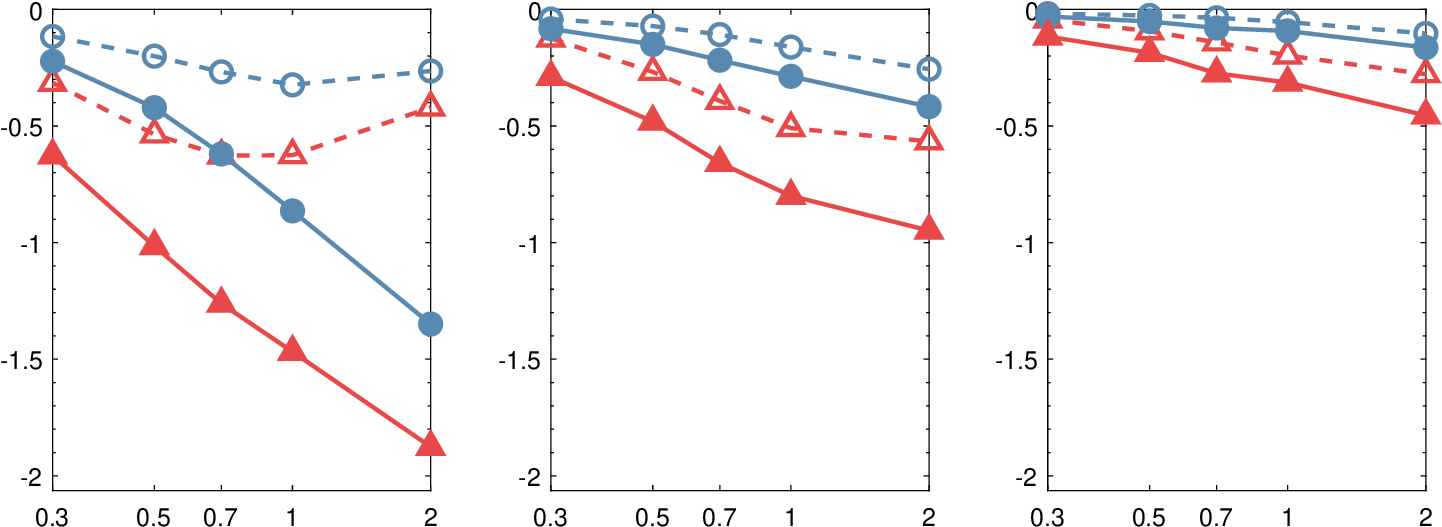}
    \put(-3,12){\rotatebox{90}{$A/u_\eta,\, B/u_\eta$}}
    \put(4.5,12){(g)} \put(41,12){(h)} \put(74,12){(i)}
    % \put(82, 7){B 2WC}  % \put(82,10){B 1WC} % \put(82,13){A 2WC}% \put(82,16){A 1WC}
\end{overpic}} 

\subfloat
{\begin{overpic}
    [trim = 0mm -5mm 0mm 0mm,
    scale=0.5,clip,tics=20]{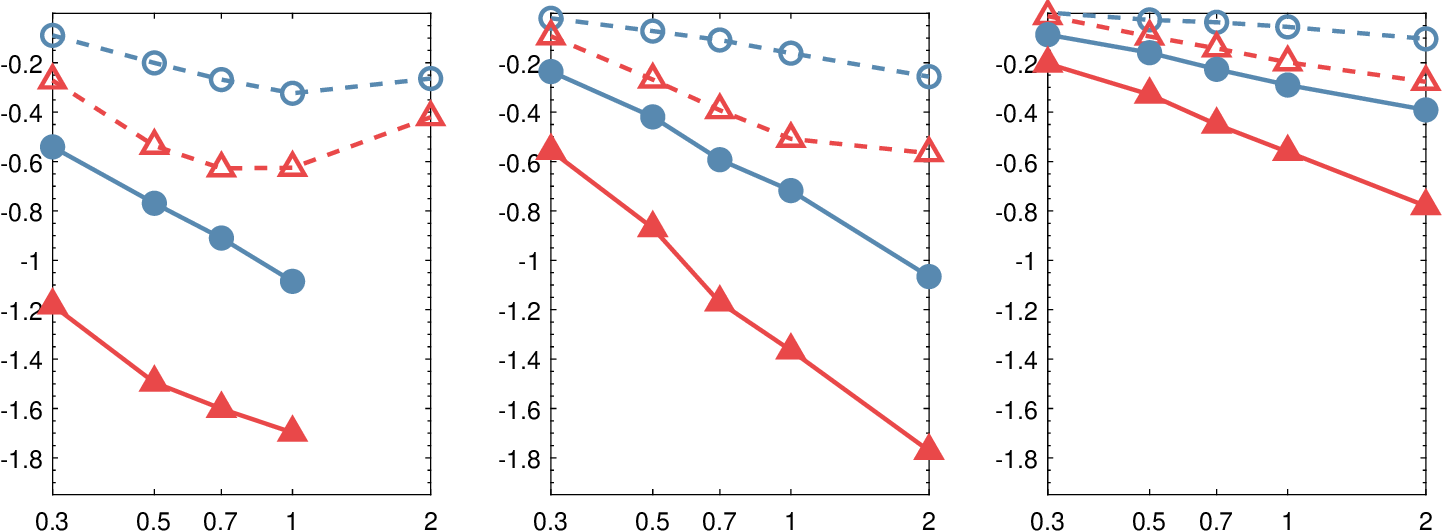}
    \put(-3,15){\rotatebox{90}{$A/u_\eta,\, B/u_\eta$}}
    \put(4.5,12){(j)} \put(41,12){(k)} \put(74,12){(l)}
    \put(15,-1){$St$} \put(50,-1){$St$} \put(84,-1){$St$} 
    % \put(82, 7){B 2WC}  % \put(82,10){B 1WC} % \put(82,13){A 2WC}% \put(82,16){A 1WC}
\end{overpic}} 

\caption{Plots of $A$ (blue) and $B$ (red) versus Stokes number $St$, for 1WC (dashed) and 2WC (solid) flows for volume fractions (a, b, c) $\Phi_1$, (d, e, f) for $\Phi_2$, (g, h, i) $\Phi_3$, and (j, k, l) $\Phi_5$.
Subplots (a), (d), (g) and (j) correspond to simulations with $Fr=0.3$; 
(b), (e), (h) and (k) correspond to simulations with $Fr=1$; 
and (c), (f), (i) and (j) correspond to simulations with $Fr=3$, respectively. 
See Fig.~\ref{fig:AB_plot} for the $St$-dependence in simulations with volume fraction $\Phi_4$.
\label{fig:App_AB_plot_vs_St}}
\end{figure}

% -------------

\end{document}